\documentclass[aps,prd,showpacs,amssymb]{revtex4}
\usepackage{graphicx} 
\usepackage{bm}
\newcommand{\beq}{\begin{equation}}
\newcommand{\eeq}{\end{equation}}
\newcommand{\ba}{\begin{array}}
\newcommand{\ea}{\end{array}}
\newcommand{\lsim}   {\mathrel{\mathop{\kern 0pt \rlap
  {\raise.2ex\hbox{$<$}}}
  \lower.9ex\hbox{\kern-.190em $\sim$}}}
\newcommand{\gsim}   {\mathrel{\mathop{\kern 0pt \rlap
  {\raise.2ex\hbox{$>$}}}
\lower.9ex\hbox{\kern-.190em $\sim$}}}
\newcommand{\etal}{\emph{et al.}}

\begin{document}
\title{On astrophysical solution to ultra high energy cosmic rays}
\author{Veniamin Berezinsky}
 \affiliation{INFN, Laboratori Nazionali del Gran Sasso, I-67010
  Assergi (AQ), Italy\\
and  Institute for Nuclear Research of the RAS, Moscow, Russia}
 \author{Askhat Gazizov}
\affiliation{B. I. Stepanov Institute of Physics of the National
Academy of Sciences of Belarus,\\
Nezavisimosti Ave.\ 68, 220062 Minsk, Belarus}
 \author{Svetlana Grigorieva}
 \affiliation{Institute for Nuclear Research of the RAS,\\
60th October Revolution prospect 7A, Moscow, Russia}


\begin{abstract}
We argue that an astrophysical solution to the ultra high energy cosmic
ray (UHECR) problem is viable. The detailed study of UHECR energy spectra 
is performed. The spectral features of extragalactic
protons interacting with Cosmic Microwave Background (CMB) are
calculated in model-independent way. Using the power-law generation
spectrum $\propto E^{-\gamma_g}$ as the only assumption, we analyze
four features of the proton spectrum: the GZK cutoff, dip, bump and the
second dip. We found the dip, induced by electron-positron production
on CMB, as the most robust feature, existing in energy range $1\times
10^{18} - 4\times 10^{19}$~eV. Its shape is stable relative to various
phenomena included in calculations: discreteness of the source
distribution, different modes of UHE proton propagation (from
rectilinear to diffusive), local overdensity or deficit of the sources,
large-scale inhomogeneities in the universe and interaction
fluctuations. The dip is well confirmed by observations of AGASA,
HiRes, Fly's Eye and Yakutsk detectors. With two free parameters
($\gamma_g$ and flux normalization constant) the dip describes about 20
energy bins with $\chi^2/{\rm d.o.f.} \approx 1$ for each experiment.
The best fit is reached at $\gamma_g =2.7$, with the allowed range 2.55
- 2.75. The dip is used for energy calibration of the detectors. For
each detector independently the energy is shifted by factor $\lambda$
to reach the minimum $\chi^2$. We found $\lambda_{\rm Ag}=0.9$,~
$\lambda_{\rm Hi}=1.2$ and $\lambda_{\rm Ya}=0.75$ for AGASA, HiRes and
Yakutsk detectors, respectively. Remarkably, that after this energy
shift the fluxes and spectra of all three detectors agree perfectly,
with discrepancy between AGASA and HiRes at $E> 1\times 10^{20}$~eV
being not statistically significant. The excellent agreement of the dip
with observations should be considered as  confirmation of UHE proton
interaction with CMB.
The dip has two flattenings. The high energy flattening at $E \approx
1\times 10^{19}$~eV automatically explains ankle, the feature observed
in all experiments starting from 1980s. The low-energy flattening at $E
\approx 1\times 10^{18}$~eV provides the transition to galactic
cosmic rays. This transition is studied quantitatively in this work.
Inclusion of primary nuclei with fraction more than $20\%$ upsets
the agreement of the dip with observations, which we interpret as
an indication to acceleration mechanism.
We study in detail the formal problems of spectra calculations: energy
losses (the new detailed calculations are presented), analytic method
of spectrum calculations and study of fluctuations with help of kinetic
equation.
The UHECR sources, AGN and GRBs, are studied in a model-dependent
way, and acceleration is discussed.
Based on the agreement of the dip with existing data, we make the
robust prediction for the spectrum at $1\times 10^{18} - 1\times
10^{20}$~eV to be measured in the nearest future by Auger detector. We
also predict the spectral signature of nearby sources, if they are
observed by Auger.
This paper is long and contains many technical details. For those who
are interested only in physical content we recommend to read
Introduction and Conclusions, which are written as autonomous parts of
the paper.
\end{abstract}


\maketitle

\section{\label{sec:introduction} Introduction}
The systematic study of Ultra High Energy Cosmic Rays (UHECR) 
started in late 1950s after construction of Volcano Ranch (USA)
and Moscow University (USSR) arrays. During the next 50 years of
research the origin of UHE particles, which hit the detectors, was
not well understood. At present due to the data of the last
generation arrays, Haverah Park (UK)\cite{HP}, Yakutsk (Russia)
\cite{Ya}, Akeno and AGASA (Japan) \cite{AGASAa,AGASAb}, Fly's
Eye \cite{FE} and HiRes \cite{HiRes} (USA) we are probably very
close to understanding the origin of UHECR. The forthcoming data
of Auger detector (see \cite{auger} for the first results) will
undoubtedly shed more light on this problem.

On the theoretical side we have an important clue to understanding the
UHECR origin: the interaction of extragalactic protons, nuclei and
photons with CMB, which leaves the imprint on UHE proton spectrum, most
notably in the form of the Greisen-Zatsepin-Kuzmin (GZK)
\cite{GZKG,GZKZK} cutoff for the protons.

We shortly summarize the basic experimental results and the
results of the data analysis, important for understanding of UHECR
origin (for a review see \cite{NaWa}). \\
{\em (i)}~~ The spectra of UHECR are measured with good accuracy at
$1\times 10^{18} - 1\times 10^{20}$~eV, and these data have a power to
reject or confirm some models. The discrepancy between the AGASA and
HiRes data at $E > 1\times 10^{20}$~eV might have the statistical
origin (see \cite{MBO} and discussion in Section
\ref{sec:Ag-Hi-discrep}), and the GZK cutoff may exist.\\
{\em (ii)}~~ The mass composition at $E\gsim 1\times 10^{18}$~eV (as
well as below) is not well known (for a review see \cite{Watson}).

Different methods result in different mass composition, and the same methods
disagree in different experiments. In principle, the most reliable
method of measuring the mass composition is given by elongation rate
(energy dependence of  maximum depth of a shower, $X_{\rm max}$)
measured by the fluorescent method. The data of Fly's Eye in 1994
\cite{FE} favored iron nuclei at $\sim 1\times 10^{18}$~eV with a
gradual transition to the protons at $1\times 10^{19}$~eV. The further
development of this method by the HiRes detector, which is the
extension of Fly's Eye, shows the transition to the proton composition
already at $1\times 10^{18}$~eV \cite{Sokola,Sokolb}. At present the
data of HiRes \cite{Sokola,Sokolb}, HiRes-MIA \cite{HiRes-MIA} and
Yakutsk \cite{Glushkov2000} favor the proton-dominant composition at
$E\gsim 1\times 10^{18}$~eV, data of Haverah Park \cite{HP} do not
contradict such composition at $E\gsim (1-2)\times 10^{18}$~eV, while
data of Fly's Eye \cite{FE} and Akeno \cite{Akeno-mass} agree
with mixed composition dominated by iron.\\
{\em (iii)}~~ The arrival directions of particles with energy $E \geq
4\times 10^{19}$~eV show the small-angle clustering within the angular
resolution of detectors. AGASA found 3 doublets and one triplet among
47 detected particles \cite{clust-AGASA} (see the discussion in
\cite{FW}). In the combined data of several arrays \cite{clust-tot}
there were found 8 doublets and 2 triplets in 92 events. The stereo
HiRes data \cite{clust-Hi} do not show small-angle clustering for 27
events at $E \geq 4\times 10^{19}$~eV, maybe due to limited statistics.

Small-angle clustering is most naturally explained in the case of
rectilinear propagation as a random arrival of two (three) particles
from a single source \cite {DTT}. This effect has been calculated in
Refs.\ \cite{FK,YNS1,Sigl1,YNS2,BlMa,KaSe,BlMaOl}. In the last five works the
calculations have been performed by the Monte Carlo (MC) simulations 
and results agree well.
According to \cite{KaSe} the density of the sources, needed to explain
the observed number of doublets is $n_s= (1 - 3)\times
10^{-5}$~Mpc$^{-3}$. In \cite{BlMa} the best fit is given by $n_s \sim
1\times 10^{-5}$~Mpc$^{-3}$ and the large uncertainties (in particular
due to ones in observational
data) are emphasized.\\
{\em (iv)}~~ Recently there have been found the statistically
significant correlations between directions of particles with energies
$(4 - 8)\times 10^{19}$~eV and directions to AGN of the special type -
~BL Lacs \cite{corr} (see also the criticism \cite{Sar} and the reply
\cite{TT}).

The items {\em (iii)} and {\em (iv)} favor rectilinear propagation of
primaries from the point-like extragalactic sources, presumably AGN.
However, the propagation in magnetic fields also exhibits clustering
\cite{LSB,Sigl1,Sato-clust}.

The quasi-rectilinear propagation of ultra-high energy protons is found
possible in MHD simulations \cite{Dolag} of magnetic fields in
large-scale structures of the universe (see however the simulations in
\cite{Sigla,Siglb} with quite different results).

There are many unresolved problems in the field of Ultra High Energy
Cosmic Rays, such as nature of primaries (protons? nuclei? or the other
particles?), transition from galactic to extragalactic cosmic rays,
sources and acceleration, but most intriguing problem remains existence
of superGZK particles with energies higher than $E \sim 1\times
10^{20}$~eV. ``The AGASA excess'', namely 11 events with energy higher
than $1\times 10^{20}$~eV, is still difficult to explain, though there
are indications that it may have the statistical origin 
combined with systematic errors in energy determination (see section
\ref{sec:Ag-Hi-discrep}). The AGASA excess, if it is real, should be 
described by another component
of UHECR, most probably connected with the new physics: superheavy dark
matter, new signal carriers, like e.g.\ light stable hadron and
strongly interacting neutrino, the Lorentz invariance violation etc.

The problem with superGZK particles is seen in other detectors, too.
Apart from the AGASA events, there are five others: the golden Fly's
Eye event with $E \approx 3\times 10^{20}$~eV, one HiRes event with $E
\approx 1.8\times 10^{20}$~eV and three Yakutsk events with $E \approx
1\times 10^{20}$~eV. No sources are observed in the direction of these
particles at the distance of order of attenuation length. The most
severe problem is for the golden Fly's Eye event: with attenuation
length $l_{\rm att}= 21$~Mpc and the homogeneous magnetic field 1~nG on
this scale, the deflection of particle is only $3.7^{\circ}$. Within
this angle there are no remarkable sources at distance $ \sim 20$ Mpc
\cite{ES}.

In this paper we analyze the status of most conservative
astrophysical solution to ultra-high energy cosmic ray problem,
assuming that primary particles are protons or nuclei accelerated in
extragalactic sources. In the first part of the paper (Sections II - V)
we analyze the signatures of ultra-high energy protons propagating
through CMB. We found that the dip, a spectral feature in energy range
$1\times 10^{18} - 4\times 10^{19}$~eV, is well confirmed by
observational data of AGASA, HiRes, Yakutsk and Fly's Eye detectors. In
Sections VI -VII we discuss in the model-dependent way the transition
from galactic to extragalactic cosmic rays and extragalactic sources:
AGN and GRBs.

We use in formulae for the flux throughout the paper, e.g.\ in
equations (\ref{Q_gen}), (\ref{Jdiff}), (\ref{Junm}),
(\ref{flux-lattice}), (\ref{broken}) and in Appendix \ref{app-Emax},
energy $E$ measured in GeV, luminosity $L_p$ - in GeV/s, emissivity
(comoving energy density production rate) $\mathcal{L}$ - 
in GeV cm$^{-3}$ s$^{-1}$ and $E_{\rm min}=1$~GeV, 
if not otherwise indicated. 

\section{\label{sec:energy-losses} Energy losses and the universal spectrum 
of UHE protons}
In this Section we present our recent calculations of energy
losses for UHE protons interacting with CMB, and calculate the spectrum
of protons, assuming the homogeneous distribution of sources in the
space and continuous energy losses. The spectrum is calculated, using
the conservation of the number of protons, interacting with CMB.
Formally it does not depend on the mode of proton propagation (e.g.\
rectilinear or diffusive), and we shall discuss when this approximation
is valid. The proton spectrum calculated in this way we call {\em
universal}.
\subsection{Energy losses}
\label{sec:en-loss} We present here the accurate calculations of 
energy losses due to pair
production, $p+\gamma_{CMB}\to p+e^++e^-$, and pion production,
$p+\gamma_{CMB}\to N+{\rm pions}$, where $\gamma_{CMB}$ is a microwave
photon.

The energy losses of UHE proton per unit time due to its interaction
with low energy photons is given by
\beq %
-\frac{1}{E}\frac{dE}{dt}=\frac{c}{2\Gamma^2}
\int_{\varepsilon_{th}}^{\infty} d\varepsilon_r \sigma(\varepsilon_r)
f(\varepsilon_r)\varepsilon_r\int_{\varepsilon_r/2\Gamma} ^{\infty}
d\varepsilon \frac{n_\gamma(\varepsilon)}{\varepsilon^2},
\label{enloss} %
\eeq %
where $\Gamma$ is the Lorentz factor of the proton,
$\varepsilon_r$ is the energy of background photon in the reference
system of the proton at rest, $\varepsilon_{th}$ is the threshold of
the considered reaction in the rest system of the proton,
$\sigma(\varepsilon_r)$ is the cross-section, $f(\varepsilon_r)$ is the
mean fraction of energy lost by the proton in one $p\gamma$ collision
in the laboratory system, (see Fig.~\ref{comp} in Appendix
\ref{app-enloss}), $\varepsilon$ is the energy of the background photon
in the laboratory system, and $n_\gamma(\varepsilon)$ is the density of
background photons.

\begin{figure}
   \begin{minipage}[ht]{8cm}
     \centering
     \includegraphics[width=8cm]{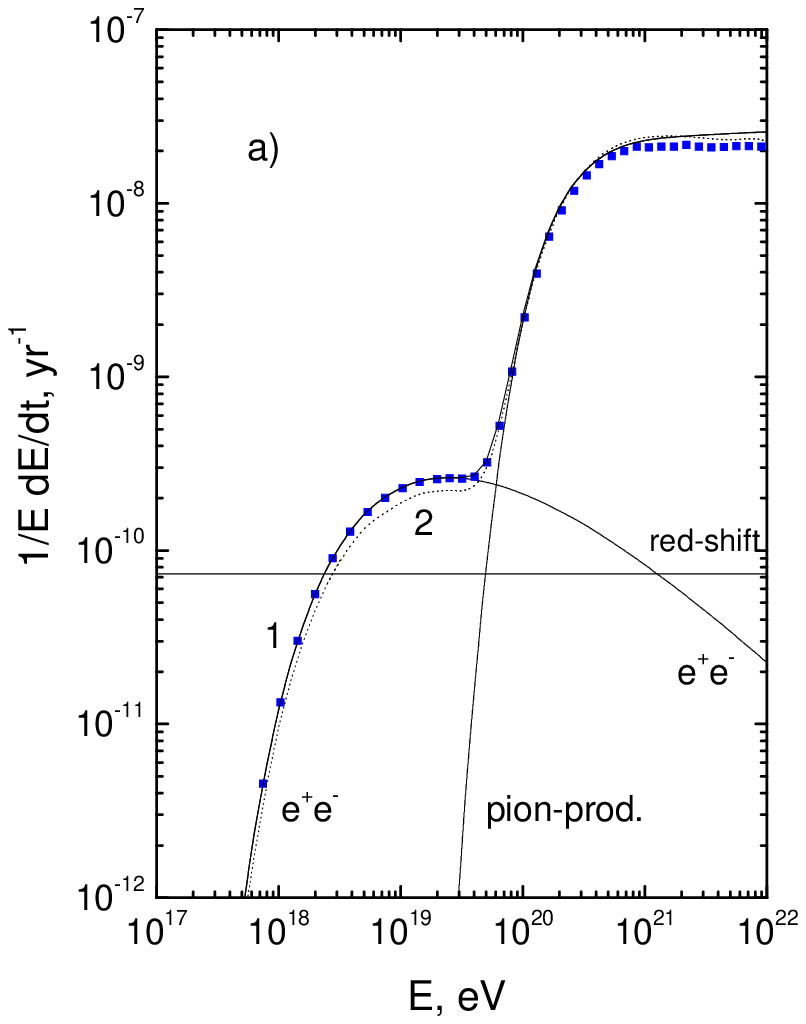}
   \end{minipage}
   \hspace{5mm}
   \vspace{-1mm}
 \begin{minipage}[ht]{8cm}
    \centering
    \includegraphics[width=8cm]{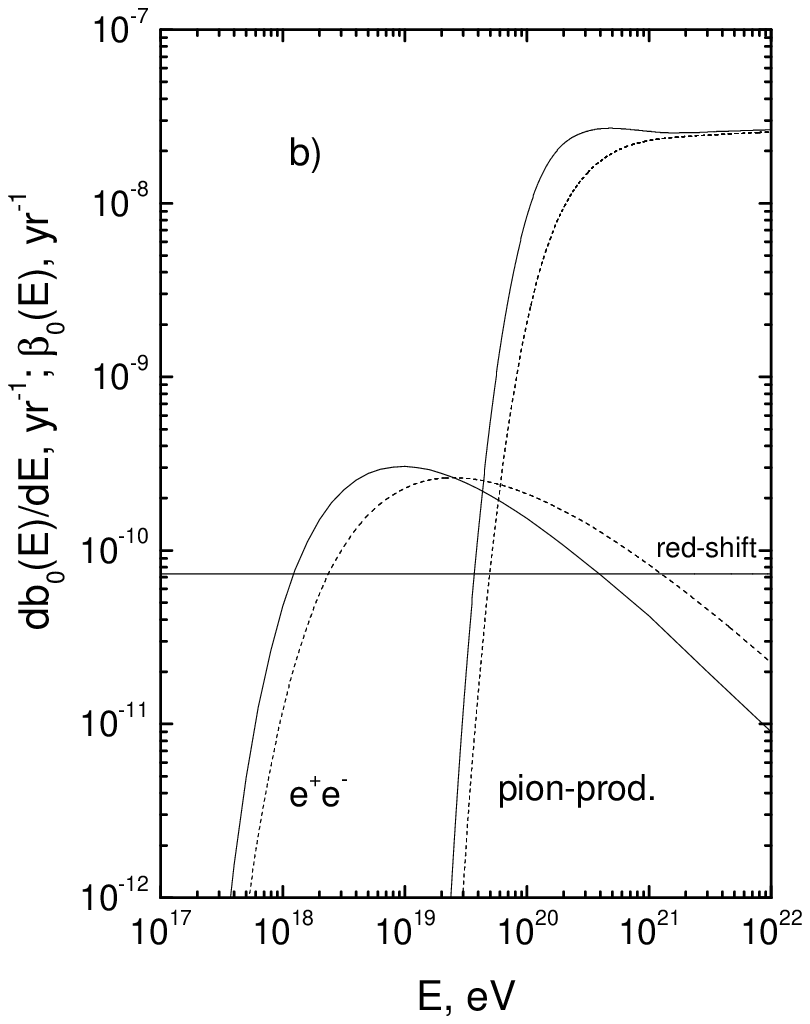}
    \label{deriv}
 \end{minipage} %
\caption{(a) UHE proton energy losses $E^{-1}dE/dt$ at $z=0$~ (present
work: curve 1; Berezinsky and Grigorieva (1988)
\protect\cite{BeGrbump}: curve 2; Stanev \etal\ 2000
\protect\cite{stanev} : black squares). The line  'red-shift'
($H_0=72$~ km/s Mpc) gives adiabatic energy losses. Note two important 
energies $E_{\rm eq1}=2.37\times 10^{18}$~eV, where adiabatic and pair- 
production energy losses become equal, and 
$E_{\rm eq2}=6.05\times 10^{19}$~eV, where pair-production and
photopion production energy losses are equal. The latter is one of the 
characteristics of the GZK cutoff. (b) The derivative
$db_0(E)/dE$, where $b_0(E)=-dE/dt$ at present epoch $z=0$, (solid
curve) in comparison with $\beta_0(E)=-E^{-1}dE/dt$ (dashed curve).
}
\label{loss}
\end{figure}

For CMB with temperature $T$ Eq.~(\ref{enloss}) is simplified
\beq %
-\frac{1}{E}\frac{dE}{dt}=\frac{c T}
{2\pi^2\Gamma^2}\int_{\varepsilon_{th}}
^{\infty}d\varepsilon_r\sigma(\varepsilon_r)
f(\varepsilon_r)\varepsilon_r
\left\{-\ln\left[1-\exp\left(-\frac{\varepsilon_r}{2\Gamma
T}\right)\right] \right\}.
\label{CMBRloss} %
\eeq %
Further on we shall use the notation
\beq %
\beta_0(E) = -\frac{1}{E}\frac{dE}{dt},~~~ b_0(E)= - \frac{dE}{dt}
\label{b_0} %
\eeq %
for energy losses on CMB at present cosmological epoch, $z=0$ and
$T=T_0$. For the epochs with red-shift $z$ one has:
\begin{eqnarray}
\beta (E,z)&=&(1+z)^3\beta_0[(1+z)E], \label{beta_z} \\
b(E,z)&=&(1+z)^2b_0[(1+z)E]. \label{b_z}
\end{eqnarray}
Another important physical quantity needed for calculations of spectra
is the derivative $db_0(E)/dE$, which can be calculated as
\beq %
\frac{db_0(E)}{dE}= - \beta_0(E)+ \frac{c}{4\pi^2\Gamma^3}
\int_{\varepsilon_{\rm th}}^{\infty}
d\varepsilon_r \sigma(\varepsilon_r)f(\varepsilon_r)
\frac{\varepsilon_r^2}{\exp \left (\frac{\varepsilon_r}{2\Gamma
T_0}\right ) -1} .
\label{db/dE} %
\eeq %
As one can see from
Fig.~\ref{loss}, $db_0(E)/{dE}$ is numerically very similar to
$\beta_0(E)$, and for approximate calculations one can use
$\beta_0(E)$ values for both functions.

From Eqs.~(\ref{enloss}) and (\ref{CMBRloss}) one can see that
apart from cross-section the mean fraction of energy lost by the
proton in laboratory system in one collision, $f(\varepsilon_r) $
(see Eq.~(\ref{averfraction})), is the basic quantity needed for
calculations of energy losses. The threshold values of these
quantities are well known:
\beq %
f_{\rm pair} = \frac{2m_e}{2m_e+m_p},~~~ ~~~~~
f_{\rm pion} = \frac{m_\pi}{m_\pi + m_p},
\label{frac} %
\eeq %
where $f_{\rm pair}$ and $f_{\rm pion}$ are the
threshold fractions for $p+\gamma \rightarrow p+e^+ +e^-$ and $p+\gamma
\rightarrow N+\pi$, respectively.

Pair production loss has been previously discussed in many papers.
All authors directly or indirectly followed the standard approach
of Ref.~\cite{Blumenthal} where the first Born approximation of
the Bethe-Heitler cross-section with proton mass
$m_p\rightarrow\infty$ was used. In contrast to
Ref.~\cite{Blumenthal}, we use here the first Born approximation
approach of Ref.~\cite{BergLinder} accounting for the finite proton
mass. This allows us to calculate the average fraction of energy
lost by the proton in laboratory system by performing a fourfold
integration over invariant mass of electron-positron pair, $M_X$,
over an angle between incident and scattered proton, and polar and
azimuthal angles of an electron in the c.m.s.\ of the $e^+
e^-$-pair (see Appendix A for further details). 
Calculating
photopion energy loss we follow methods developed in papers
\cite{BeGa,GNPCS}. Total cross-sections are taken according to
Ref.~\cite{totcrs}.
At low c.m.s.\ energy, $E_c$, we consider the binary reactions
$p+\gamma \rightarrow \pi^0 +p$, $p+\gamma \rightarrow \pi^+ + n$ (they
include the resonance $p+\gamma \rightarrow \Delta^+$). Differential
cross-sections of binary processes at small energies are taken from
\cite{pioncrsa,pioncrsb,pioncrsc}. At $E_c > 4.3$~GeV we assume the
scaling behavior of differential cross-sections, the latter being taken
from Ref.~\cite{scala,scalb,scalc}. In the intermediate energy range we
interpolate between the angular distribution of these two regimes with
the cross-section being the difference between total cross-section and
cross-section of the all binary processes. Angular distributions for
this part of cross-section vary from isotropic at threshold to those
imposed by inclusive pion photoproduction data at high energies. The
overall differential cross-sections coincide with low-energy binary
description and high-energy scaling distributions and join smoothly
these two regimes in the intermediate region.

\noindent The results of our calculations are presented in
Fig.~\ref{loss} in terms of
$E^{-1}dE/dt$ as function of energy (curve 1). Also plotted is the
derivative $db_0(E)/dE$ (Fig.~\ref{loss}b). This quantity is needed for
calculation of differential energy spectrum.
In Fig.~\ref{loss} we plot for comparison the
energy losses as calculated by Berezinsky and Grigorieva 1988
\cite{BeGrbump} (dashed curve 2). The difference in energy losses due
to pion production is very small, not exceeding 5\% in the energy
region relevant for comparison with experimental data ($E\leq
10^{21}~$eV). The difference with energy losses due to pair production
is larger and reaches maximal value 15\%. The results of calculations
by Stanev \etal\ \cite{stanev} are shown by black squares. These
authors have performed the detailed calculations for both
aforementioned processes, though their approach is different from ours,
especially for photopion process. Our new energy losses are practically
indistinguishable from Stanev \etal\ \cite{stanev} for pair production
and pion production at low energies, and differ by 15-20\% for pion
production at highest energies (see Fig.~\ref {loss}). Stanev \etal\
claimed that energy losses due to pair production is underestimated by
Berezinsky and Grigorieva \cite{BeGrbump} by 30-40\%. Comparison of
{\it data files} of Stanev \etal\ and Berezinsky and Grigorieva (see
also Fig.~\ref{loss}) shows that this difference is significantly less.
Most probably, Stanev \etal\ scanned inaccurately the data from the
journal version of the paper \cite{BeGrbump}.

\subsection{Universal spectrum}
\label{sec:universal} 
To calculate the spectrum one should first of all
evolve the proton energy $E$ from the time of observation $t=t_0$ (or
$z=0$) to the cosmological epoch of generation $t$ (or red-shift $z$),
using the the adiabatic energy losses $EH(z)$ and $b(E,z)$ given by
Eq.~(\ref{beta_z}):
\beq %
-dE/dt=EH(z)+(1+z)^2b_0\left [(1+z)E \right ],
\label{evol} %
\eeq %
where $H(z)= H_0 \sqrt{\Omega_m(1+z)^3 +
\Omega_{\Lambda}}$ is the Hubble parameter at cosmological epoch $z$,
with $H_0=72$~km/s Mpc, $\Omega_m=0.27$ and $\Omega_{\Lambda}=0.73$
\cite{WMAP}.

We calculate the spectrum from conservation of number of particles in
the comoving volume (protons change their energy but do not disappear).
For the number of UHE protons per unit comoving volume, $n_p(E)$, one
has:
\beq %
n_p(E)dE=\int_{t_{\rm min}}^{t_0}dt~Q_{\rm gen}(E_g,t)~dE_g ,
\label{conserv} %
\eeq %
where $t$ is an age of the universe, $E_g=
E_g(E,t)$  is a generation energy at age $t$, calculated according to
Eq.~(\ref{evol}) and $Q_{\rm gen}(E_g,t)$ is the generation rate per
unit comoving volume, which can be expressed through {\em emissivity} 
${\cal L}_0$, the energy release per unit time and unit of comoving volume, at
$t=t_0$, as
\beq %
Q_{\rm gen}(E_g,t)={\cal L}_0 (1+z)^m Kq_{\rm
gen}(E_g),
\label{Q_gen} %
\eeq %
where $(1+z)^m$ describes the possible
cosmological evolution of the sources. In the case of the power-law
generation, $q_{\rm gen}(E_g)=E_g^{-\gamma_g}$, with normalization
constant $K= \gamma_g-2$ for $\gamma_g>2$ and $K=(\ln E_{\rm
max}/E_{\rm min})^{-1}$ for $\gamma_g=2$. We recall that in these
formulae and everywhere below energies $E$ are given in GeV, emissivity
$\cal{L}$ in GeV cm$^{-3}$s$^{-1}$ and source luminosity $L$ in GeV s$^{-1}$.

From Eq.~(\ref{conserv}) one obtains the diffuse flux as
\beq %
\label{Jdiff}%
J_p(E)=\frac{c}{4\pi}~{\cal L}_0~K \times
\int_0^{z_{max}} dz \left |\frac{dt}{dz} \right | (1+z)^m q_{\rm
gen}(E_g) \frac{dE_g}{dE},
\eeq %
where $|dt/dz|= H^{-1}(z)/(1+z)$ and analytic expression for
$dE_g/dE$ is given by Eq.~(\ref{dE_g/dE}) in
Appendix~\ref{app-dEg/dE}.

The spectrum (\ref{Jdiff}) is referred to as {\em universal spectrum}.
Formally it is derived from conservation of number of particles and
does not depend on propagation mode (see Eq.~(\ref{conserv})). But in
fact, the homogeneity of the particles, tacitly assumed in this
derivation, implies the homogeneity of the sources, and thus the
condition of validity of universal spectrum is a small separation of
sources. The homogeneous distribution of particles in case of
homogeneous distribution of sources and {\em inhomogeneous} magnetic
fields follows from the Liouville theorem (see Ref.~\cite{AB}).

Several effects could in principle modify the shape of universal
spectrum. They include propagation in magnetic fields, discreteness in
distribution of the sources, large-scale inhomogeneous distribution of
sources, local source overdensity or deficit,  and
fluctuations in interaction. These effects will be studied in the next
sections. With exception of energies beyond GZK cutoff and energies
below $1\times 10^{18}$~eV, the universal spectrum is only weakly
modified by aforementioned effects. Here we shortly comment on role of
magnetic fields.
\begin{figure}[ht]
\includegraphics[width=8.0cm]{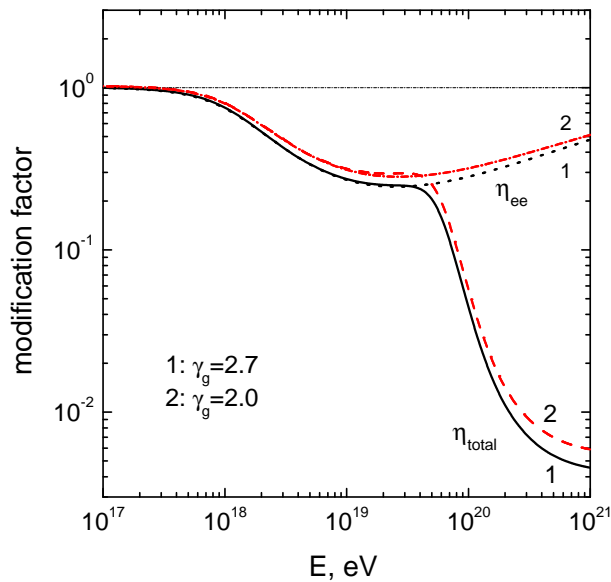}
\vspace{2mm} \caption{ Modification factor for the power-law generation
spectra with $\gamma_g$ in the range 2.0 - 2.7. Curve $\eta=1$
corresponds to adiabatic energy losses, curves $\eta_{ee}$ - to
adiabatic and pair production energy losses and curves $\eta_{\rm tot}$
~-~ to all energy losses. The dip, seen at $1\times 10^{18}\leq E \leq
4\times 10^{19}$~eV, has two flattenings: at low energy
$E \approx 1\times 10^{18}$~eV and at high
energy $E \approx 1\times 10^{19}$~eV. The second flattening explains
well the observed spectrum feature, known as ankle. }
\label{mfactor}
\end{figure}
As numerical simulations (see e.g.\ \cite{SLB,Sato}) show, the
propagation of UHE protons in strong magnetic fields changes the energy
spectrum (for physical explanation of this effect see \cite{AB}). The
influence of magnetic field on spectrum depends on the separation of
the sources, $d$. For uniform distribution of sources with separation
$d$ much less than characteristic lengths of propagation, such as
attenuation length $l_{\rm att}$ and the diffusion length $l_{\rm
diff}$, the diffuse spectrum of UHECR is the universal one independent
of mode of propagation \cite{AB}. This statement has a status of the
theorem. For the wide range of magnetic fields 0.1 - 10~nG and
distances between sources $d \lsim 50$~Mpc the spectrum at $E >
10^{18}$~eV  is close to the universal one \cite{AB1}.
\subsection{Modification factor}
\label{sec:mfactor} The analysis of spectra is very convenient to
perform in terms of
the {\em modification factor}.\\
Modification factor is defined as a ratio of the spectrum $J_p(E)$ (see
Eq.~(\ref{Jdiff})) with all energy losses taken into account, to
the unmodified spectrum $J_p^{\rm unm}(E)$, where only adiabatic energy losses
(red shift) are included.
\beq %
\eta(E)=\frac{J_p(E)}{J_p^{\rm unm}(E)}.
\label{modif} %
\eeq %
For the power-law generation spectrum for $\gamma_g >2$ without
evolution one has
\beq %
J_p^{\rm unm}(E)=\frac{c}{4\pi}(\gamma_g
-2){\cal L}_0 E^{-\gamma_g} \int_0^{z_{\rm max}}dz
\left |\frac{dt}{dz}\right |(1+z)^{-\gamma_g+1}.
\label{Junm} %
\eeq %
The modification factor
is a less model-dependent quantity than the spectrum. In particular, it
should depend weakly on $\gamma_g$, because both numerator and
denominator in Eq.~(\ref{modif}) include $E^{-\gamma_g}$. In the next Section
we consider the non-evolutionary case $m=0$ (see Section \ref{sec:evolution} 
for discussion of evolution). In
Fig.~\ref{mfactor} the modification factor is shown as a function of
energy for two spectrum indices $\gamma_g=2.0$ and $\gamma_g=2.7$. As
expected above, they do not differ much from each other. Note that by
definition $\eta(E) \leq 1$.

\section{\label{sec:signatures} Signatures of UHE protons interacting with CMB}

The extragalactic protons propagating through CMB have signatures in
the form of three spectrum features: GZK cutoff, dip and bump. The dip
is produced due to $e^+e^-$-production and bump -- by pile-up protons
accumulated near beginning of the GZK cutoff. We add here the fourth
signature: the second dip.

The analysis of these features, especially dip and bump, is convenient
to perform in terms of {\em modification factor} \cite{BeGrbump,St}.
For the GZK cutoff we shall use the traditional spectra.

\subsection{GZK cutoff}
\label{sec:GZKcutoff} The GZK cutoff \cite{GZKG,GZKZK} is most
remarkable phenomenon, which describes the sharp steepening of the
spectrum due to pion production. The GZK cutoff is a model--dependent
feature of the spectrum, e.g.\ the GZK cutoff for a single source
depends on the distance to the source.
\begin{figure}[ht]
\includegraphics[width=8.0cm]{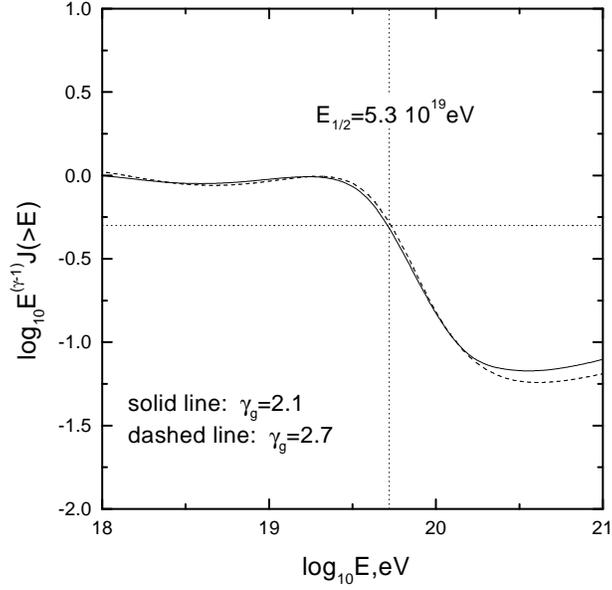}
\vspace{2mm} \caption{$E_{1/2}$ as characteristic of the GZK cutoff.
The calculated integral spectra are multiplied to factor $E^{\gamma -
1}$, where $\gamma-1$ found to fit the spectra in the interval $1\times
10^{18} - 3\times 10^{19}$~eV. $E_{1/2}$ is shown by the vertical line.
This value is valid for generation spectra with  $2.1\leq \gamma_g \leq
2.7$. } \label{Ehalf}
\end{figure}
A common convention is that the GZK cutoff is defined for diffuse flux
from the sources uniformly distributed over the universe. In this case
one can give two definitions of the GZK-cutoff position. In the first
one it is determined as the energy, $E_{GZK} \approx
4\times 10^{19}~eV$, where the steep increase in the energy losses
starts (see Fig.~\ref{loss}). The GZK cutoff starts at this energy. The
corresponding path length of a proton is $R_{GZK}\approx
(E^{-1}dE/cdt)^{-1}\approx 1.3\times 10^3$~Mpc. The advantage of this
definition of the cutoff energy is independence of a  spectrum index, but
this energy is too low to judge about presence or absence of the cutoff
in the measured spectrum. More practical definition is $E_{1/2}$, where
the flux with cutoff becomes lower by factor 2 than power-law
extrapolation. This definition is convenient to use for the integral
spectrum, which is better approximated by power-law function, than the
differential one. 
\begin{figure}[ht]
\begin{minipage}[h]{8cm}
\centering
\includegraphics[width=7.6cm,clip]{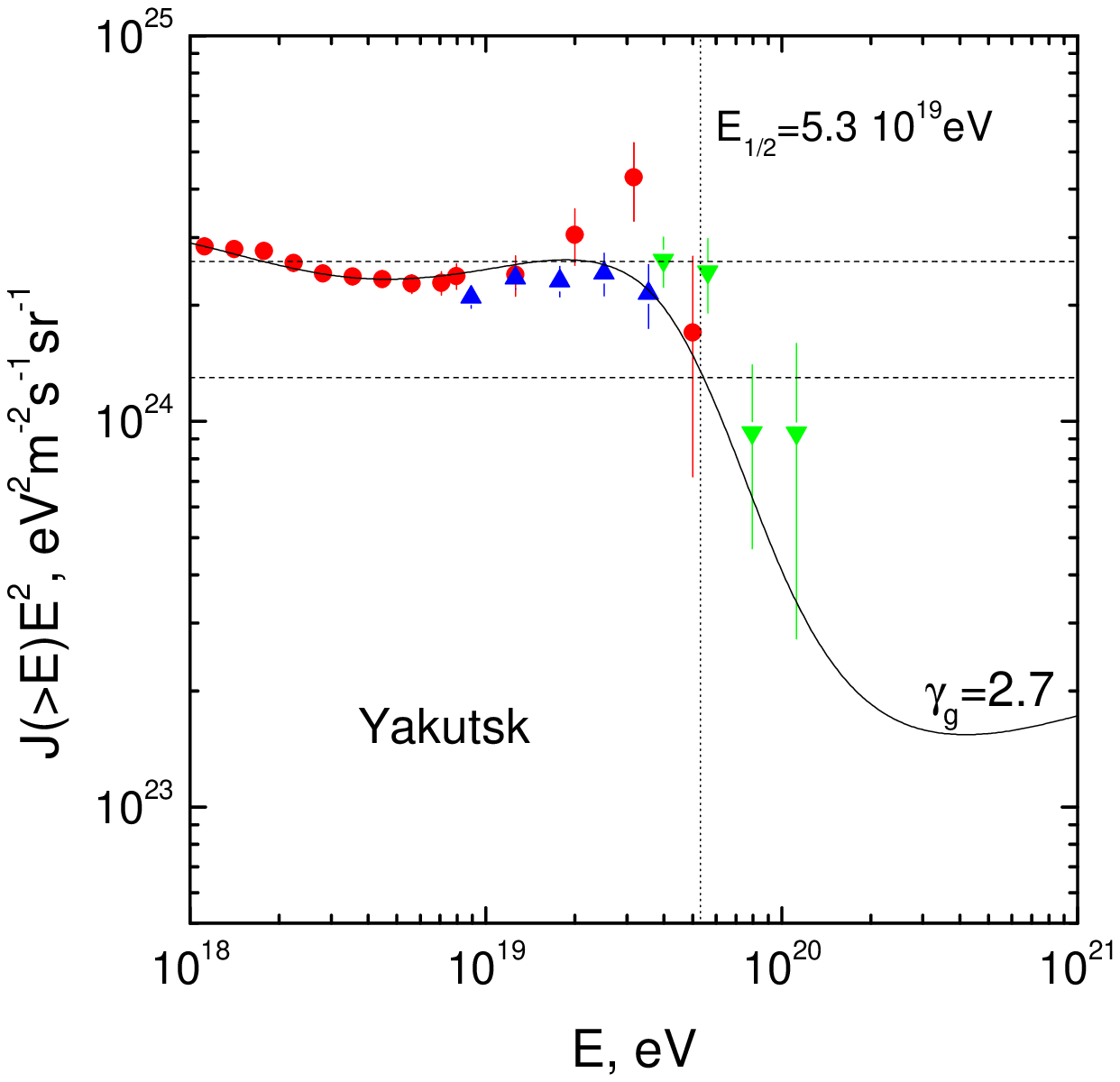}
\end{minipage}
\hspace{5mm}
\begin{minipage}[h]{8cm}
\centering
\includegraphics[width=7.6cm,clip]{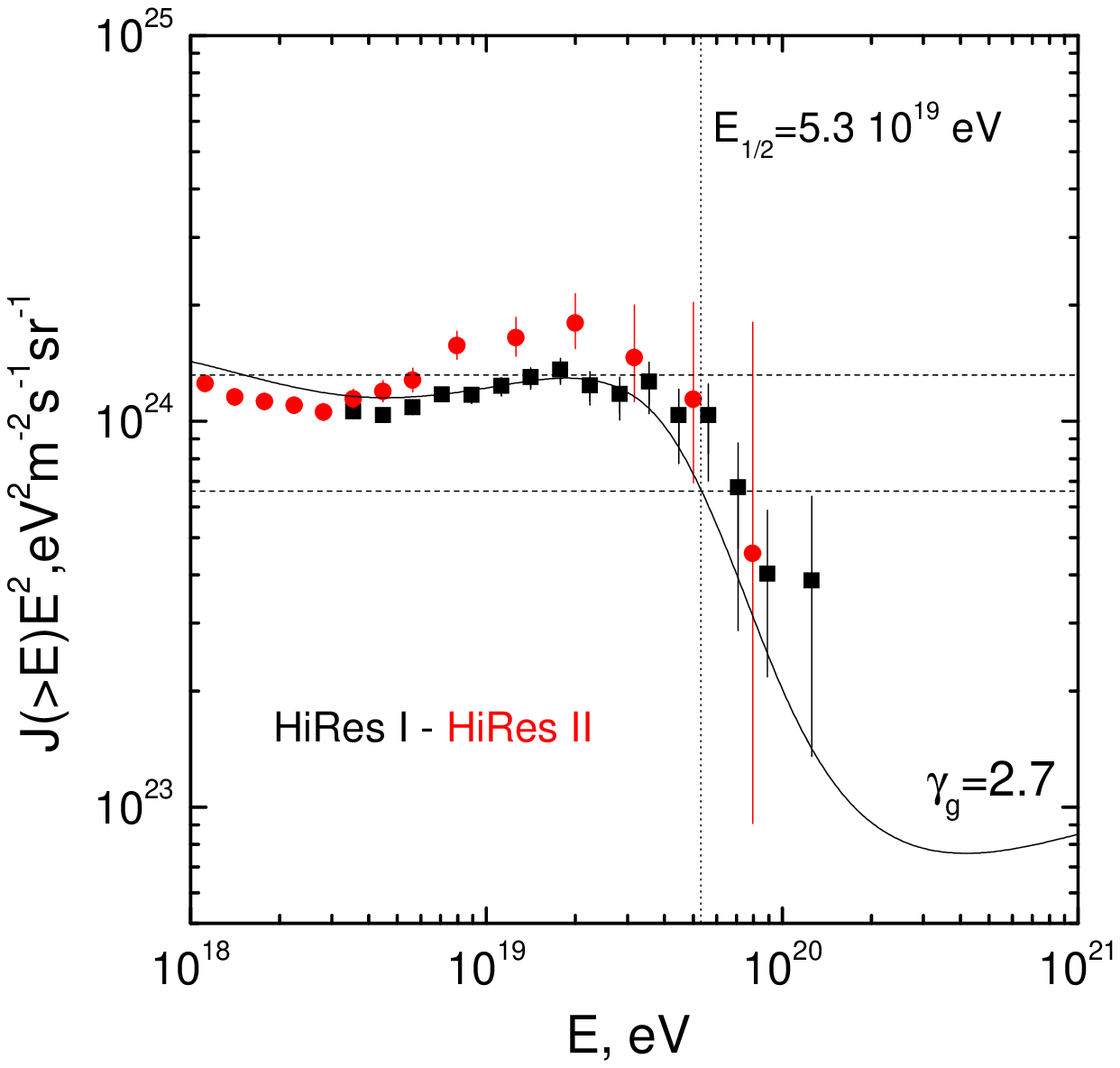}
\end{minipage}
\caption{\label{Ehalf-comp} Predicted $E_{1/2}$ value in comparison
with integral spectrum of Yakutsk array (left panel) and HiRes (right
panel). One can see the agreement of the Yakutsk data with the
theoretical value  $E_{1/2} \approx 5.3\times 10^{19}$~eV (vertical
line), while in the case of HiRes data this value is about $7\times
10^{19}$~eV with large uncertainties due to difference between  HiRes I
and HiRes II data. HiRes collaboration found \cite{e1/2}
$E_{1/2}=(5.9^{+2.4}_{-0.8})\times 10^{19}$~eV. 
}
\end{figure}
In Fig.~\ref{Ehalf} the function
$E^{(\gamma-1)}J(>E)$, where $J(>E)$, the calculated integral diffuse
spectrum, is plotted as function of energy. Note, that
$\gamma>\gamma_g$ is an effective index of the power-law approximation
of the spectrum modified by energy losses. For wide range of generation
indices $2.1 \leq \gamma_g \leq 2.7$ the cutoff energy is the same,
$E_{1/2} \approx 5.3\times 10^{19}$~eV. The corresponding proton path
length is $R_{1/2} \approx 800$~Mpc.
In panel a) of Fig.~\ref{Ehalf-comp} $E_{1/2}$ is found from the
integral spectrum of the Yakutsk array in the reasonable agreement with
theoretical prediction. The HiRes data are shown in panel b). These
data have large uncertainties, which prevent the accurate determination
of $E_{1/2}$. However, they agree with the predicted value $E_{1/2}
\approx 5.3\times 10^{19}$~eV. In the recent paper \cite{e1/2} the
Hires collaboration found $E_{1/2}=(5.9^{+2.4}_{-0.8})\times 10^{19}$~eV 
in  better agreement with the predicted value. 

\begin{figure}
\begin{minipage}[h]{8cm}
\centering
\includegraphics[width=7.6cm]{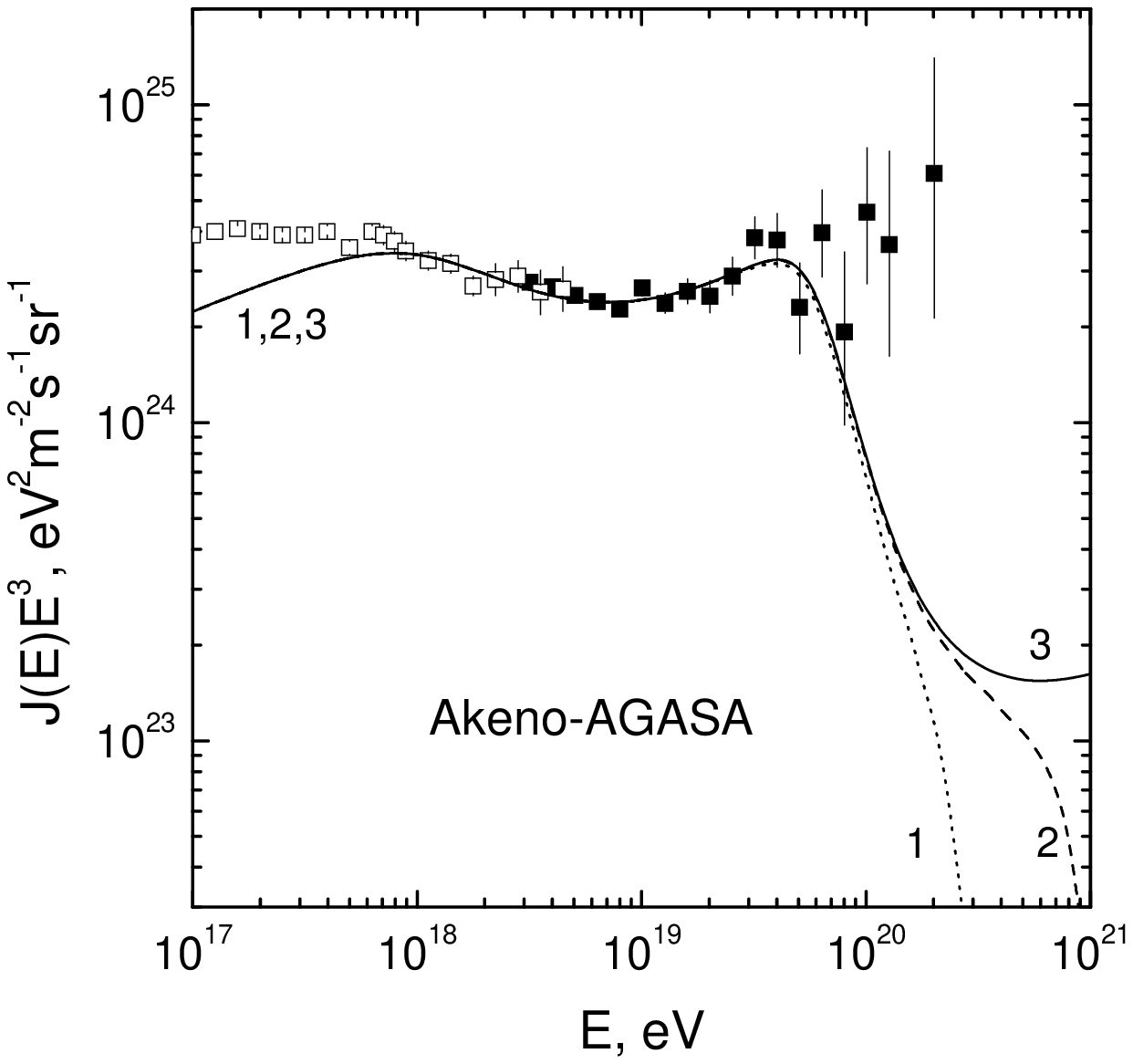}
\end{minipage}
\hspace{5mm} %
\begin{minipage}[h]{8cm}
\centering
\includegraphics[width=7.6cm]{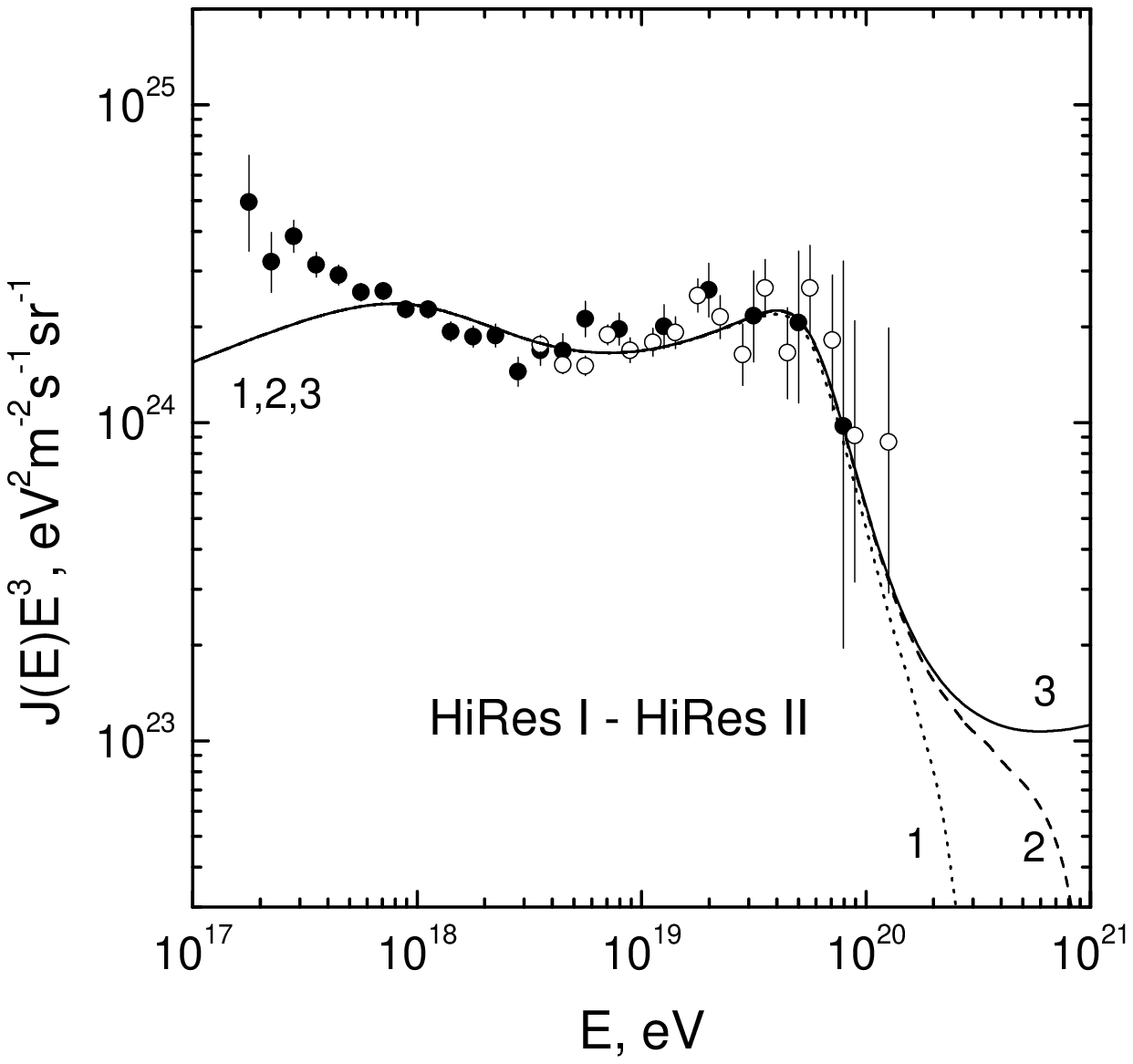}
\end{minipage}
\caption{UHECR differential spectra  calculated with $\gamma_g=2.7$ and
with  different $E_{max}$ are compared with AGASA data (left panel) and
HiRes data (right panel). The curves (1),(2),(3) correspond to  maximum
acceleration energies $E_{max}= 3\times 10^{20}~ {\rm eV},~ 1\times
10^{21}~ {\rm eV}$ and $\infty$, respectively.} \label{diff-Emax}
\end{figure}

In Fig.~\ref{diff-Emax} the calculated
universal spectra with the GZK cutoff are compared with AGASA  and HiRes data
\cite{AGASAa,AGASAb,HiRes}. While the HiRes data agree with the
predicted GZK cutoff, the AGASA data show significant excess over this
prediction at $E > 1\times 10^{20}$~eV. 
\begin{figure}
\begin{minipage}[h]{8cm}
\centering
\includegraphics[width=7.6cm]{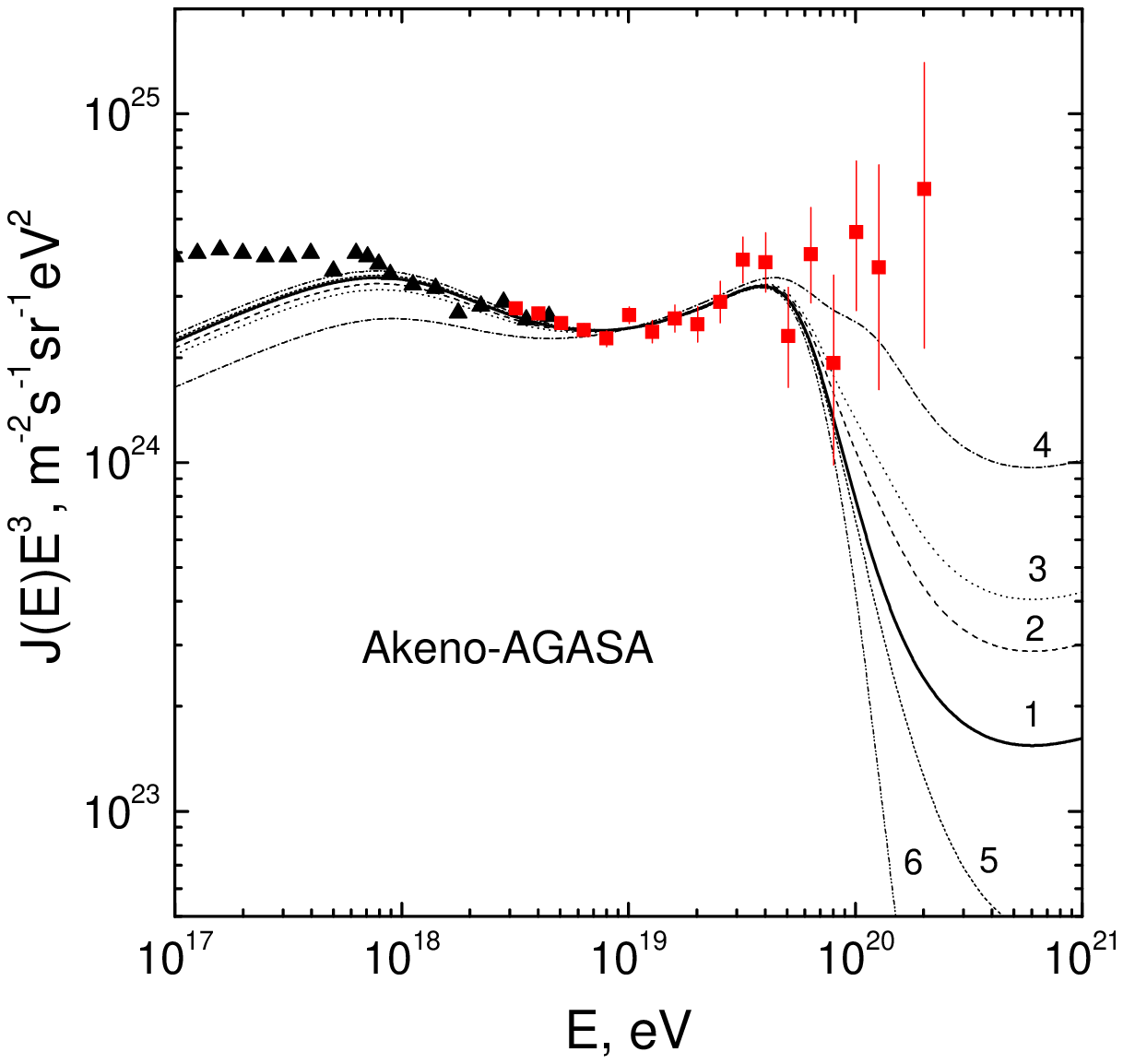}
\end{minipage}
\hspace{5mm}
\begin{minipage}[h]{8cm}
\centering
\includegraphics[width=7.6cm]{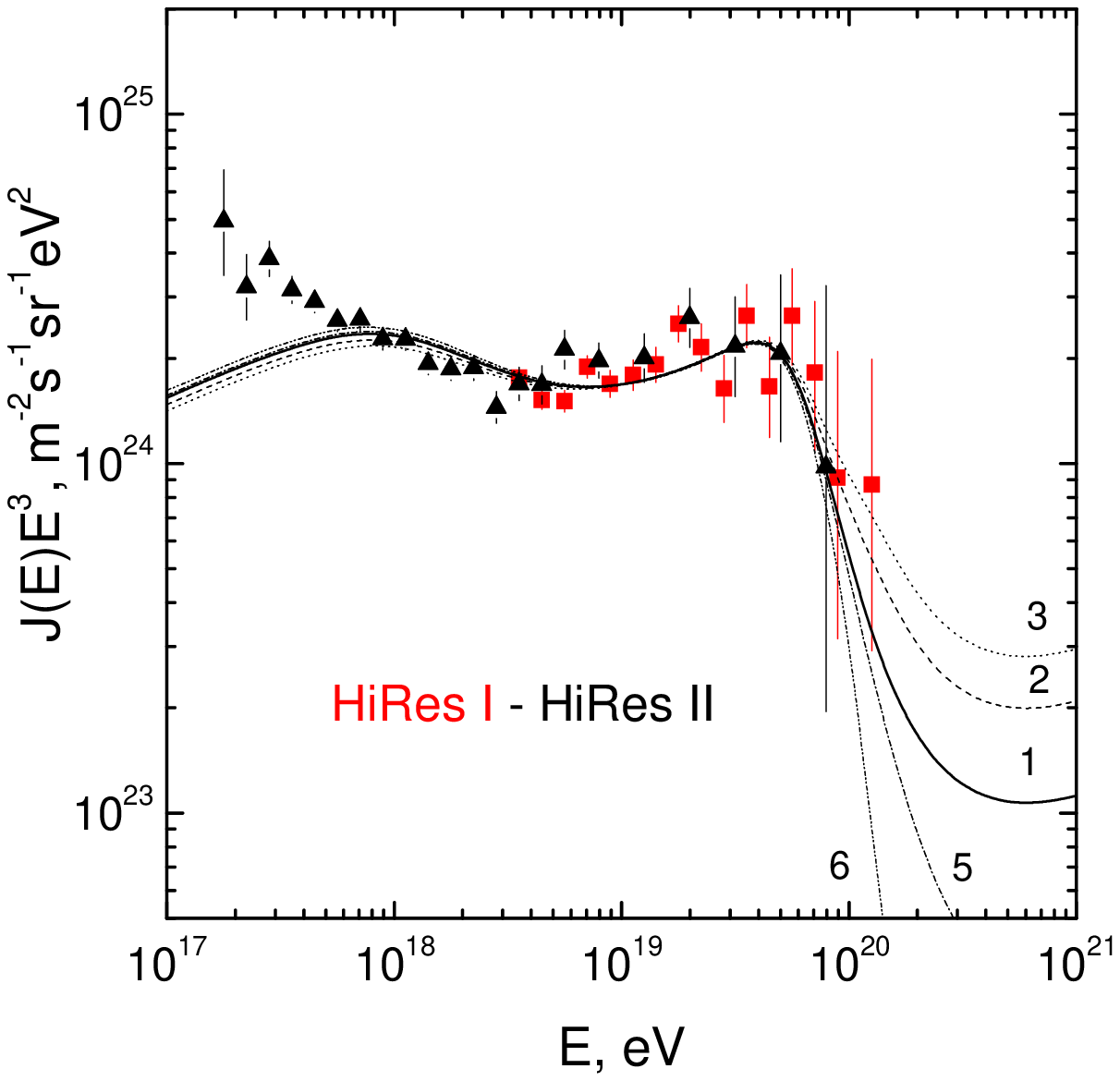}
\end{minipage}
\caption{Effect of local source overdensity and deficit on the shape of
the GZK cutoff. The curve 1 is for the universal spectrum, the curves
2, 3  are for overdensity in the region of 30~Mpc with overdensity
factor $n/n_0$ equal to 2 and 3, respectively. The curves 5 and 6 show
the deficit $n/n_0=0$ in the region of 10~Mpc and 30~Mpc, respectively.
The calculated spectra are compared with the AGASA data (left panel)
and with HiRes (right panel). The description of the AGASA excess,
curve 4, needs unrealistic overdensity $n/n_0=10$.} %
\label{overdensity} %
\end{figure}

The GZK cutoff as calculated above has many uncertainties.

The energy shape of the GZK feature is model dependent. The local
excess of sources makes it flatter, and the deficit -- steeper. The
shape is affected by $E_{\rm max}^{\rm acc}$ (see Fig.~\ref{diff-Emax})
and by fluctuations of source luminosities and distances between the
sources. The cutoff, if discovered, can be produced as the acceleration
cutoff (steepening below the maximum energy of acceleration in the
generation spectrum). Since the shape of both, the GZK cutoff and
acceleration cutoff, is model-dependent, it will be difficult to argue
in favor of any of them, in case a cutoff is discovered.

To illustrate the effect of local overdensity/deficit of the UHECR
sources we calculate here the UHECR spectra with  different local
ratios $n/n_0$, where $n$ is the local density of the UHECR sources and
$n_0$ is mean extragalactic source density. We use the various sizes of
overdensity/deficit regions $R_{overd}$, equal to 10, 20 and 30 Mpc.
The results of our calculations for the overdensity case are presented in
Fig.~\ref{overdensity}   for $\gamma_g=2.7$, ~ $m=0$ and for four
values of overdensity $n/n_0$ equal to 1,~ 2,~ 3 and 10, assuming the
size of overdensity region 30~Mpc (the results for $R_{overd}=20~$Mpc
are not much different). The spectra for the case of the deficit are
shown by curves 5 and 6 in the both panels. The theoretical spectra
shown in Fig.~\ref{overdensity} illustrate uncertainties in the
prediction of the shape of the GZK feature.

An interesting question is whether the local overdensity of the sources
can explain the AGASA excess. This problem has been already addressed
in \cite{Bl-overd} for the realistic distribution of visible galaxies,
considering them as the UHECR sources. Our calculations show (see left
panel of Fig.~\ref{overdensity}) that unrealistic overdensity $n/n_0
\gtrsim 10$ is needed to explain the AGASA excess.

Both effects, $E_{\rm max}$ and local source overdensity (deficit)
affect weakly the shape of the GZK cutoff at $E \lsim 1 \times
10^{20}$~eV. Thus, {\em the precise measurements of the spectrum in
this energy region, as well as measurement of $E_{1/2}$, give the best
test of the GZK cutoff.} At higher energies theoretical predictions
have large model-dependent uncertainties.

\subsection{Bump in the diffuse spectrum}
\label{sec:bump} Protons do not disappear in the photopion
interactions, they only loose energy and are accumulated near beginning
of the GZK cutoff in the form of a bump.

\begin{figure}[ht]
\begin{minipage}[h]{8cm}
\centering
\includegraphics[width=7.8cm]{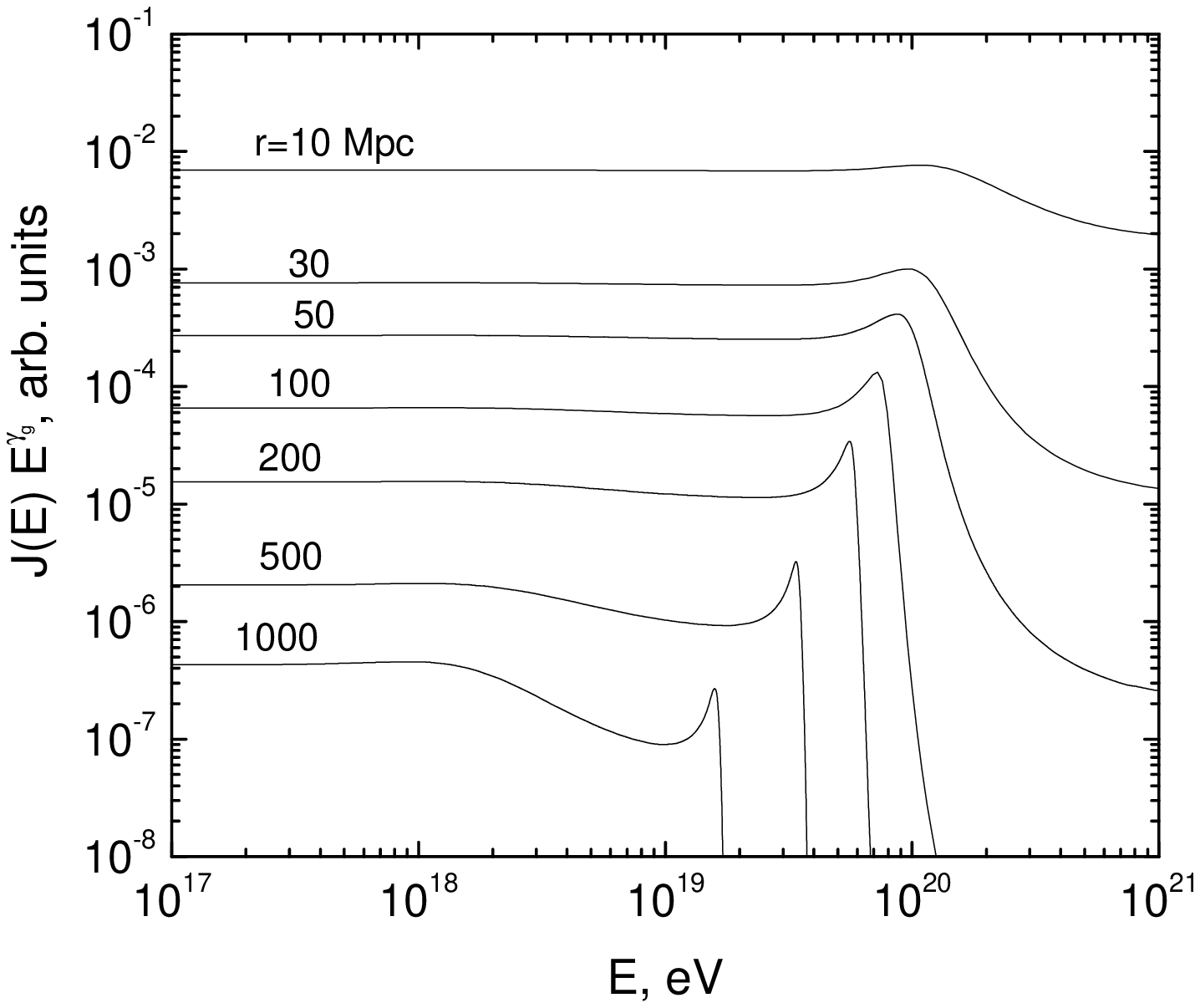}
\end{minipage}
\hspace{5mm}
\begin{minipage}[h]{8cm}
\centering
\includegraphics[width=7.8cm]{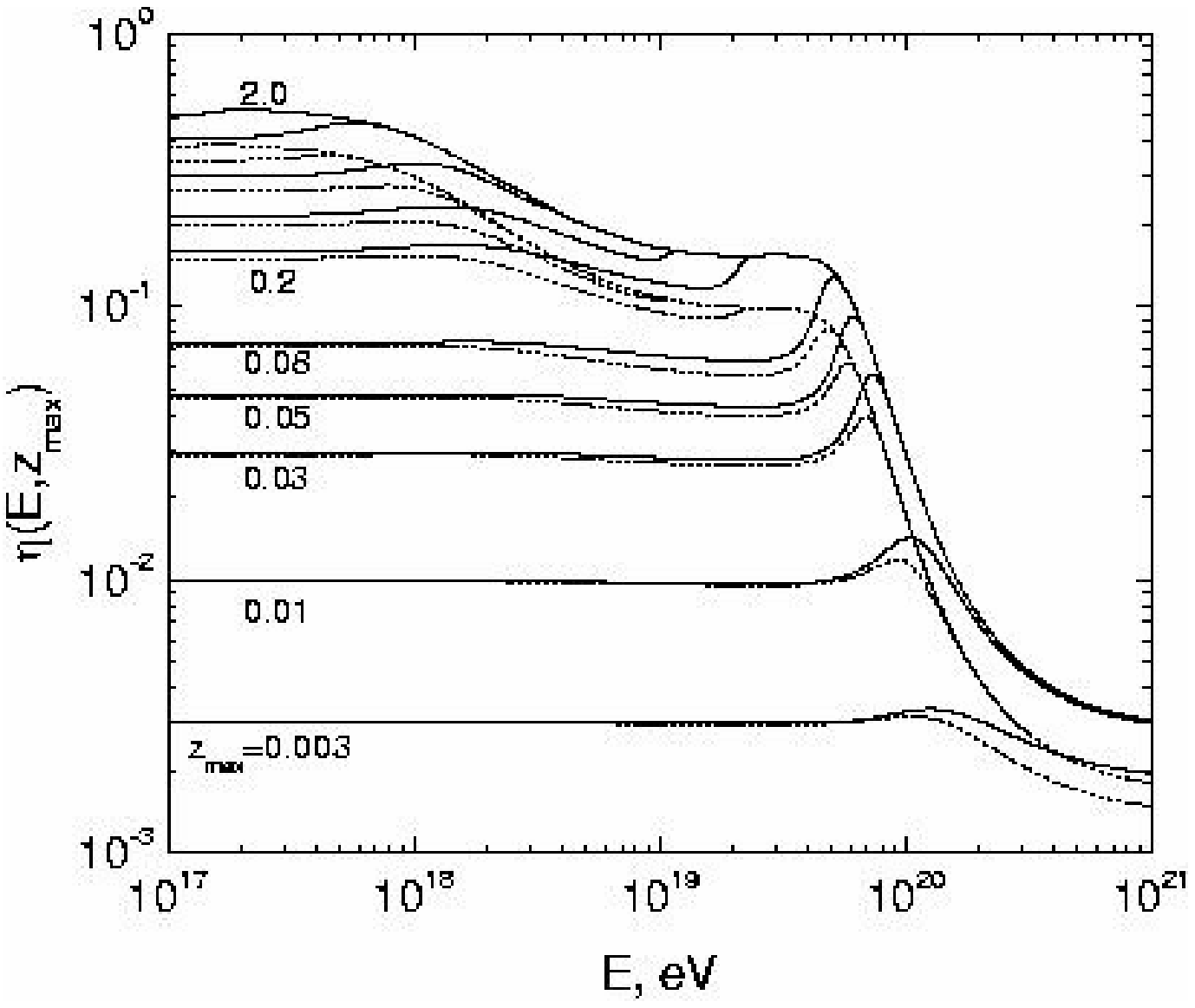}
\end{minipage}
\caption{Bumps for single sources at different distances (left panel)
and disappearance of bumps in diffuse spectra (right panel). The
calculations are performed for cosmological parameters \cite{WMAP}
$H_0=72$~km/sMpc, $\Omega_m=0.27$ and $\Omega_{\Lambda}=0.73$, and
numerically they are different from  Ref.~\cite{BeGrbump}. For the
diffuse spectra (right panel) the sources are distributed uniformly in
the sphere of radius $R_{\rm max}$, corresponding to $z_{\rm max}$. The
solid and dashed curves are for $\gamma_g=2.7$ and $\gamma_g=2.0$,
respectively. The curves between $z_{\rm max}=0.2$ and $z_{\rm
max}=2.0$  have  $z_{\rm max}=0.3,~0.5,~1.0$. } \label{bump}
\end{figure}

We see no indication of the bump in the modification-factor energy
dependence in Fig.~\ref{mfactor}. As explained above, it should have
been located  at merging of $\eta_{ee}(E)$ and $\eta_{\rm tot}(E)$
curves. The absence of the bump in the diffuse spectrum can be easily
understood. The bumps are clearly seen in the spectra of the single
remote sources (see left panel in Fig.~\ref{bump}). These bumps,
located at different energies, produce a flat feature, when they are
summed up in the diffuse spectrum. This effect is illustrated by
Fig.~\ref{bump} (right panel). The diffuse flux there is calculated in
the model where sources are distributed uniformly in the sphere of
radius $R_{\rm max}$ (or $z_{\rm max}$). When $z_{\rm max}$ are  small
(between 0.01 and 0.1) the bumps are seen in the diffuse spectra. When
radius of the sphere becomes larger, the bumps merge producing the flat
feature in the spectrum.  If the diffuse spectrum is plotted as
$E^3J_p(E)$ this flat feature looks like a pseudo-bump.
\subsection{The dip}
\label{sec:dip} The {\em dip} is more reliable signature of
interaction of protons with CMB than GZK feature. The shape of the
GZK feature is strongly model-dependent (see
Section~\ref{sec:GZKcutoff}), while the shape of the {\em dip} is
fixed and has a specific form which is difficult to mimic by other
mechanisms, unless they have many free parameters. The protons in
the dip are collected from the large volume with the radius about
1000~Mpc and therefore the assumption of the uniform distribution
of sources within this volume is well justified, in contrast to
the GZK cutoff, which strongly depends on local overdensity or
deficit of the sources. The GZK cutoff can be mimicked by
acceleration cutoff, and since the shape of the GZK cutoff is not
reliably predicted, these two cases could be difficult to
distinguish.

The problem of identification of the dip depends on the accuracy of
observational data, which should confirm the specific (and well
predicted) shape of this feature. Do the present data have the needed
accuracy?

\begin{figure}[ht]
\begin{minipage}[h]{8cm}
\centering
\includegraphics[width=7.6cm,clip]{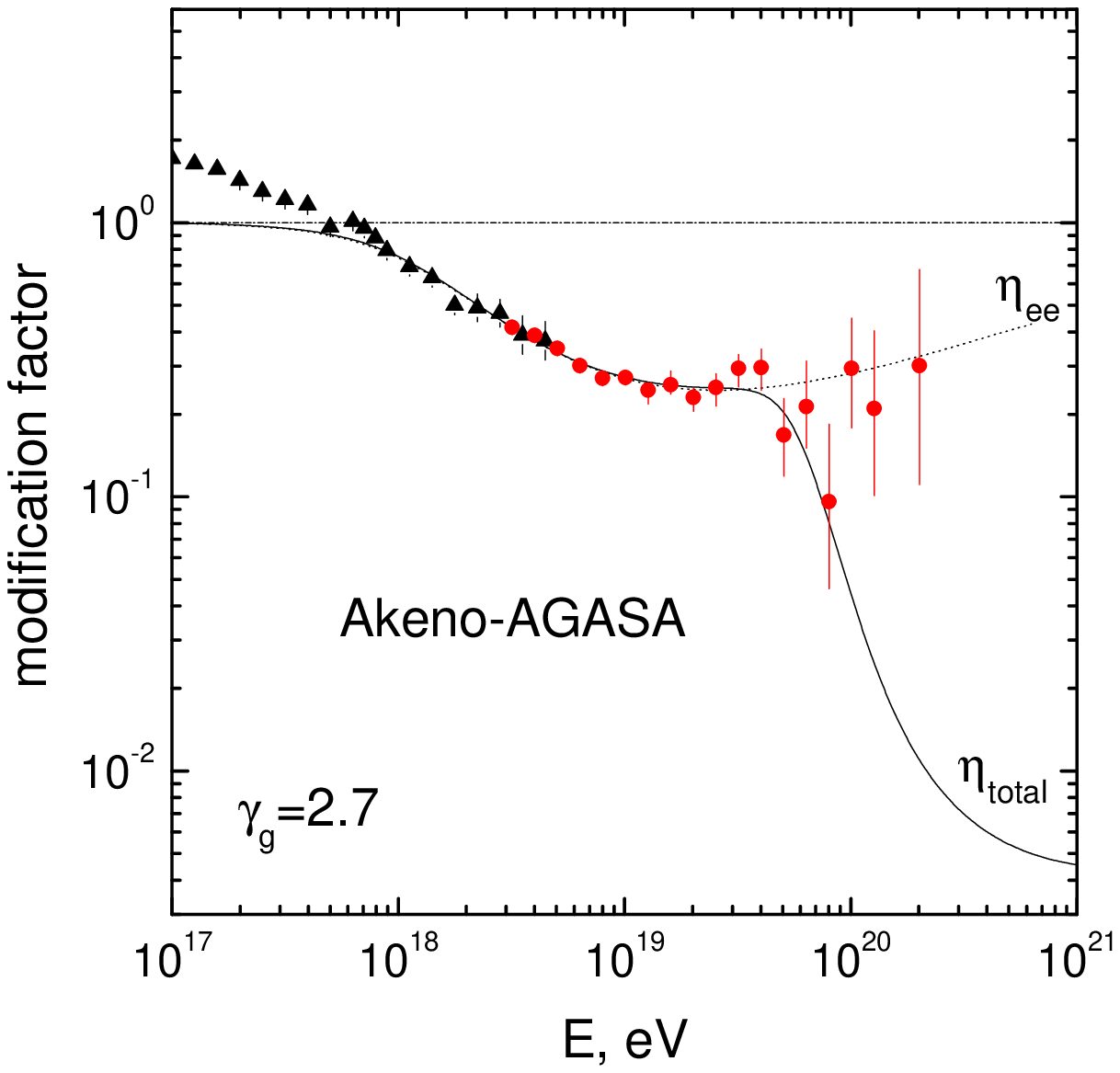}
\end{minipage}
\hspace{2mm}
\begin{minipage}[h]{8cm}
\centering
\includegraphics[width=7.6cm,clip]{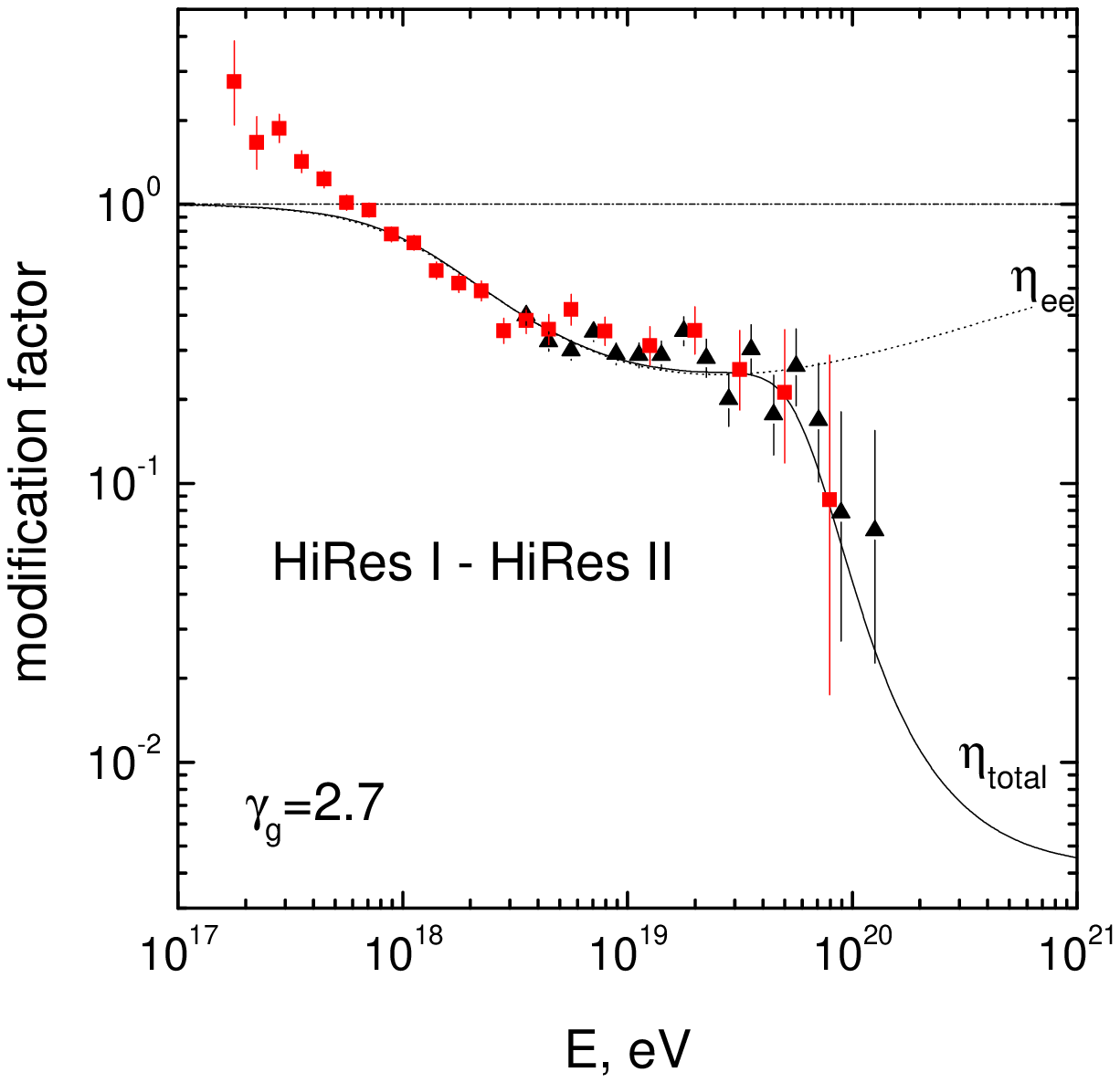}
\end{minipage}
\vspace{2mm}
\begin{minipage}{7.8cm}
\centering
\includegraphics[width=7.6cm,clip]{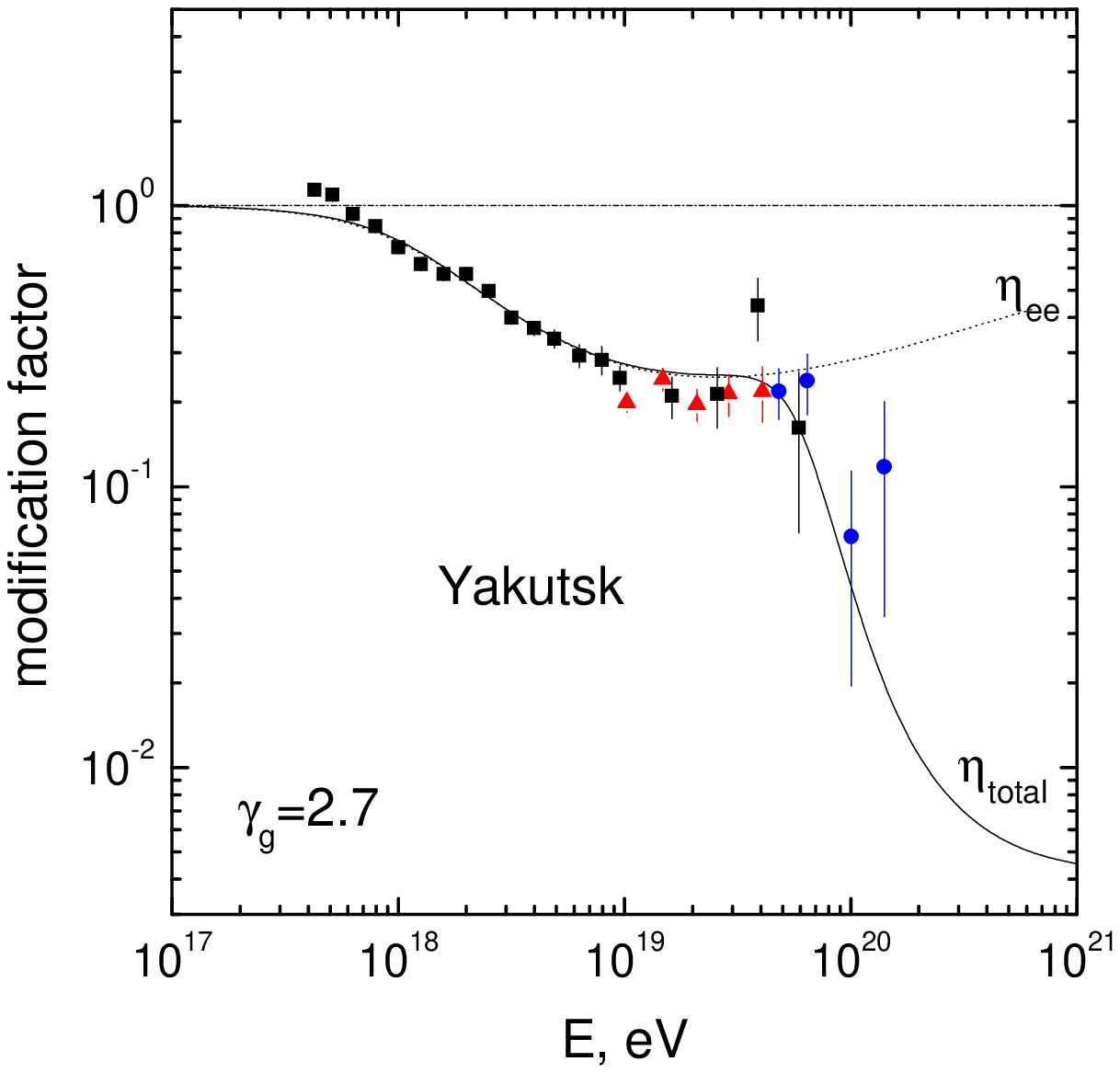}
\end{minipage}
\hspace{5mm} %
\begin{minipage}[h]{7.8cm}
\centering
\includegraphics[width=7.3cm,height=7.4cm,clip]{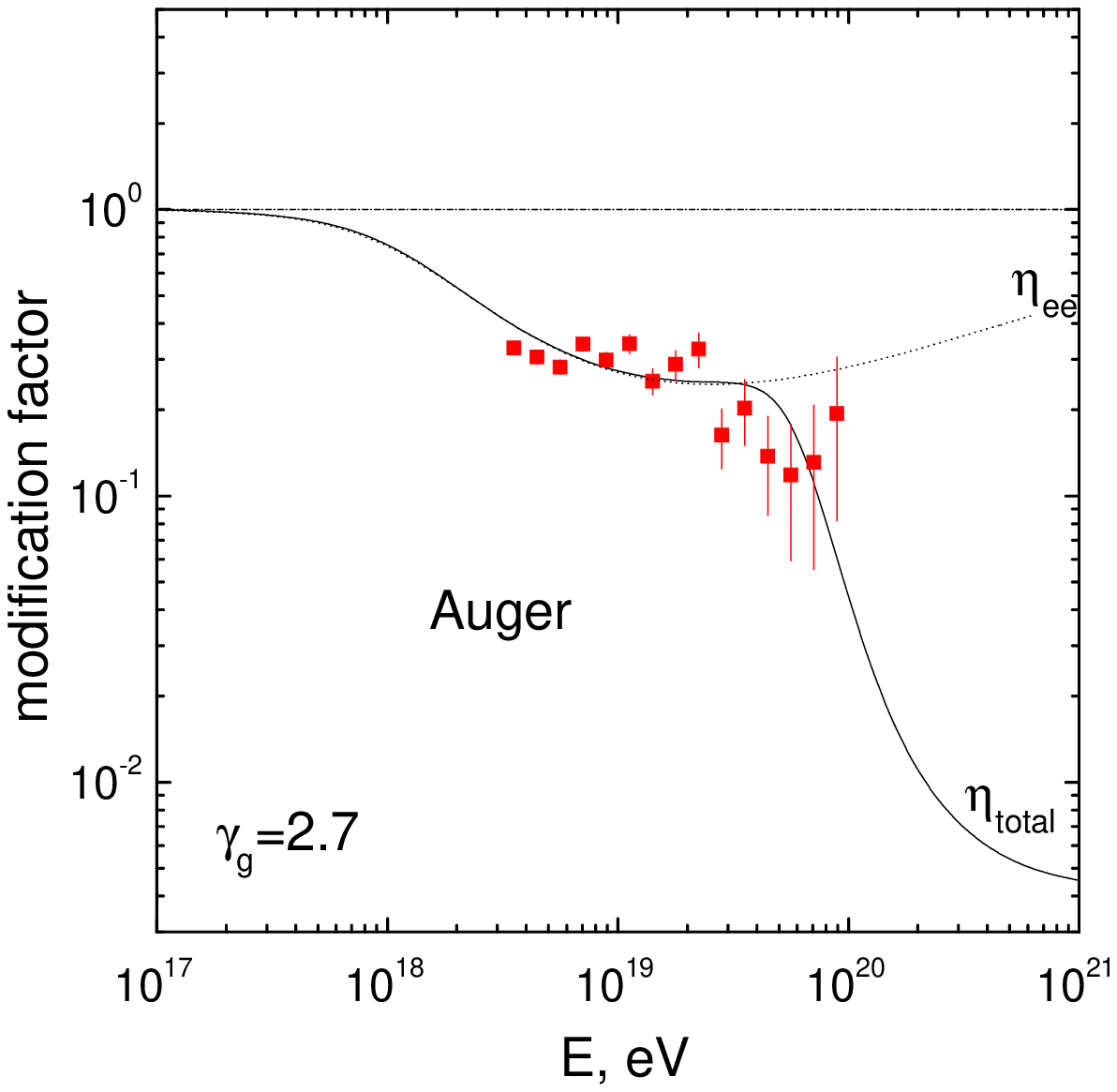}
\end{minipage}
\caption{\label{dips} Predicted dip in comparison with AGASA, HiRes,
Yakutsk and Auger\protect\cite{auger} data.}
\end{figure}

The comparison of the calculated modification factor with that obtained
from the Akeno-AGASA data, using $\gamma_g=2.7$, is given in
Fig.~\ref{dips}. It shows the excellent agreement between predicted and
observed modification factors for the dip.
In Fig.~\ref{dips} one observes that at $E < 1\times 10^{18}$~eV the
agreement between calculated and observed modification factors becomes
worse and the observational modification factor becomes larger than 1.
Since by definition $\eta(E)\leq 1$, it signals the appearance of
another component of cosmic rays, which is almost undoubtedly given by
galactic cosmic rays. The condition $\eta > 1$ implies the dominance of
the new (galactic) component, the transition occurs at $E < 1\times
10^{18}$~eV.

To calculate $\chi^2$ for the confirmation of the dip by
Akeno-AGASA data, we choose the energy interval between $1\times
10^{18}$~eV and $4\times 10^{19}$~eV (the energy of intersection
of $\eta_{ee}(E)$ and $\eta_{\rm tot}(E)$). In calculations we
used the Gaussian statistics for low-energy bins, and the Poisson
statistics for the high energy bins of AGASA. It results in
$\chi^2=19.06$. The number of Akeno-AGASA bins is 19. We use in
calculations two free parameters: $\gamma_g$ and the total
normalization of spectrum. In effect, the confirmation of the dip
is characterized by $\chi^2=19.06$ for d.o.f.=17, or
$\chi^2$/d.o.f.=1.12, very close to the ideal value 1.0 for the
Poisson statistics.

In the right-upper panel of Fig.~\ref{dips} the comparison of
modification factor with the HiRes data is shown. The agreement is also
very good: $\chi^2= 19.5$ for $d.o.f.= 19$ for the Poisson statistics.
The Yakutsk and Fly's eye data (not shown here) agree with dip as well.
The Auger spectrum \cite{auger} at this preliminary stage does not
contradict the dip.

The good agreement of the shape of the dip $\eta_{ee}(E)$ with
observations is a strong evidence for extragalactic protons interacting
with CMB. This evidence is confirmed by the HiRes data on the mass
composition \cite{Sokola,Sokolb}. While the data of the Yakutsk array
\cite{Glushkov2000} and HiRes-MIA \cite{HiRes-MIA} support this mass
composition, and Haverah Park data \cite{HP} do not contradict it at $E
\gtrsim (1 - 2)\times 10^{18}$~eV, the data of Akeno \cite{Akeno-mass}
and Fly's Eye \cite{FE} favor the mixed composition dominated by heavy
nuclei.

The observation of the dip should be considered as independent evidence
in favor of proton-dominated primary composition in the energy range
$1\times 10^{18} - 4\times 10^{19}$~eV.
\subsection{The second dip}
\label{sec:2dip}
The second dip in the spectrum of extragalactic UHE protons
appears at energy $E=6.3 \times 10^{19}$ due to interplay between pair
production and photopion production. It is the direct consequence
of energy $E_{\rm eq2}=6.05\times 10^{19}$~eV, where energy losses due
to $e^+e^-$-production and pion-production become equal
(see Fig.~\ref{loss}). This spectrum feature is explained as follows.

The pion-production energy loss increases with energy very fast,
and at energy slightly below $E_{\rm eq2}$  $e^+e^-$-production
dominates and spectrum can be with high accuracy described in
continuous energy-loss approximation. At energy slightly higher
than  $E_{\rm eq2}$ the pion-production dominates and the precise
calculation of spectrum should be performed in the
kinetic-equation approach.  In this method the evolution of number
of particles in interval $dE$ is given by two compensating terms,
describing the particle exit and regeneration due to $p\gamma$
collisions. The small continuous energy losses affect only the
exit term and break this compensation, diminishing the flux. The
exact calculations are given in Appendix~\ref{app-2dip}. The
second dip is very narrow and its amplitude at maximum reaches 
$\sim 10\%$ (see Fig.~\ref{fluct-ratio}). This feature can be observed by
detectors with very good energy resolution, and it gives the
precise mark for energy calibration of a detector. It can be observed 
only marginally by the Auger detector. 
\section{\label{sec:robustness} Robustness of the dip prediction}
We calculated the dip for the universal spectrum, i.e.\ for the case
when distances between sources are small enough, and the spectrum does
not depend on the propagation mode. In this Section we shall study
stability of the dip relative to other possible phenomena, namely,
discreteness in the source distribution, propagation in magnetic fields
etc. We shall consider also some phenomena related to existence of the
dip.

\subsection{Discreteness in the source distribution}
\label{sec:discreteness} 
As it follows from analysis of the small-scale anisotropy
(see \cite{DTT,FK,YNS1,Sigl1,YNS2}), the average distance 
between UHECR sources is $d \sim 30 - 50$~Mpc \cite{BlMa,KaSe}.
Such discreteness affects the spectrum, especially at highest energies,
when energy attenuation length is comparable with $d$.
\begin{figure}[htb]
\includegraphics[width=9.5cm]{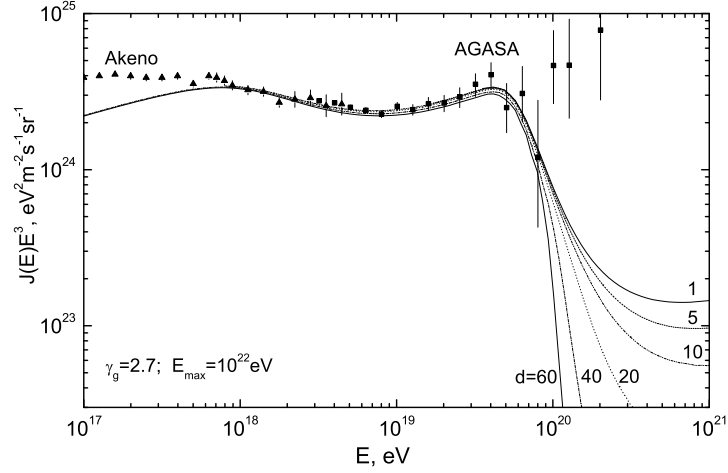}
\vspace{2mm} \caption{Proton spectra for rectilinear propagation from
discrete sources. Sources are located in vertices of 3D cubic grid with
spacing $d=60, 40, 20, 10, 5$ and $1$ Mpc. The calculations are
performed for $z_{max} = 4$, $E_{max}=1\times 10^{22}$ eV and $\gamma_g
= 2.7$. } \label{discretness}
\end{figure}
In this subsection we demonstrate the stability of the dip relative to
discreteness of the sources. We illustrate the effect of discreteness
by an example of UHE protons propagating rectilinearly from sources
located in the vertices of a 3D cubic lattice with spacing $d$.
Positions of sources are given by coordinates $x=id$, $y=jd$ and
$z=kd$, where $i, j, k = 0, \pm 1, \pm 2 \ldots$. The observer is assumed at
$x=y=z=0$ with no source there.
The diffuse flux for the power-law generation spectrum $\propto
E_g^{-\gamma_g}$ is given by summation over all vertices. The maximum
distance is defined by the maximum red-shift $z_{max}$. Then the
observed flux is given by
\begin{equation}
J_p(E)= \frac{(\gamma_g-2){\cal L}_0 d}{(4\pi)^2} \sum \limits_{i,j,k}
\frac{E_g(E,z_{ijk})^{-\gamma_g}}{(i^2+j^2+k^2)(1+z_{ijk})}
\frac{dE_g}{dE}, \label{flux-lattice}
\end{equation}
where ${\cal L}_0=L_p/d^3$ is emissivity, $z_{ijk}$ is the red-shift
for a source with coordinates $i, j, k$, and factor $(1+z_{ijk})$ takes
into account the time dilation.

The calculated spectra for $d=1, 5, 10, 20, 40$ and $60$~Mpc  are shown
in Fig.~\ref{discretness} in comparison with the AGASA-Akeno data. In
calculations we used $E_{\rm max}=1\times 10^{22}$~eV, $m=0$ (no
evolution), $z_{\rm max}=4$ and $\gamma_g = 2.7$. Emissivity ${\cal
L}_0$ is chosen to fit the AGASA data. One can see that discreteness in
the source distribution affects weakly the dip, but the effect is more
noticeable for the shape of the GZK cutoff.

With $d$ decreasing, the calculated spectra regularly converge to
the universal one, as it should be according to propagation theorem \cite{AB}.
This theorem ensures also that the spectra from Fig.~\ref{flux-lattice}
are valid for the case of weak magnetic field, when the diffusion
length $l_{\rm diff} \gsim d$.
\subsection{Dip in the case of diffusive propagation} %
\label{sec:dip-diffusion} The dip, seen in the universal spectrum, is
also present in the case of diffusive propagation in magnetic field
\cite{AB1}. The calculations are performed for diffusion in random
magnetic fields with the coherent magnetic field $B_0= 1$~nG
\begin{figure}[htb]
\includegraphics[width=8.0cm]{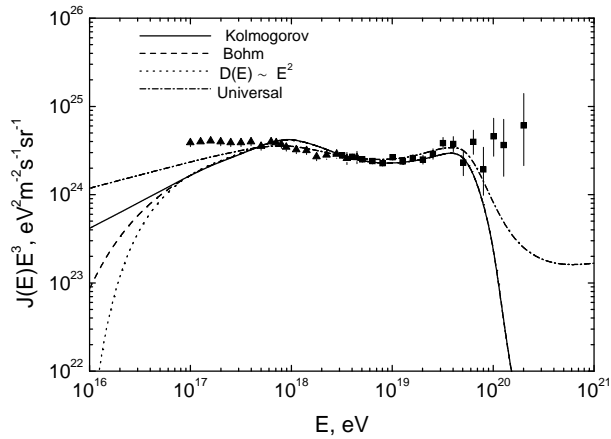}
\vspace{2mm} 
\caption{Diffusive energy spectrum in the case of $B_0=1$
nG, $l_c=1$ Mpc and for the diffusion regimes: Kolmogorov (continuous
line), Bohm (dashed line) and $D(E)\propto E^2$ (dotted line). The
separation between sources is $d=50$ Mpc and the injection spectrum
index is $\gamma_g=2.7$. The universal spectrum (dash-dotted line) 
and the AGASA-Akeno data  are also shown.} 
\label{dip-diff}
\end{figure}
and up to $B_0= 100$~nG on the basic scale $l_c= 1$~Mpc. The calculated
spectrum is shown in Fig.~\ref{dip-diff} for the case $(B_0, l_c)=
(1~{\rm nG}, 1~{\rm Mpc})$ and distance between sources $d=50$~Mpc. The
critical energy, where diffusion changes its regime is  $E_b \approx
1\times 10^{18}$~eV. The spectra are shown for three different regimes
at $E < E_b$ as indicated  in Fig.~\ref{dip-diff}. The universal
spectrum is also presented. One can see that the dip agrees well with
universal spectrum and observational data, while the shape of the GZK
cutoff differs considerably from the universal spectrum.

These calculations demonstrate stability of the dip relative to
changing of the propagation mode, and sensitivity of the GZK cutoff to
the way of propagation.
\subsection{Energy calibration of the detectors using the dip}
\label{sec:Ag-Hi-discrep} 

Since the position and shape of the dip is robustly fixed by proton
interaction with CMB, it can be used as energy calibrator for the
detectors. We use the following procedure for the calibration. 
Assuming the energy-independent systematic error, we shift the 
energies in each given experiment by factor $\lambda$ to
reach minimum $\chi^2$ for comparison with the calculated dip. The 
systematic errors 
in energy determination of existing detectors exceed 20\%, and it
determines the expected value of $\lambda$. The described procedure
results in $\lambda_A=0.9$, $\lambda_{\rm Ya}=0.75$ and 
$\lambda_{\rm Hi}=1.2$ for AGASA, Yakutsk and HiRes detectors,
respectively. After this energy calibration 
the fluxes in all experiments agree with each other.
\begin{figure}[ht]
\vspace{-2mm}
\begin{minipage}[h]{8cm}
\centering \vspace{-1mm}
\includegraphics[width=7.4cm,height=6.9cm]{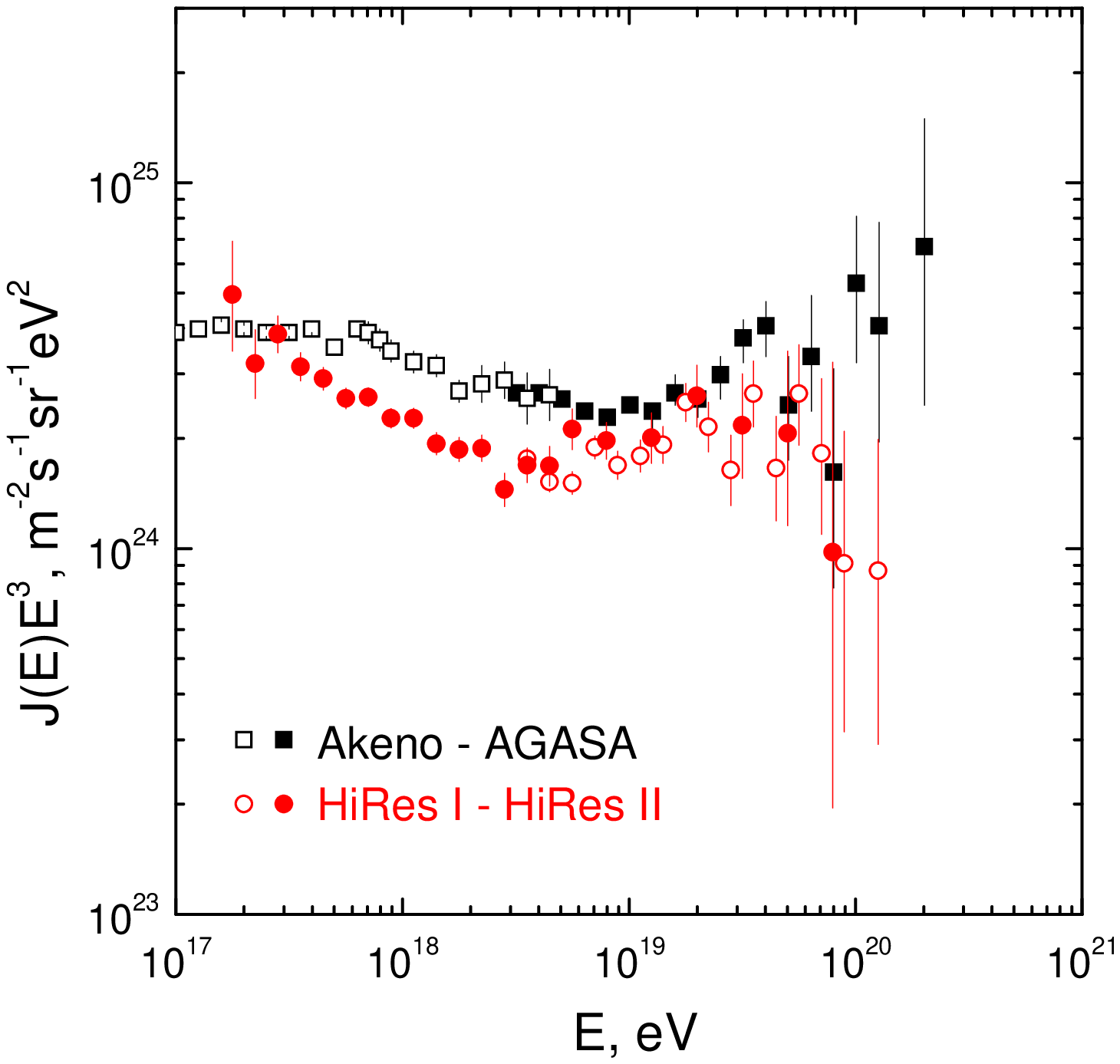}
\end{minipage}
\hspace{2mm} \vspace{1mm}
\begin{minipage}[h]{8cm}
\centering
\includegraphics[width=7.3cm,clip]{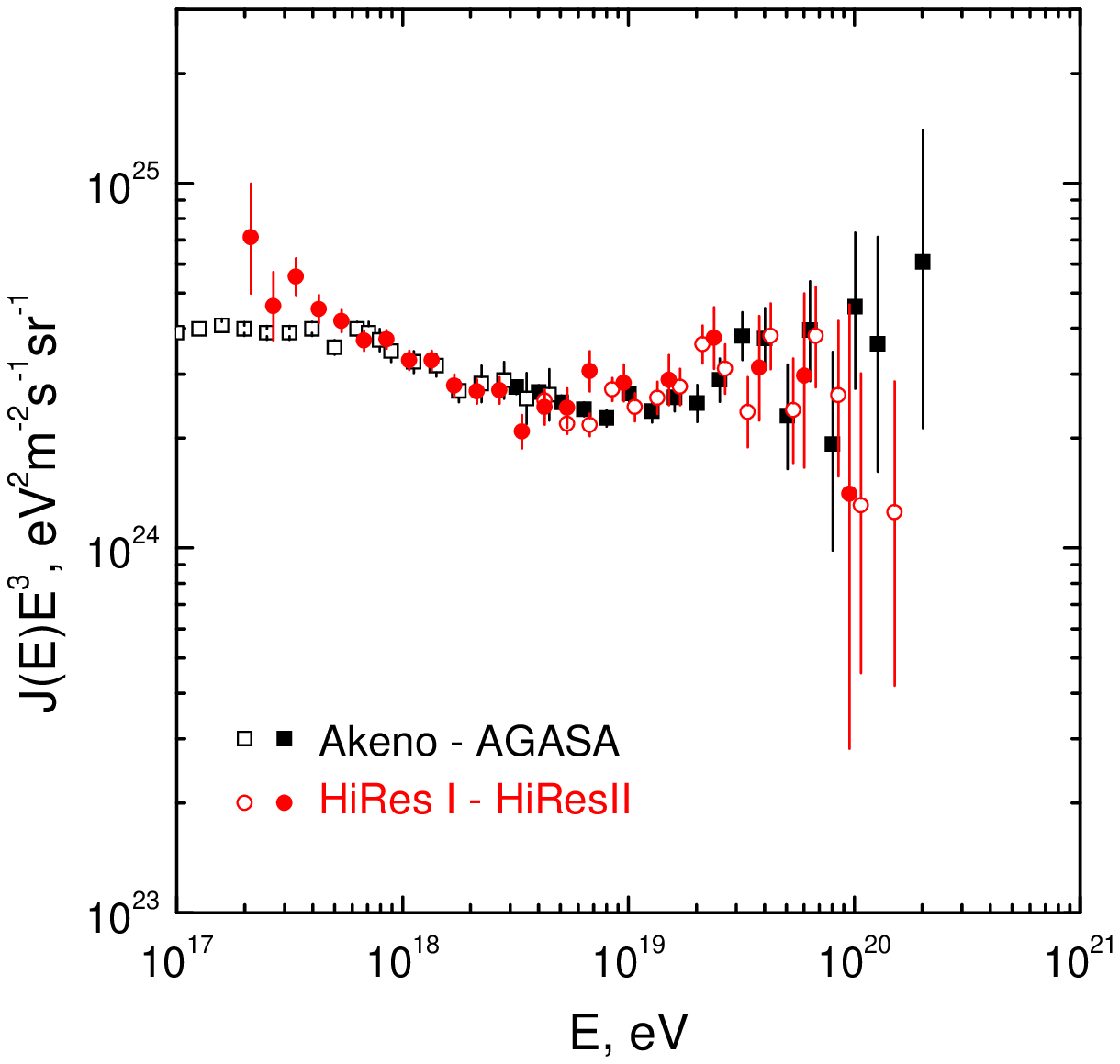}
\end{minipage}
\begin{minipage}{8cm}
\centering
\includegraphics[width=7.5cm,clip]{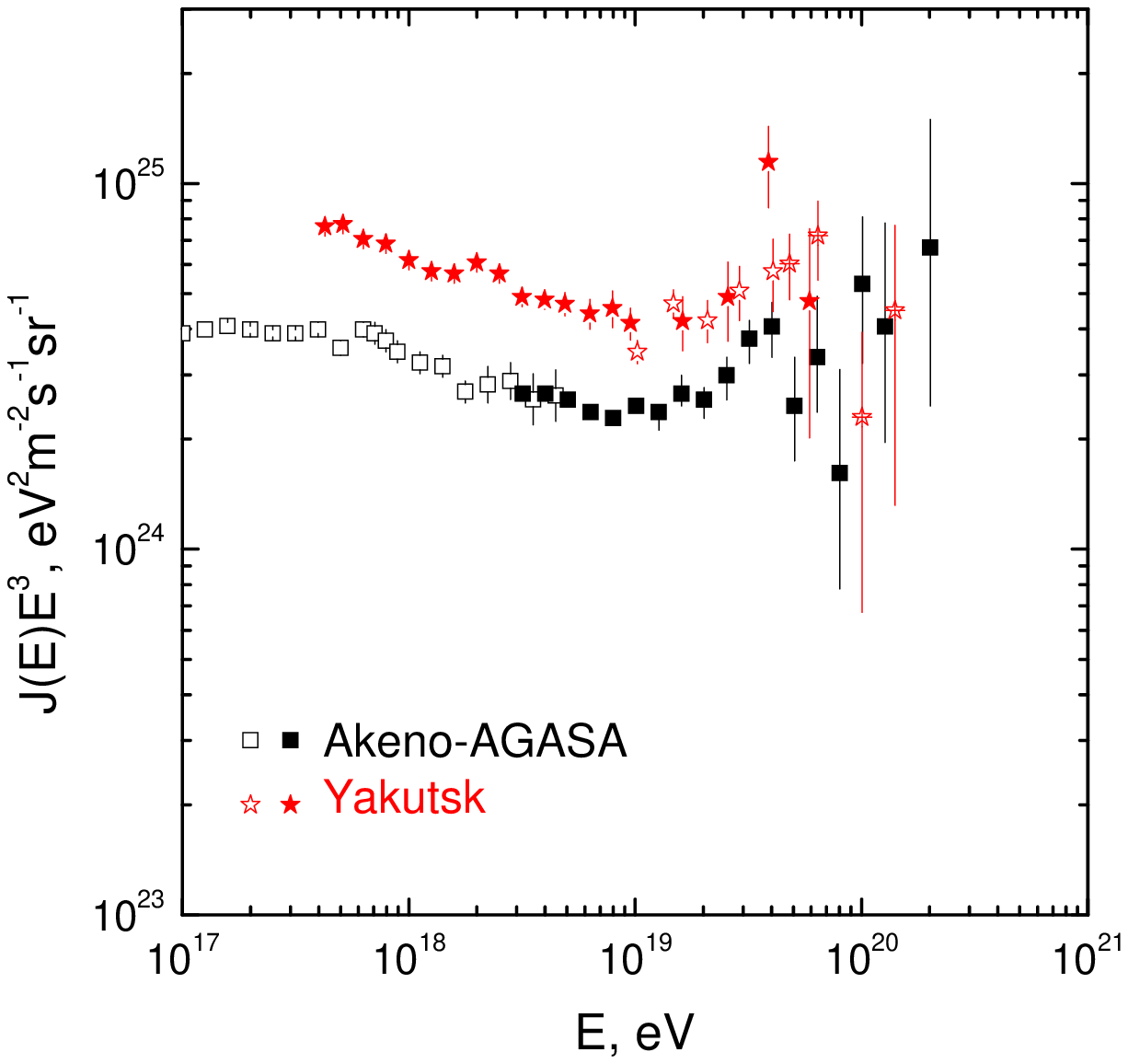}
\end{minipage}
\hspace{2mm}
\begin{minipage}[h]{8cm}
\centering
\includegraphics[width=7.3cm,height=7.1cm,clip]{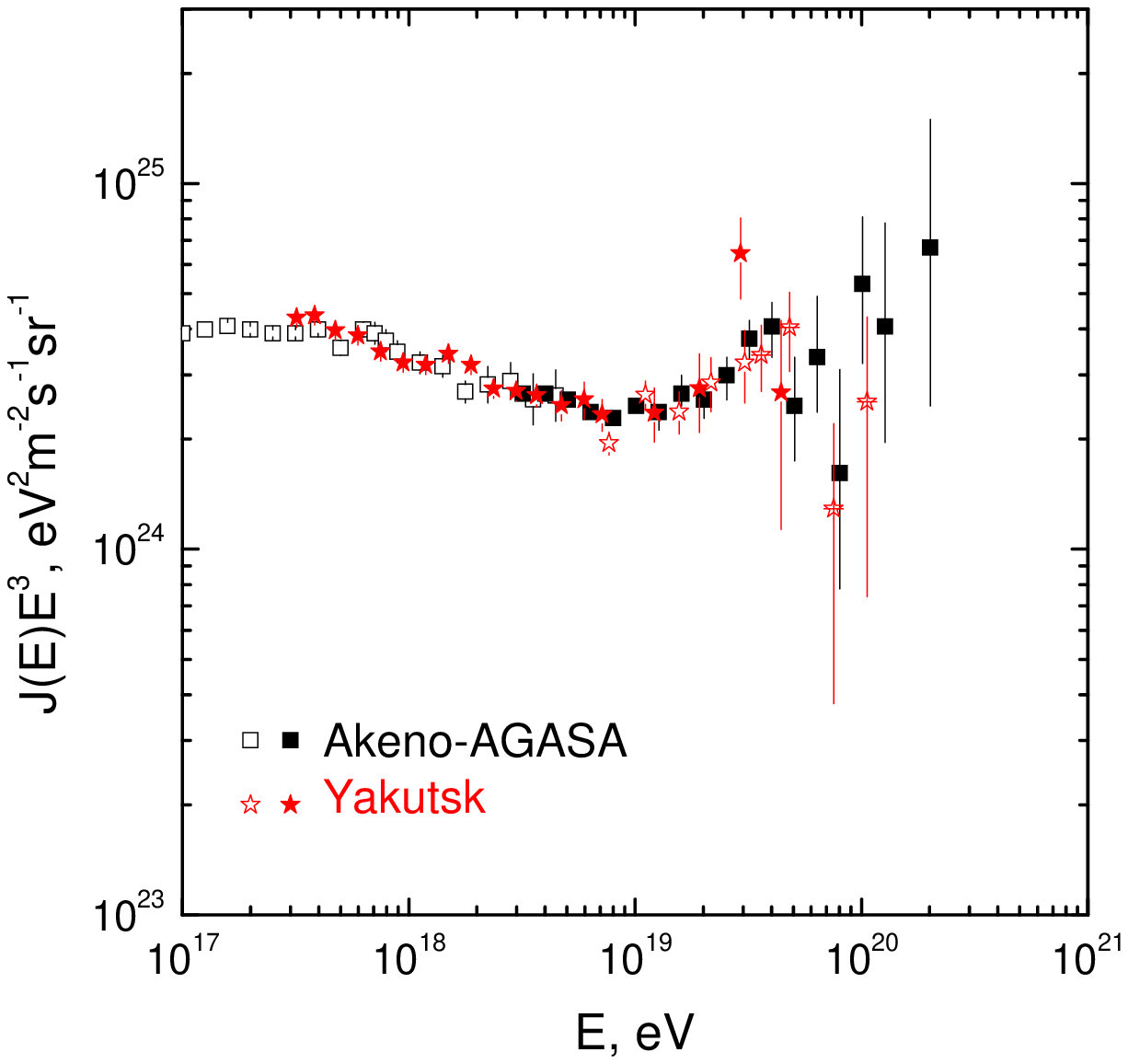}
\end{minipage}
\caption{\label{Ag-Hi-Ya} Spectra of Akeno-AGASA, HiRes and  Yakutsk
before and after energy calibration by the dip. The spectra with energy
shift are shown in the right panels. The energy shifts needed for the
best fit of the dip are $\lambda_{\rm Ag}=0.9$ ,  $\lambda_{\rm Hi}=1.2$
and $\lambda_{\rm Ya}=0.75$ for AGASA, HiRes and Yakutsk, respectively.
}
\end{figure}
First we consider the AGASA and HiRes
data. There are two discrepancies between these data (see the upper-left
panel of Fig.~\ref{Ag-Hi-Ya}): one is described by factor 1.5 - 2.0 in
energy region $1\times 10^{18} - 8\times 10^{19}$~eV, and the second -- 
at $E \geq 1\times 10^{20}$~eV. In Fig.~\ref{Ag-Hi-Ya} the spectra of
Akeno-AGASA and HiRes are shown before and after the energy
calibration. One can see the good agreement of the calibrated data at $E <
1\times 10^{20}$~eV and their consistency at $E > 1\times
10^{20}$~eV. This result should be considered together with
calculations \cite{DBO}, where it was demonstrated that 11
superGZK AGASA events can be simulated by the spectrum with GZK
cutoff in case of 30\% error in energy determination. We may
tentatively conclude that existing discrepancy between AGASA and
HiRes spectra at all energies are due to systematic energy errors and 
statistics.

In Fig.~\ref{Ag-Hi-Ya} the energy spectra are shown for AGASA and
Yakutsk spectra before and after  
energy calibration. Again, the best fit to the
dip shape results in excellent agreement in the absolute values of
fluxes.

The agreement between spectra of all three detectors after energy
calibration by the dip confirms the dip as the spectrum feature
produced by interaction of the protons with CMB, and demonstrates
compatibility of fluxes measured by AGASA, HiRes and Yakutsk detectors.

\subsection{Dip and extragalactic UHE nuclei}
\label{sec:nuclei} 
The proton dip has very good agreement with
observations (see Fig.~\ref{dips}). However, in all astrophysical
sources the nuclei must be also accelerated to the energies, naively, Z
times higher than that for protons. Do  UHE nuclei in primary flux
upset the good agreement seen in Fig.~\ref{dips}? This problem has been
recently considered in \cite{BGG-PL,Allard,Sigl05} (for study of
propagation UHE nuclei through CMB see
\cite{photodisinta,photodisintb,photodisintc}).
\begin{figure}[ht]
\begin{minipage}[h]{8cm}
\centering
\includegraphics[width=7.6cm,clip]{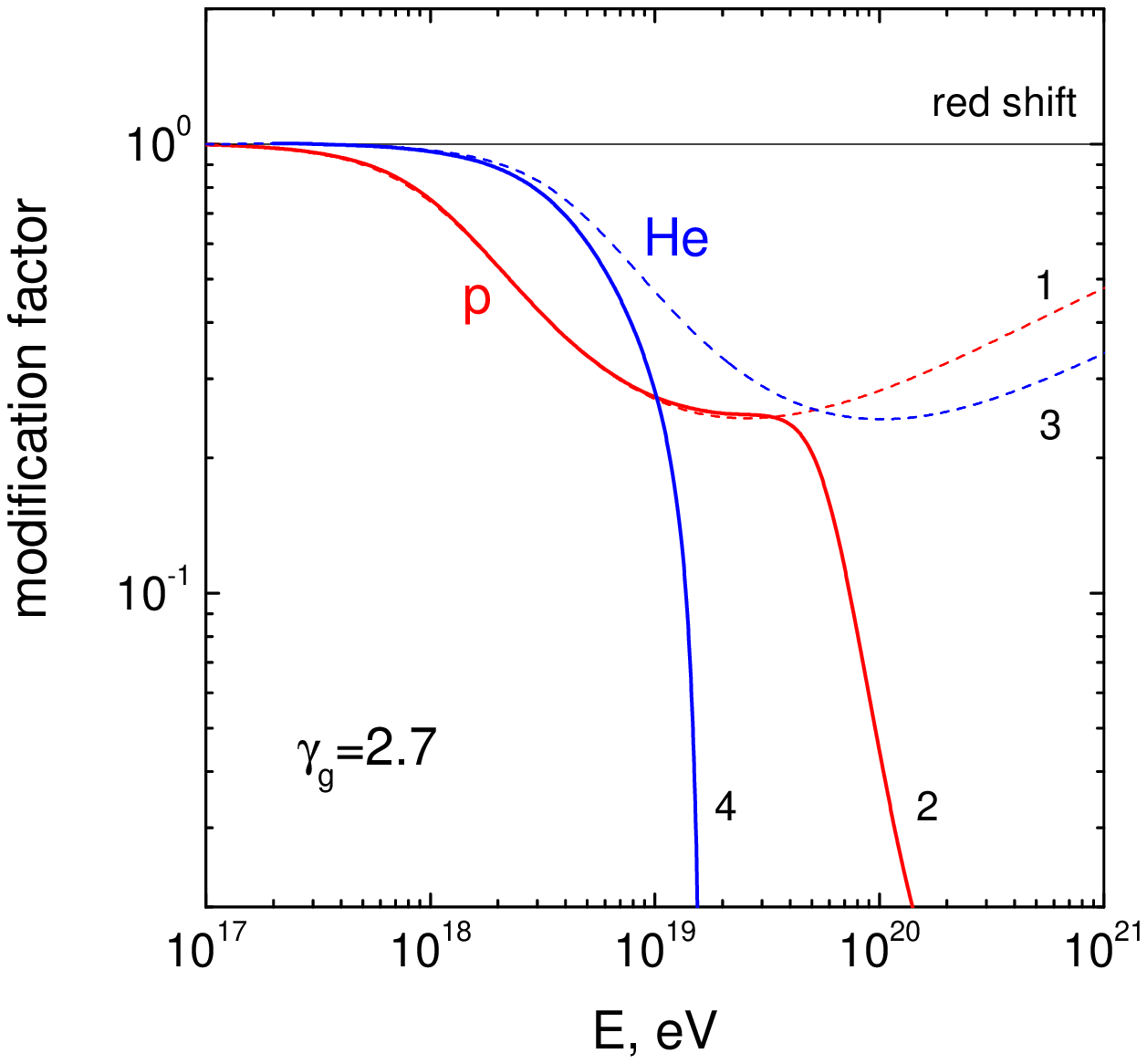}
\end{minipage}
\hspace{5mm}
\begin{minipage}[h]{8cm}
\centering
\includegraphics[width=7.6cm,clip]{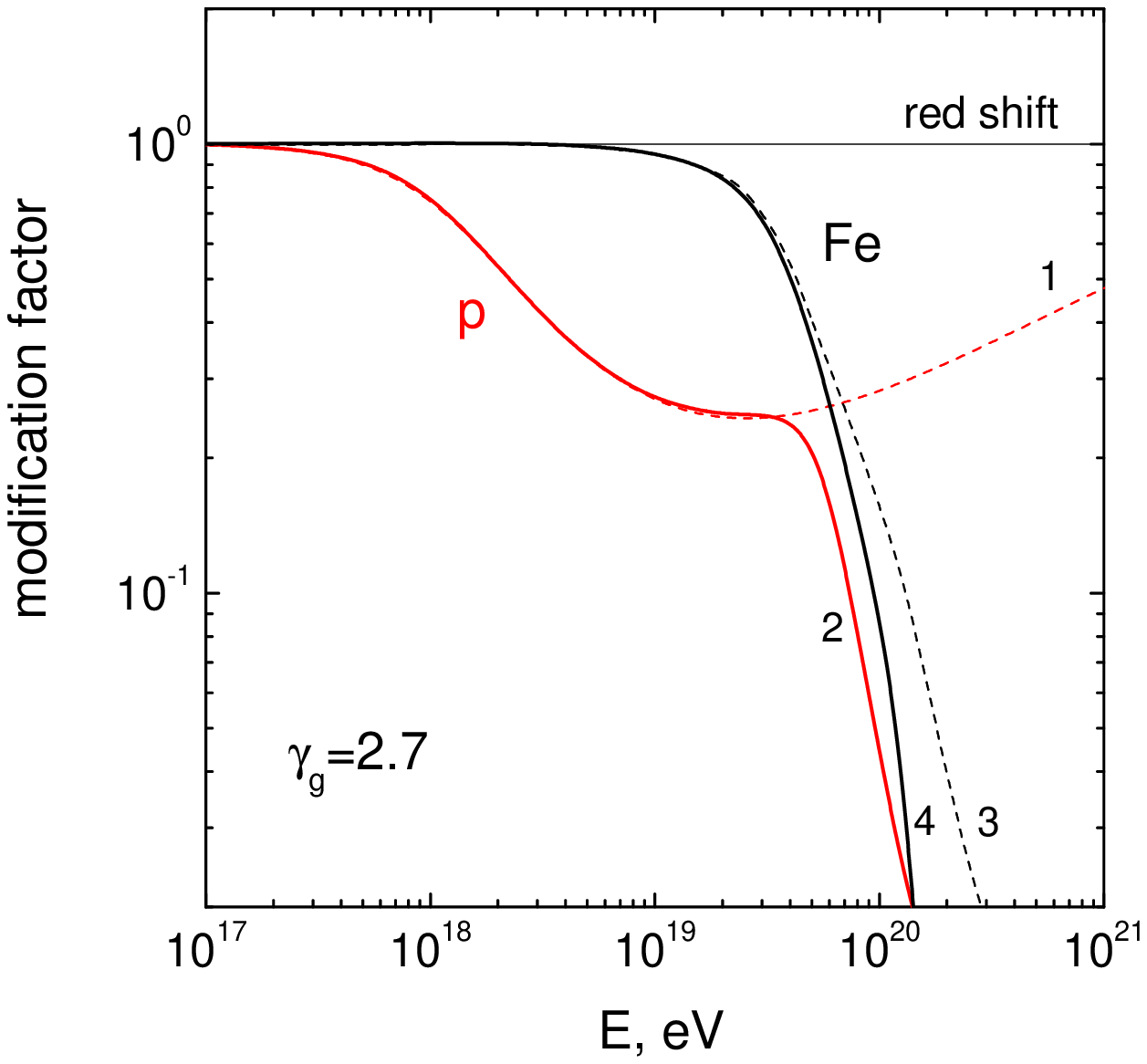}
\end{minipage}
\caption{
\label{dip-nucl} %
Modification factors for helium and iron
nuclei in comparison with that for protons. Proton modification
factors are given by curves 1 and 2. Nuclei modification factors
are given by curves 3 (adiabatic and pair production energy
losses) and by curves 4 (with photodisintegration included).
 }
\end{figure}
\begin{figure}[ht]
\begin{minipage}[h]{8cm}
\centering
\includegraphics[width=7.6cm,clip]{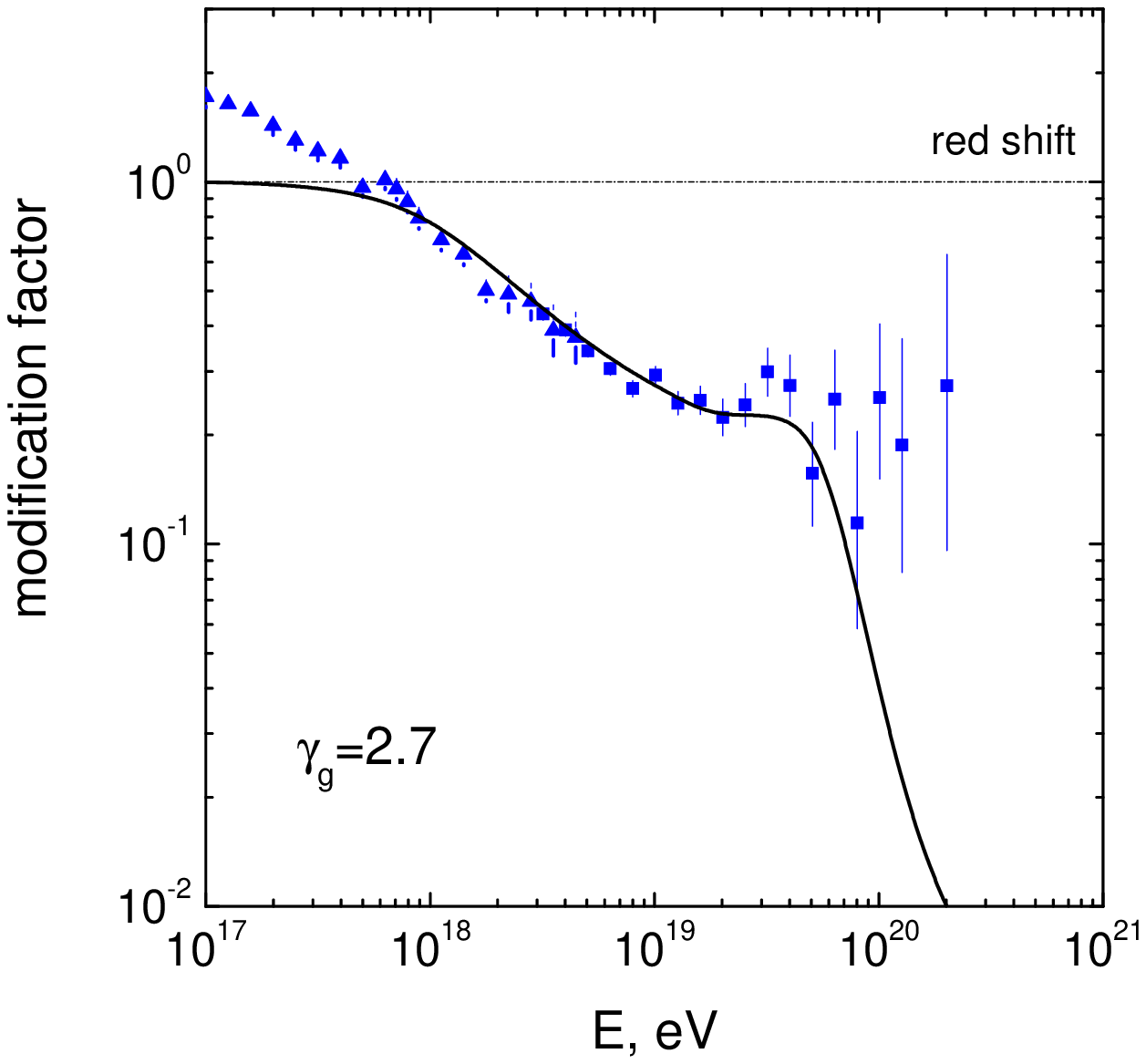}
\end{minipage}
\hspace{5mm}
\begin{minipage}[h]{8cm}
\centering
\includegraphics[width=7.6cm,clip]{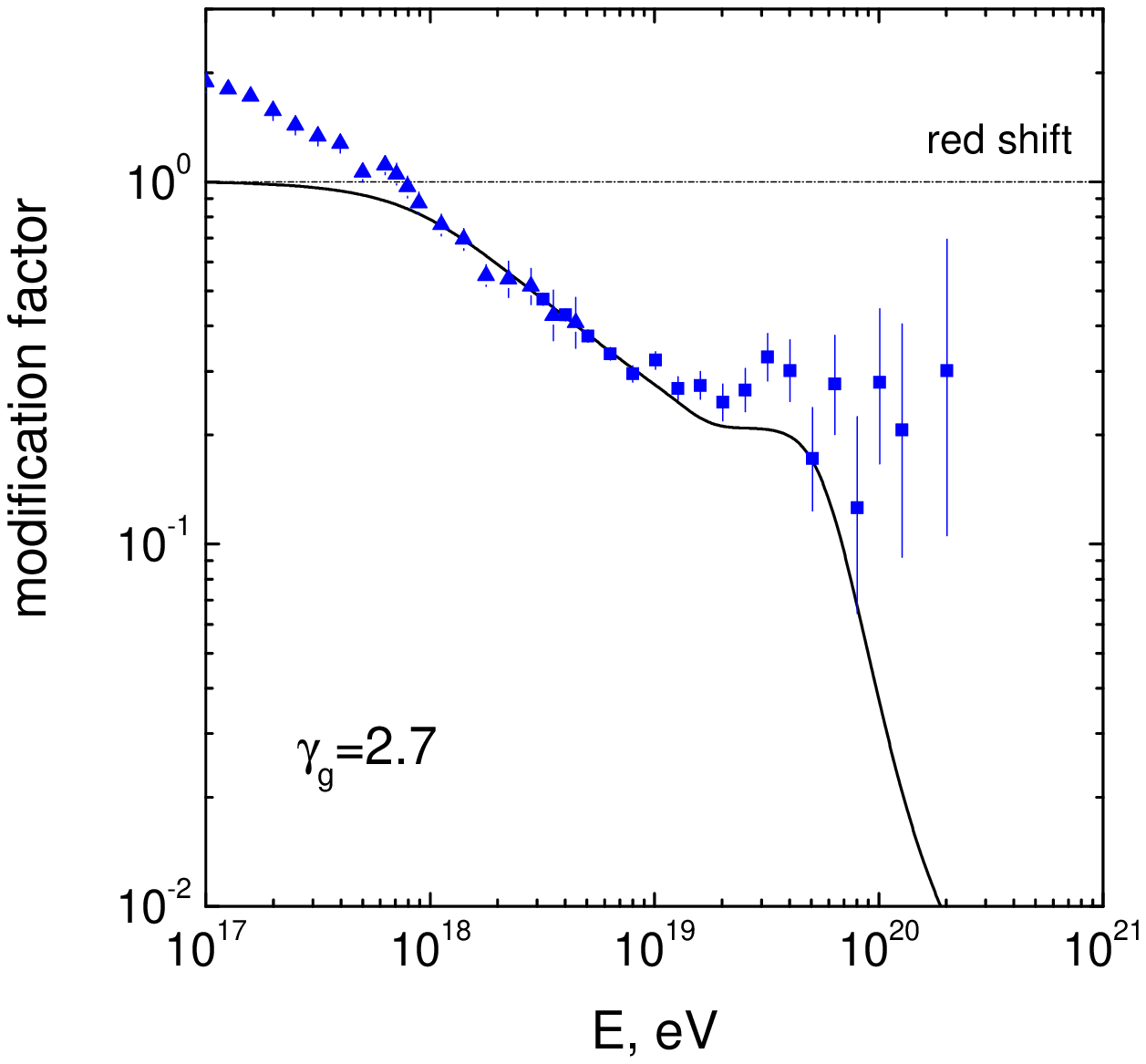}
\end{minipage}
\caption{
\label{dip-mix} %
Modification factors for the mixed composition
of protons and helium nuclei in comparison with AGASA data. The left
panel corresponds to mixing parameter $\lambda=0.1$, and the right
panel to $\lambda=0.2$. }
\end{figure}
The presence of nuclei in primary extragalactic spectrum modifies the
dip \cite{BGG-PL}.  In Fig.~\ref{dip-nucl} the dips for helium and iron
nuclei are shown in comparison with that for protons. From this figure
one can see that presence of 15 - 20 \% of nuclei in the primary flux
breaks the good agreement of proton deep with observations. The
modification factor for cosmic rays composed of protons and nuclei with
the fraction $\lambda_A= Q_A(E)/Q_p(E)$, where $Q(E)$ is generation
function for nuclei (A) and protons (p), can be easily calculated as
\beq %
\eta_{\rm tot}(E)=\frac{\eta_p(E)+ \lambda\eta_A(E)}{1+ \lambda}.
\label{mfactor-A} %
\eeq %
The fraction $\lambda$ is a model-dependent value, which depends
on ratio of number densities of gas components $n_A/n_H$ in media,
where acceleration operates, on ionization of the gas and on the
injection mechanism of acceleration \cite{Berez05}. Besides, the 
UHE nuclei can be destroyed inside the source or in its vicinity 
\cite{Sigl05}.

The strongest distortion of proton
modification factor is given by helium nuclei, for which $n_{\rm
He}/n_{\rm H} \approx 0.08$ corresponding to the helium mass
fraction $Y_p=0.24$. In Fig.~\ref{dip-mix} the modification
factors are shown for the mixed composition of protons and helium
with the mixing parameter $\lambda=0.1$ (left panel) and
$\lambda=0.2$ (right panel). One can judge from these graphs about
allowed values of mixing parameter $\lambda$. If agreement of the
proton dip with observations is not incidental (the probability of
this is small according to small $\chi^2$/d.o.f.),
Fig.~\ref{dip-mix} should be interpreted as indication to possible
acceleration mechanism \cite{Berez05}.

\subsection{Dip and cosmological evolution of the sources}
\label{sec:evolution}
The cosmological evolution of the sources, i.e. increase of 
the luminosities and/or space densities with red-shift $z$, is 
observed for many astronomical populations.  
The evolution  is reliably observed for star
formation rate in the normal galaxies, but this case is irrelevant for 
UHECR, because neither stars nor normal galaxies can be the  UHECR
sources due to low cosmic-ray luminosities $L_p$ and 
maximum energy of acceleration $E_{\rm max}$.  
\begin{figure}[ht]
\centering
\includegraphics[width=85mm]{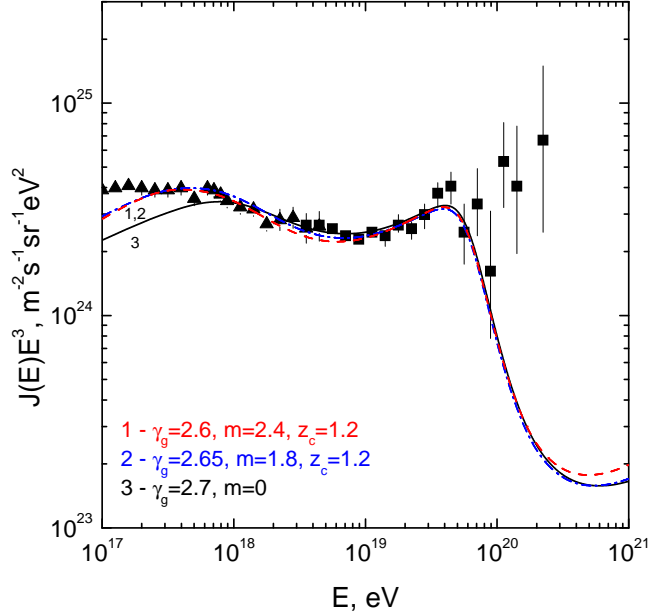}
\caption{The dip in evolutionary models in comparison with the
AGASA data. The parameters of evolution used in the calculations for
curves 1 and 2 are similar to those observed for AGN. The curve 3 is 
for $m=0$.} 
\label{evolution} %
\end{figure}
AGN, which satisfy these 
requirements, also exhibit the evolution seen in radio, optical and X-ray 
observations. The X-ray radiation is  probably most relevant tracer for 
evolution of UHECR because both radiations are feed by the energy release 
provided by accretion to a massive black hole:
X-rays -- through radiation of accretion disk, and UHECR -- through 
acceleration in the jets. According to 
recent detailed analysis in \cite{Ueda} and \cite{Barger} the
evolution of AGN seen in X ray radiation can be  described by factor 
$(1+z)^m$ up to $z_c \approx 1.2$ and is saturated at larger
$z$.  In \cite{Ueda} the pure luminosity evolution and pure density
evolution are allowed with $m=2.7$ and $m=4.2$, respectively, and 
with $z_c \approx 1.2$ for both cases. In \cite{Barger} the
pure luminosity evolution is considered as preferable with $m=3.2$ 
and $z_c =1.2$. These authors do not distinguish between
different morphological types of AGN. It is possible that some AGN 
undergo weak cosmological evolution, or no evolution at all. 
BL Lacs \cite{corr}, which are important as potential UHECR sources,  
show no signs of positive cosmological evolution \cite{BLLac}.

In case of UHECR there is no need to distinguish between luminosity
and density evolution, because the diffuse flux is determined by 
the emissivity, which includes both luminosity and space evolution, 
as it follows from  Eq.~(\ref{Q_gen}).

In Fig.~\ref{evolution} we present the calculated dip spectrum in
evolutionary models, inspired by the data cited above. For
comparison we show also the case of absence of evolution $m=0$, as 
can be valid for BL Lacs. 
From Fig.~\ref{evolution} one can see  that the spectra with evolution up 
to $z_c > 1$ can explain the observational data down to 
(a few)$\times 10^{17}$~eV and even below, in accordance with early
calculations \cite{BGG1,Steckerb,Fodor} (see \cite{Bergman} for recent 
analysis). However, for any
reasonable magnetic fields protons with these energies have small 
diffusion lengths and the spectrum 
acquires the diffusion 'cutoff' at energy $E_b=1\times 10^{18}$~eV
(see Section \ref{sec:dip-diffusion}).

We conclude that for many reasonable evolution regimes the dip agrees
with observational data as well as the non-evolutionary case $m=0$.   

\section{\label{sec:fluctuations} Role of interaction fluctuations}
UHE proton spectrum is affected by fluctuations in the photopion
production. These fluctuations may change the proton spectrum only at
energy substantially higher than $E= 4\times 10^{19}$~eV. At this
energy the half of energy losses is caused by $e^+e^-$ production which
does not fluctuate. Up to energy $1\times 10^{20}$~eV the
photoproduction of pions occurs at the threshold in collisions with
photons from high-energy tail of the Planck distribution, and fraction
of energy lost does not fluctuate, being fixed by the threshold value.
Indeed, for $E_p=1\times 10^{20}$~eV the minimal energy of CMB photon
needed for pion production is $\epsilon=3\times 10^{-3}$~eV to be
compared with energy of photon in the Planck distribution maximum
$\epsilon_m=3.7\times 10^{-4}$~eV. The only fluctuating value is the
interaction length.

The noticeable effect of fluctuations is expected for protons with
energies $E > 1\times 10^{20}$~eV.

As it is well known \cite{LL10,GiSy}, the kinetic equations give an
adequate method to account for the fluctuations in interaction.
Neglecting the conversion of proton to neutron (neutron decays back to
proton with small energy loss) the kinetic equation for UHE protons
with adiabatic energy losses and with $p+\gamma \rightarrow p+e^+ +
e^-$ and $p+\gamma \rightarrow N + pions$ scattering in collisions with
CMB photons can be written down as follows:

\begin{eqnarray} %
\label{flucteqn} %
\frac{\partial n_p(E,t)}{\partial t} = -3 H(t) n_p(E,t) +
\frac{\partial}{\partial E}
\left \{ \left [ H(t)E+ b_{pair}(E,t) \right ] n_p(E,t) \right\} -  \\ %
P(E,t)n_p(E,t)+ \int_E^{E_{max}} dE' P(E',E,t) n_p(E',t) +
Q_{gen}(E,t), \nonumber %
\end{eqnarray} %
where $n_p(E,t)$ is the number  density of UHE protons per unit energy,
$Q_{gen}(E,t)$ is the generation rate, given by Eq.~(\ref{Q_gen}) with
$m=0$ and $H(t)$ is the Hubble parameter. The first term in the
r.h.s.\ of Eq.~(\ref{flucteqn}) describes expansion of the universe. The
energy loss $b_{pair}(E)$ due to $e^+ e^-$-pair production is treated
as continuous energy loss. The photopion collisions are described with
help of probability $P(E,t)$ for
proton exit from energy interval $(E,E+dE)$ due to $p\gamma$-collisions
and with help of their regeneration in the same energy interval
described by probability $P(E',E,t)$. These two probabilities  describe
fluctuations in the interaction length and in fraction of energy lost
in the interaction; the interaction length is equivalent in this
picture to time of proton exit from energy interval $dE$.
\begin{figure}[ht]
\begin{minipage}[h]{8cm}
\centering
\includegraphics[width=76mm,height=76mm,clip]{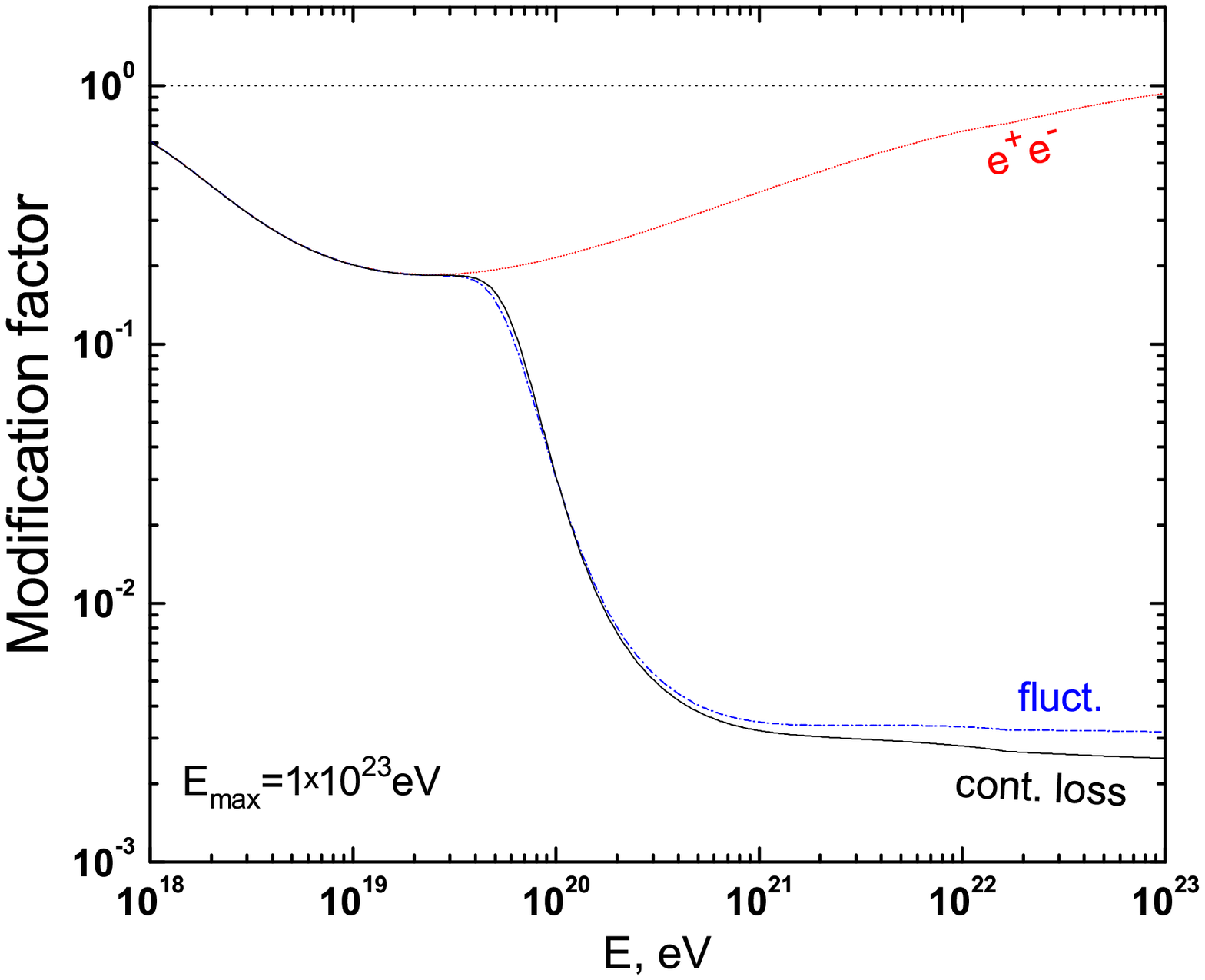}
\end{minipage}
\hspace{5mm}
\begin{minipage}[h]{8cm}
\centering
\includegraphics[width=76mm,height=76mm,clip]{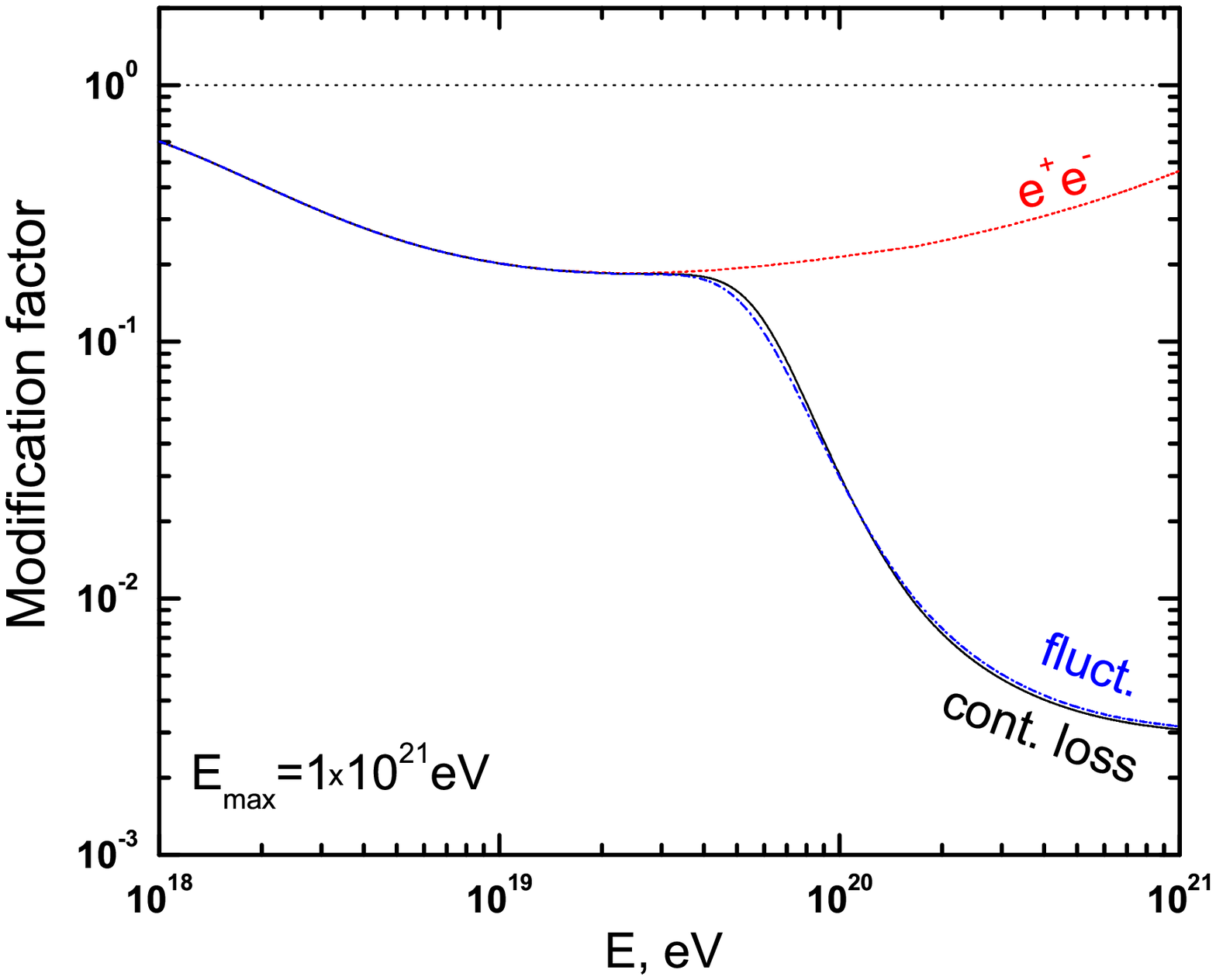}
\end{minipage}
\caption{Modification factor for the power-law generation spectra
with $\gamma_g=2.7$ and $z_{\rm max}=5$. Upper dotted line
corresponds to the case of adiabatic energy loss. Curve '$e^+e^-$' 
gives the modification factor for the case 
of adiabatic and pair-production energy
losses. Curve 'cont.loss' corresponds to total energy
losses in continuous loss approximation. The 'fluct.' curve
describes the case of numerical solution to the kinetic equation
(\ref{flucteqn}). The left and right panels correspond to 
$E_{\rm max}=1\times 10^{23}$~eV and $E_{\rm max}=1\times 10^{21}$~eV, 
respectively.
} %
\label{fluct} %
\end{figure}
The exit probability $P(E,t)$ due to collisions with CMB photons with
temperature $T$ can be written as:
\begin{eqnarray}
\label{P(E,t)} %
P(E,t) = -\frac{c T}{4\pi^2 E^2}\int_{m_\pi +
m_p}^\infty dE_c E_c (E_c^2 - m_p^2)\sigma(E_c) \ln \left[ 1-\exp
\left( -\frac{E_c^2-m_p^2}{4ET} \right) \right],
\end{eqnarray}
where $E_c$ is the total c.m.s.\ energy of colliding proton and photon,
$E_c^{min}=m_\pi + m_p$, $\sigma(E_c)$ is the photopion cross-section
and $T$ is the CMB temperature at cosmological epoch $t$.

Similarly, in regeneration term of Eq.~(\ref{flucteqn}) the probability
for a proton with energy $E'$ to produce a proton with energy $E$ is
given by
\begin{eqnarray}
P(E',E,t) = - \frac{c T}{4\pi^2(E')^2} \int
\limits_{E_c^{min}(x)}^\infty dE_c E_c (E_c^2 - m_p^2)
\frac{d\sigma(E_c,E',E)}{dE} \ln \left[ 1-\exp
\left(-\frac{E_c^2-m_p^2}{4E'T} \right) \right], %
\label{P(E',E,t)} %
\end{eqnarray}
where $E'$ and $E$ are the energies of primary and secondary protons,
respectively, and $x=E/E'$. The minimum value of the allowed c.m.s.\
energy in this case is given by
\begin{equation}
E_c^{min}(x) = [m_{\pi}^2/(1-x)+m_p^2/x]^{1/2}.
\end{equation}
This bound corresponds to the process with minimum invariant mass,
namely to $p+\gamma \rightarrow \pi^0+p$.

We have solved Eq.~(\ref{flucteqn}) numerically. The calculated
spectrum $J_p(E)= (c/4\pi)n_p(E,t_0)$, where $t_0$ is the age of the
universe at red-shift $z=0$, is presented in Fig.~\ref{fluct} as
modification factor $\eta(E)= J_p(E)/J_p^{\rm unm}(E)$ for generation
spectrum with $\gamma_g = 2.7$ and $E_{max}=1\times10^{23}$~eV (left
panel) and $E_{max}=1\times10^{21}$~ eV (right panel). For comparison
the modification factors for universal spectrum with continuous energy
losses is also shown. The difference in these two spectra at highest
energies must be due to fluctuations in energy losses, though formally
we have to say that this is the difference between solution to kinetic
equation  (\ref{flucteqn}) and the continuous energy loss
approximation. For $E_{max}=1\times10^{23}$~eV one can see the
difference in the spectra about 25\%  at highest energies and tiny
difference above intersection of $\eta_{ee}$ and $\eta_{\rm tot}$
curves. For $E_{max}=1\times10^{21}$~ eV the difference is small.

Note, that modification factors do not vanish at $E_{\rm max}$ even
when generation function goes abruptly to zero, since both solutions
vanish keeping the same value of ratios $\eta(E)=J_p(E)/J_{\rm
unm}(E)$. It is easy to demonstrate analytically that ratio of flux in
continuous loss approximation $J_{\rm cont}(E)$ to unmodified flux
$J_{\rm unm}(E)$, given by Eq.~(\ref{Junm}) tends to $H_0/\beta_{\rm
coll}(E_{\rm max})$, when $E \rightarrow E_{\rm max}$. 
From Fig.~\ref{fluct} one can see that this ratio coincides
exactly with our numerical calculations (e.g. the analytical value is  
$2.45\times 10^{-3}$ for $E_{\rm max}=1\times 10^{23}$~eV),
and this gives a proof that our numerical calculations are correct.

The effect of interaction fluctuations is usually taken into account
with help of Monte-Carlo simulations. The method of kinetic equation
corresponds to averaging over large number of Monte-Carlo simulations,
and if all other assumptions are the same, the results must coincide
exactly. These assumptions include $E_{\rm max}$ and parameters of
$p\gamma$-interaction. However, the existing Monte-Carlo simulations in
most cases include some other assumptions in comparison with kinetic
equations, which modify spectrum stronger than interaction
fluctuations. One of them is discreteness in the source distribution
(in kinetic equations the homogeneous distribution is assumed), the
other is fluctuations of distances to the nearby sources.

It is possible however to make the comparison with Monte-Carlo
simulations for homogeneously distributed sources and using identical
interaction model. Such a comparison is discussed in
Appendix~\ref{app-fluct}.

As to results presented here, it is necessary to emphasize that the
difference of kinetic-equation solution and continuous-energy-loss
approximation presented in Fig.~\ref{fluct} includes fluctuations, {\em
but not only fluctuations}. The transition of kinetic equation to
continuous-energy-loss equation depends on some other conditions which
can fail. What the presented calculations demonstrate (see
Fig.~\ref{fluct}) is that continuous-energy-loss approximation
describes with very good accuracy more reliable kinetic-equation
solution, which in particular includes interaction fluctuations.

\section{\label{sec:transition} Transition from extragalactic to galactic cosmic rays}
\begin{figure}[ht]
\begin{minipage}[h]{8cm}
\centering
\includegraphics[width=7.6cm,clip]{fig8a.eps}
\end{minipage}
\begin{minipage}[h]{8cm}
\centering \vspace{-3mm}
\includegraphics[width=72mm,height=72mm,clip]{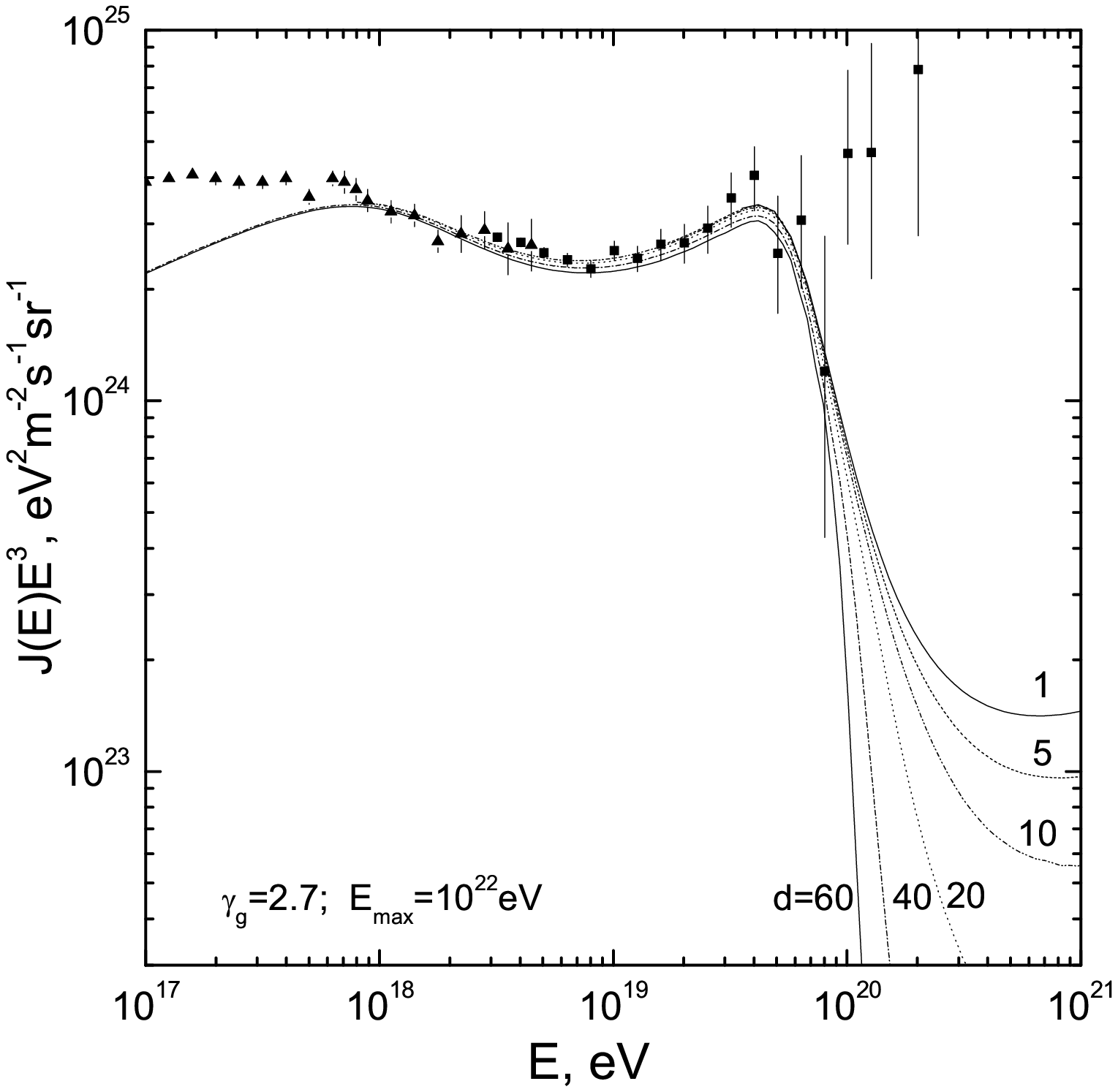}
\end{minipage}
\vspace{2mm}
\begin{minipage}{8cm}
\centering
\includegraphics[width=7.6cm,clip]{fig10.eps}
\end{minipage}
\begin{minipage}[h]{8cm}
 \caption{\label{trans} Appearance of transition energy
$E_b \approx 1\times 10^{18}$~eV in the modification factor compared
  with AGASA data (left panel), in the spectrum for
  rectilinear propagation from the sources with separation $d$
  indicated in the figure (right panel) and in the spectrum
  for diffusive propagation (lower panel)
  }
\end{minipage}
\end{figure}

In the analysis above we obtained several indications that transition
from extragalactic to galactic cosmic rays occurs at $E \approx 1\times
10^{18}$~eV. These evidences are summarized in Fig.~\ref{trans}.

The predicted spectrum above $1\times 10^{18}$~eV describes perfectly
well the observed spectra: see the modification factor in the
upper-left panel of Fig.~\ref{trans} compared with AGASA data and in
Fig.~\ref{dips} with HiRes. However, at $E \lsim 1\times 10^{18}$
experimental modification factor becomes $\eta > 1$, in contrast to
definition $\eta \leq 1$. It signals the  appearance of a new component,
which can be nothing but the galactic cosmic rays.
In the right panel the spectrum for rectilinear propagation from the
sources with different separation $d$ and with $\gamma_g=2.7$ is
compared with AGASA data. One can see that at $E < 1\times 10^{18}$~eV
the calculated extragalactic spectrum becomes less than that observed.

In the lower panel the similar comparison is shown for diffusive
propagation with different diffusion regimes at lower energies
\cite{AB1,Lemoine}. The random magnetic field with basic scale 
$l_c = 1$~Mpc and magnetic field on this scale $B_0 = 1$~nG is assumed. The
dash-dotted curve (universal spectrum) corresponds to the case
when the separation between sources $d \rightarrow 0$.

In all cases the transition from extragalactic to galactic component
begins at $E_b \approx 1\times 10^{18}$~eV, with index $b$ for
``beginning''.

What is the reason of this universality? We study the transition,
moving from high towards low energies. $E_b$ is the beginning of
transition (or its end, if one moves from low energies). $E_b$ is
determined by energy $E_{\rm eq1}= 2.37\times 10^{18}$~eV, where
adiabatic and  pair-production energy losses become equal. The
quantitative analysis of this connection is given in \cite{AB1}. We
shall give here the semi-quantitative explanation.

The flattening of the spectrum occurs at energies $E \leq E_b$, where
$E_b=E_{\rm eq}/(1+z_{\rm eff})^2$ and $z_{\rm eff}$ should be
estimated as  red-shift up to which the main contribution to unmodified
spectrum occurs. The simplified analytic estimate for $\gamma_g=2.6 -
2.8$ gives $1+z_{\rm eff} \approx 1.5$ and hence $E_b \approx 1\times
10^{18}$~eV. In fact, the right and lower panels of Fig.~\ref{trans}
present the exact calculations of this kind.

In experimental data the transition is searched for as a feature
started at some low energy $E_{\rm 2kn}$ - the second knee. Its
determination depends on experimental procedure, and all we can predict
is $E_{\rm 2kn} < E_b$. Determined in different experiments $E_{2kn}
\sim (0.4 - 0.8)\times 10^{18}$~eV.

The transition at the second knee appears also in the study of
propagation of cosmic rays in the Galaxy (see e.g.\
\cite{2kneea,2kneeb,2kneec}).

Being thought of as purely galactic feature, the position of the second
knee in our analysis appears as direct consequence of extragalactic
proton energy losses.
\begin{figure}[ht]
\centering
\includegraphics[width=160mm,height=75mm,clip]{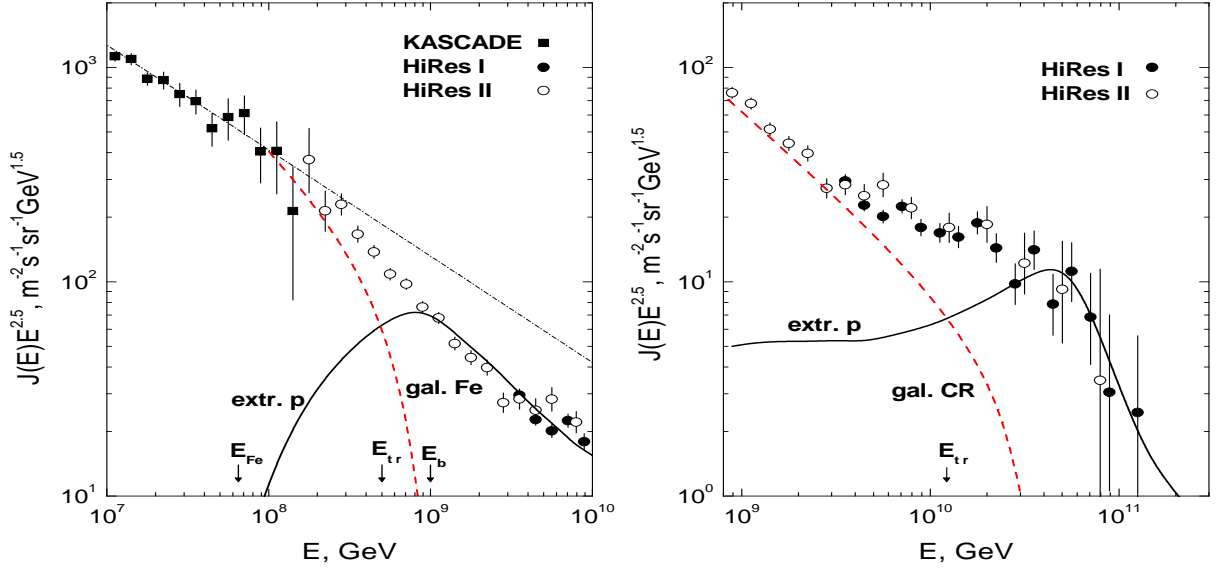}
\caption{\label{transition} Transition from extragalactic to galactic
cosmic rays in  the second knee (left panel) and ankle (right panel)
models. In the left panel are shown: KASCADE total spectrum, which
above the iron knee $E_{\rm Fe}$ is composed mostly by iron nuclei
('gal.Fe' curve), below $E_b$ the extragalactic proton spectrum ('extr.p'
curve) is calculated for diffusive propagation (see Section
\ref{sec:AGN}) and $E_{\rm tr}$ is the energy of transition from
galactic cosmic rays to extragalactic protons. The dot-dash line shows
power-law extrapolation of low-energy KASCADE spectrum. In the right
panel extragalactic proton spectrum is calculated for generation
spectrum $\propto E^{-2}$, while galactic spectrum (curve 'gal.CR') is
taken as difference between the observed total spectrum and calculated
spectrum of extragalactic protons. }
\end{figure}

The transition at the second knee is illustrated by
Fig.~\ref{transition}. The clue to understanding of this transition is
given by the KASCADE data \cite{kascadea,kascadeb}. They confirm the
rigidity model, according to which position of a knee for nuclei with
charge Z is connected with the position of the proton knee $E_p$ as
$E_Z=Z E_p$. There are two versions of this model. One is the
confinement-rigidity model (bending above the knee is due to
insufficient confinement in galactic magnetic field), and the other is
acceleration-rigidity model ($E_{\rm max}$ is determined by rigidity).
In both models the heaviest nuclei (iron) start to disappear at $E >
E_{\rm Fe}= 6.5\times 10^{16}$~eV, if the proton knee is located at
$E_p \approx 2.5\times 10^{15}$~eV. The shape of the spectrum above the
iron knee ($E> E_{\rm Fe}$) is model dependent, with two reliably
predicted features: it must be steeper than the spectrum below the iron
knee ($E < E_{\rm Fe}$), shown by the dash-dot curve, and iron nuclei 
must be the dominant
component there (see Fig.~\ref{transition}). The high energy part of
the spectrum has a characteristic energy $E_b$, below which the
spectrum becomes more flat, i.e.\ drops down when multiplied to
$E^{2.5}$ (see Fig.~\ref{transition}). This part of the spectrum is
shown for the diffusive propagation described in section \ref{sec:AGN}.
These two falling parts of the spectrum inevitably intersect at some
energy $E_{\rm tr}$, which can be defined as transition energy from
galactic to extragalactic cosmic rays. The 'end' of galactic cosmic rays
$E_{\rm Fe}= 6.5\times 10^{16}$~eV and the beginning of full dominance
of extragalactic component $E_b \approx 1\times 10^{18}$~eV differ by
an order of magnitude. Note, that power-law extrapolation of the total
galactic spectrum, shown by dot-dash line, beyond the iron knee $E_{\rm
Fe}$ has no physical meaning in the rigidity models and must not be
discussed.

The second-knee transition gives an alternative possibility in
comparison with ankle-transition hypothesis known from end of 1970s. It
is inspired by flattening of the spectrum at $E_a \approx 1\times
10^{19}$~eV seen in the AGASA and Yakutsk data (left panels in
Fig.~\ref{dips}) and possibly at $(0.5 - 1)\times 10^{19}$~eV in the
Hires data (the right panel in Fig.~\ref{dips}). Being multiplied to
factor $E^{2.5}$, as in Fig.~\ref{transition}, the ankle transition
looks very similar to that at the second knee. Note that in the latter
case the ankle is just an intrinsic part of the dip.

The ankle transition has been recently discussed in
Refs.~\cite{Hillas,WiWo,Allard,Allard1,DeMSt,Hillas1}.

In the ankle model it is assumed that galactic cosmic ray spectrum has
a power-law shape $\propto E^{-\gamma}$ from the proton knee $E_p
\approx 2.5\times 10^{15}$~eV to about $E_a \sim 1\times 10^{19}$~eV
where it becomes steeper and crosses the more flat extragalactic
spectrum (see the right panel of Fig.~\ref{transition}). The ankle
transition in Fig.~\ref{transition} is shown for extragalactic proton
spectrum with generation index $\gamma_g=2$, while galactic spectrum,
given by curve ``gal.CR'' is calculated as difference of the observed
total spectrum and calculated extragalactic proton spectrum.

The ankle model has the problems with galactic component of cosmic
rays. The spectrum at $1\times 10^{18} - 1\times 10^{19}$~eV is taken
ad hoc to fit the observations, while in the second knee model this
part of the spectrum is calculated with excellent agreement with the
data. In the rigidity models the heaviest nuclei (iron) start to
disappear at $E>E_{\rm Fe}= 6.5\times 10^{16}$~eV. How the gap between
$1\times 10^{17}$~eV and  $1\times 10^{19}$~eV is filled?

Galactic protons start to disappear at $E > 2.5\times 10^{15}$~eV.
Where they came from at $E> 1\times 10^{17}$~eV to be seen e.g.\ in the
Akeno detector with fraction $\sim 10\%$?

The ankle model needs acceleration by galactic sources up to $1\times
10^{19}$~eV (at least for iron nuclei), which is difficult to afford.
The second knee model ameliorates this requirement by one order of
magnitude.

The second-knee model predicts the spectrum shape down to $1\times
10^{18}$~eV with extremely good accuracy ($\chi^2$/d.o.f.= 1.12 for
Akeno-AGASA and $\chi^2$/d.o.f.= 1.03 for HiRes). In the ankle model
one has to consider this agreement as accidental, though such
hypothesis has very low probability, determined by $\chi^2$ cited
above. As an alternative the ankle model-builders can suggest only
hopes for future development of galactic propagation  models to be as
precisely calculated as the dip.

\section{\label{sec:sources} Astrophysical sources of UHECR}
In the sections above we have performed the model-independent analysis
of spectra of extragalactic protons interacting with CMB. We have
calculated the features of the proton spectrum assuming the power-law
generation spectrum $\propto E^{-\gamma_g}$ valid at $E \geq 1\times
10^{18}$~eV, and compared predicted features with observations.  We
found that proton dip, a model-independent feature at energy between
$1\times 10^{18}$~eV and $4\times 10^{19}$~eV, is well confirmed by
observations. Only two free parameters are involved in fitting of
observational data: $\gamma_g = 2.7$ (the allowed range is 2.55 - 2.75)
and the flux normalization constant. The various physical phenomena
included in calculations, such as discreteness in the source
distribution, the different  modes of propagation (rectilinear and
diffusive), cosmological evolution with parameters similar to AGN 
evolution, fluctuations in $p\gamma$ interaction etc, do not upset
this agreement.

The transition of extragalactic to galactic cosmic rays is also
discussed basically in model-independent manner.

In this Section we shall discuss the models: realistic energy spectra,
the sources and the models for transition from extragalactic to
galactic cosmic rays.

The UHECR sources have to satisfy two conditions: they must be
very powerful and must accelerate particles to large $E_{\rm
max}\gsim 1\times 10^{21}$~eV. There is one more restriction
\cite{DTT,FK,YNS1,Sigl1,YNS2,BlMa,KaSe}, coming from observation
of small-scale clustering: the space density of the sources should
be $(1 - 3)\times 10^{-5}$~Mpc$^{-3}$ probably with noticeable
uncertainty in this value. Thus, these sources are more rare, than
typical representatives of AGN, e.g. the Seyfert galaxies, whose space
density is $\sim 3\times 10^{-4}$~Mpc$^{-3}$. The sources could be
the rare types of AGN, and indeed the analysis of \cite{corr} show
statistically significant correlation between directions of
particles with energies $(4 - 8)\times 10^{19}$ eV and directions
to AGN of the particular type -- BL Lacs (see also criticism
\cite{Sar} and reply \cite{TT}). The acceleration in AGN can
provide the maximum energy of acceleration up to $\sim 10^{21}$~eV
for non-relativistic shock acceleration (see e.g.\ \cite{Bier}).

The relativistic shock acceleration can occur in AGN jets. Acceleration
to $E_{\rm max} \sim  10^{21}$~eV in the AGN relativistic shocks is
questionable (see discussion below).

Gamma Ray Bursts (GRBs) are another potentially possible sources 
of UHECR \cite{Usov,acc-GRB-V,acc-GRB-W}. They have very large energy 
output and can accelerate particles up to $\sim 10^{21}$~eV
\cite{acc-GRB-W,acc-GRB-V}. These sources have, however, the problems
with explaining small-angle anisotropy and with energetics (see
discussion below).

\subsection{\label{sec:spectra} Spectra}
The assumption of the power-law generation spectrum with $\gamma_g =
2.7$ extrapolated to $E_{\rm min} \sim 1$~GeV results in too large
emissivity required for observed fluxes of UHECR. To avoid this problem
the {\em broken generation spectrum} has been suggested in
Refs.~\cite{BGG1,BGG2}:
\begin{equation}
Q_{\rm gen}(E)=\left\{ \begin{array}{ll}
\propto E^{-2}   ~& {\rm at}~~ E \leq E_c \\
\propto E^{-2.7} ~& {\rm at}~~ E \geq E_c
\end{array}
\right. , %
\label{broken} %
\end{equation}
where $Q_{\rm gen}(E)$ is the generation function (rate of particle
production per unit of comoving volume), defined by
Eqs.~(\ref{conserv}) and (\ref{Q_gen}), and $E_c$ was considered as a
free parameter.

Recently it was demonstrated that broken generation spectrum can 
naturally emerge under most reasonable physical assumption. In
Ref.~\cite{KaSe05} it has been argued that while spectrum $E^{-2}$ is
universal for non-relativistic shock acceleration, the maximum
acceleration energy $E_{\rm max}$ is not, being dependent on the
physical characteristics of a source, such as its size, regular
magnetic field etc. Distribution of sources over $E_{\rm max}$ results
in steepening of the generation function, so that the distribution of
the sources $dn/dE_{\rm max} \propto E_{\rm max}^{-\beta}$  explains
the observational data, if $\beta = 1.5 - 1.6$ \cite{KaSe05}.

In Appendix \ref{app-Emax} we address the generalized problem: what
should be the distribution of spectral emissivity over $E_{\rm max}$ to
provide the generation function with the broken spectrum. We use there
the notation $\varepsilon \equiv E_{\rm max}$ and introduce the
spectral emissivity
$\mathcal{L}(\varepsilon)=n_s(\varepsilon)L_p(\varepsilon)$, where
$L_p(\varepsilon)$ is particle luminosity of a source and
$n_s(\varepsilon) \equiv n_s(\varepsilon,L_p(\varepsilon))$ is the
space density of the sources. The total emissivity is given by
$\mathcal{L}_0= \int \mathcal{L}(\varepsilon)d\varepsilon$.
Distribution of spectral emissivity $\mathcal{L}(\varepsilon)$ over
maximal energies $\varepsilon$ determines the energy steepening of the
generation function $Q_{\rm gen}(E)$ at energy $E=\varepsilon_{\rm
min}$ in the distribution. This function is calculated analytically for
arbitrary $\mathcal{L}(\varepsilon)$, assuming that $\varepsilon$ is
confined to the interval ($\varepsilon_{\rm min}\;,\;\varepsilon_{\rm
max})$. It is demonstrated that $Q_{\rm gen}(E)$ can be the power-law
function $\propto E^{-\gamma_g}$ {\em exactly}, only if
$\mathcal{L}(\varepsilon)$ distribution is power-law, too $(\propto
\varepsilon^{-\beta})$. For the source generation function $q_{\rm
gen}(E) \propto E^{-(2+\alpha)}$ at $E \leq \varepsilon$, the
generation index in the interval $\varepsilon_{\rm min} \leq E \leq
\varepsilon_{\rm max}$ is found to be $\gamma_g=1+\alpha+\beta$,
including the case $\alpha=0$. The steepening of the generation
spectrum from $2+\alpha$ to $\gamma_g$ occurs approximately at energy
$E_c=\varepsilon_{\rm min}$. At energy $E> \varepsilon_{\rm max}$ the
spectrum is suppressed as $\exp(-E/\varepsilon_{\rm max})$ or stronger.

In the applications we are interested in two cases: $\alpha=0$
(non-relativistic shocks) and $\alpha= 0.2 - 0.3$
(ultra-relativistic shocks). In the latter case the term $E^{-2}$
in Eq.~(\ref{broken}) should be substituted by $E^{-(2+\alpha)}$.
The energy $E_c=\varepsilon_{min}$ in Eq.~(\ref{broken}) is
considered as a free parameter.

\subsection{Active Galactic Nuclei}
\label{sec:AGN} The AGN as sources of UHECR meet the necessary
requirements: {\em (i)} to accelerate particles to $E_{\rm max} \sim
10^{21}$~eV, {\em (ii)} to provide the necessary energy output and {\em
(iii)} to have the space density $n_s \sim (1 - 3)\times
10^{-5}$~Mpc$^{-3}$, required by small-scale clustering. We shall
discuss below these problems in some details.

\subsubsection{Acceleration and spectra}
\label{sec:acceleration} The flow of the gas in AGN jet can be
terminated by the non-relativistic shock which accelerates protons or
nuclei in the radio lobe up to $E_{\rm max} \sim 10^{21}$~eV with spectrum
$\propto E^{-2}$ \cite{Bier}.

In some cases the observed velocities in AGN jets are
ultra-relativistic with Lorentz factor up to $\Gamma \sim 5 - 10$. It
is natural to assume there the existence of  internal and external
ultra-relativistic shocks. Acceleration in relativistic shocks relevant
for UHECR has been recently studied in
Refs.~\cite{acc-GRB-V,acc-GRB-W,GaAch,Kirk,Vietri1,BlVi,acc-Lemoine1,
acc-Lemoine2}.
The acceleration spectrum is $\propto E^{-\gamma_g}$ and in the case of
isotropic scattering of particles upstream and downstream,  the spectrum
index is $\gamma_g=2.23 \pm 0.01$ \cite{Kirk}. However, recently it was
understood that this result depends on scattering properties of the
medium \cite{BlVi,acc-Lemoine2}, and the spectrum can be steeper. In
the regime of large angle scattering in \cite{BlVi}, $\gamma_g=2.7$ was
found possible for the shock with velocity $u \sim (0.8 - 0.9)c$ and
compression ratio $r=2$.  In Monte-Carlo simulation \cite{acc-Lemoine2}
it is demonstrated that effect of compression of upstream magnetic
field results in increasing of $\gamma_g$ up to the limiting value 2.7
in ultra-relativistic case $\Gamma_{\rm sh} \gg 1$.

The maximal energy of acceleration $E_{\rm max}$ is a controversial
issue.  While in most works (very notably \cite{GaAch}) it is obtained
that  $E_{\rm max}$ cannot reach $\sim 10^{21}$~eV for all realistic
cases of relativistic shocks, the authors of Ref.~\cite{Vie-DeM} argue
against this conclusion.

We shall divide a problem with $E_{\rm max}$ into two: the reliably
estimated energy gain in relativistic shocks and model-dependent
absolute value of $E_{\rm max}$. The energy gain in the first full
cycle of particle reflection upstream-downstream-upstream ($u \to d \to
u$) is about $\Gamma_{\rm sh}^2$. The next reflections are much less
effective, as was first observed in \cite{GaAch}: a particle lives a
short time in the upstream region, before it is caught up by the shock. As
a result, a particle is deflected by upstream magnetic field only to a
small angle, and thus it occurs in the downstream region with approximately
the same energy as in the first cycle. Then the energy upstream will be
also almost the same as in the first cycle. According to Monte-Carlo
simulation \cite{acc-Lemoine2} the average energy gain per each
successive $u \to d \to u$ cycle is only 1.7 (see Fig.~4 in
\cite{acc-Lemoine2} with clear explanation). With these energy gains
($\sim \Gamma_{\rm sh}^2$ in the first $u \to d \to u$ cycle and with $
\sim 2$ in each successive cycle) it is not possible to get $E_{\rm max} \sim
10^{21}$~eV in the conservative approach for AGN and GRBs, but there
are some caveats in this conclusion as indicated in
\cite{GaAch,Vie-DeM}. The magnetic field in the upstream region can be
large, and then the deflection angle of a particle after the shock
crossing is large, too. Relativistic shock acceleration can operate in
medium filled by pre-accelerated particles, and thus initial energy can
be high.

There are some other mechanisms of acceleration to energies up to
$\sim 10^{21}$~eV relevant for AGN: unipolar induction and acceleration
in strong electromagnetic waves (see \cite{book} for description and 
references). The mechanisms of jet acceleration have very special status.

The observed correlations between arrival directions of particles with
energies $(4 - 8)\times 10^{19}$~eV and BL Lacs \cite{corr} imply the
jet acceleration. This is because BL Lacs are AGN with jets directed
towards us. For this correlation the propagation of particles (most
probably protons) with energies above $4\times 10^{19}$~eV must be
(quasi)rectilinear, that can be realized in magnetic fields found in MHD
simulations in \cite{Dolag} (see however the simulations in
\cite{Sigla,Siglb} with quite different results).

\begin{figure}[ht]
\centering
\includegraphics[width=3cm,height=6cm,angle=90]{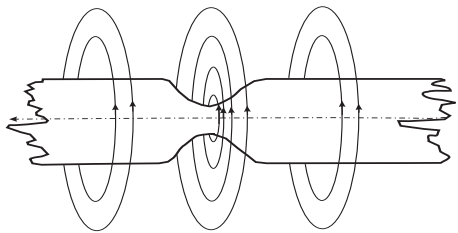}  %
\caption{Pinch acceleration in the jet with electric current }
\label{pinch}
\end{figure}

An interesting mechanism of jet acceleration, called pinch
acceleration, was suggested and developed in plasma physics
\cite{Trub}. It is based on pinch instability well known in plasma
physics, both theoretically and observationally. The electric current
along jet produces the toroidal magnetic field which stabilizes the jet
flow. The pinch instability is caused by squeezing the tube of flow by
magnetic field of the current. It results in increasing the electric
current density and magnetic field. The magnetic field compresses 
further the tube
and thus instability develops.  The acceleration of particles in
\cite{Trub} is caused by hydrodynamical increase of velocity of the
flow and by longitudinal electric field produced in the
pinch. This process is illustrated by Fig.~\ref{pinch}.
The pinch acceleration has been developed for tokamaks and was
confirmed there by observations. The generation spectrum is
uniquely predicted as 
\beq q_{\rm gen}(E) \propto E^{-\gamma_g},
~~~{\rm with} ~~~\gamma_g=1+\sqrt{3}= 2.73. 
\eeq 
Scaling a size of
the laboratory tube to the cosmic jet, one obtains $E_{\rm max}$
exceeding $10^{22}$~eV. The particle beam undergoes a few pinch
occurrences during traveling along cosmic jet (a few kpc in case
of AGN), and thus spectrum obtains a low energy cut at high value
of $E_{\rm min}$. This mechanism needs more careful study.

We finalize this subsection concluding, that there are at least three
possibilities for the broken generation spectrum with $q_{\rm
gen}\propto E^{-2.7}$ at $E>E_c$: The acceleration by non-relativistic
shocks with distribution of sources (more precisely emissivity) over
$E_{\rm max}$, the acceleration by relativistic shocks, where
$E_c=E_{\rm min}$ naturally occurs due to the first $u\to d\to u$ cycle
as $\Gamma_{\rm sh}^2 E_{\rm in}$ with $E_{\rm in}$ being particle
energy before acceleration, and in the pinch acceleration where
$\gamma_g=1+\sqrt{3}$ is rigorously predicted and $E_{\rm min}$ appears
due to several pinch occurrences. The latter mechanism provides
acceleration along a jet, necessary for correlations with BL Lacs.

\subsubsection{Spectra of UHECR from AGN}
\label{sec:AGNspectra} 
We discuss here the diffuse energy spectra
from AGN. They are model-dependent because one should specify 
the distance between sources $d$, the mode of propagation 
(e.g. rectilinear or diffusive) and the critical energy $E_c$ in
the broken spectrum of generation. The universal spectrum is valid when
characteristic distance $d$ between sources is smaller than
other propagation lengths, most notably energy attenuation length
$l_{\rm att}(E)$ and diffusion length $l_{\rm diff}(E)$. As was
discussed in Section \ref{sec:robustness} the dip, seen at energies
$1\times 10^{18} \lsim E \lsim 4\times 10^{19}$~eV, is the robust
feature, which is not modified by any known phenomenon except presence
of extragalactic nuclei with fraction higher than $20\%$. This fraction
depends on model of acceleration and can be small for some acceleration
mechanisms, e.g.\ for acceleration by relativistic shocks.
\begin{figure}[ht]
   \begin{minipage}[h]{8cm}
     \centering
     \includegraphics[width=8cm]{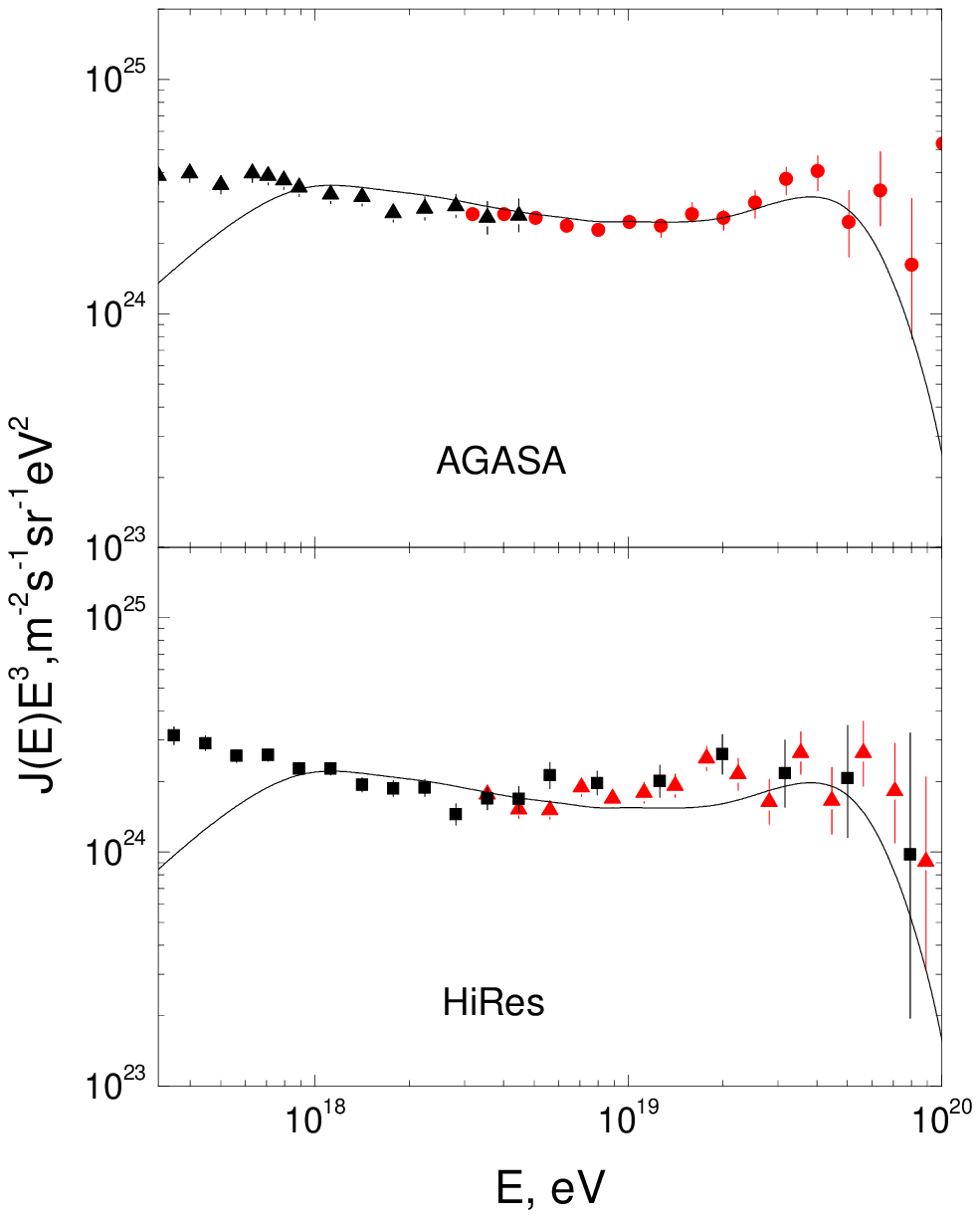}
   \end{minipage}
   \hspace{5mm}
 \begin{minipage}[h]{8cm}
    \centering
    \includegraphics[width=8cm]{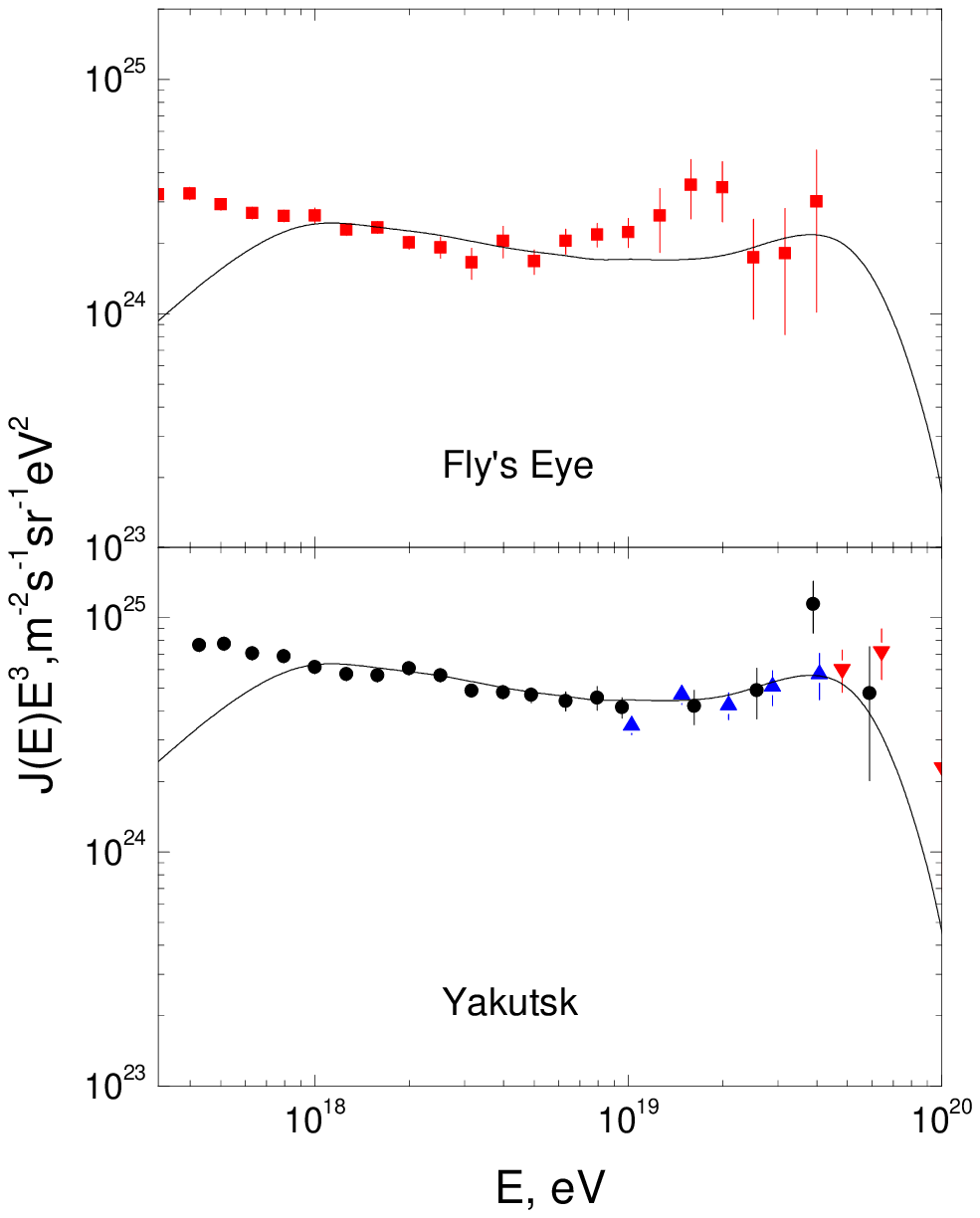}
 \end{minipage}
\caption{Comparison of calculated AGN spectra with the data of AGASA
  and HiRes (left panel) and with data of Fly's Eye and Yakutsk (right
  panel). 
The generation spectrum is $\propto E^{-2.7}$ with $E_c=1\times 10^{18}$~eV 
and with $m=0$. The propagation is diffusive with parameters as in 
Fig.~\ref{dip-diff} and with the Bohm diffusion in low-energy regime.
 }
\label{spectra}
\end{figure}

The prediction of the GZK cutoff is model dependent: the shape of this
feature can be modified by different values of $E_{\rm max}$, by local
overdensity/deficit of the sources (see Section~\ref{sec:GZKcutoff})
and by discreteness in the source distribution (see Fig.~\ref{discretness}). It
can also be mimicked by the shape of acceleration cutoff near $E_{\rm
max}$.

The spectrum predicted for $E < 1\times 10^{18}$~eV is also model
dependent. At these energies for all distances $d$ reasonable for AGN 
the universal spectrum is modified due to propagation in magnetic fields,
because $l_{\rm diff}$ is small.  
In Fig.~\ref{transition}  (left panel) the spectrum for
diffusion in random magnetic field is shown.  At energies 
$E> 1\times 10^{18}$~eV the spectrum remains universal. 

We have to introduce in calculations the broken generation spectrum 
(\ref{broken}) to provide the reasonable luminosities
of AGN. Normalized by the AGASA flux, the emissivity $\mathcal{L}_0$ is
given by $3.5\times 10^{46}$~ergs Mpc$^{-3}$yr$^{-1}$, $1.6\times
10^{47}$~ergs Mpc$^{-3}$yr$^{-1}$ and $7.0\times 10^{47}$~ergs
Mpc$^{-3}$yr$^{-1}$ for $E_c$ equals to $1\times 10^{18}$~eV,~ $1\times
10^{17}$~eV and $1\times 10^{16}$~eV, respectively.

To calculate the luminosities of AGN, $L_p=\mathcal{L}_0/n_s$, we take
the density of the sources from small-scale clustering as $n_s=2\times
10^{-5}$~Mpc$^{-3}$ \cite{BlMa,KaSe}. The luminosities are found to be
quite reasonable for AGN: $5.6\times 10^{43}$,~ $2.5\times 10^{44}$ and
$1.1\times 10^{45}$~erg s$^{-1}$ for $E_c$ equals to $1\times
10^{18}$~eV,~ $1\times 10^{17}$~eV and $1\times 10^{16}$~eV,
respectively.

The calculated spectra are shown in Fig.~\ref{spectra} for propagation
in magnetic field, as described in Fig.~\ref{dip-diff} for the Bohm 
diffusion in the low-energy regime, for $d=50$~Mpc, $\gamma_g=2.7$, 
$E_c=1\times 10^{18}$~eV and $m=0$,
inspired by BL Lacs, which show no positive evolution. The spectra for AGN
evolutionary models are shown in Fig.~\ref{evolution} in comparison
with case $m=0$.  For calculations of spectra with different 
values of $E_c$ see \cite{BGH}.

\subsubsection{Transition to galactic cosmic rays}
\label{sec:AGNtransition} For the diffusive propagation the transition
is shown in Fig.~\ref{transition}. In our calculations we followed
Ref.~\cite{AB1}, considering the same case of diffusive propagation in
random magnetic fields with basic coherent scale $l_c=1$~Mpc and
$B_0=1$~nG on this scale. These parameters define the diffusion
coefficient at very high energies, while at energies $E \lsim 1\times
10^{18}$~eV at interest we assume the Bohm diffusion. The distance
between sources is fixed as $d=50$~Mpc. As Fig.~\ref{transition} shows the
iron galactic spectrum given by curve ``gal.Fe'' describes well the
KASCADE flux at $E \approx 1\times 10^{17}$~eV and fits the Hall
(drift) diffusion spectrum at $E \gsim 1\times 10^{17}$~eV. The
predicted fraction of iron in the total flux is given in
Table~\ref{Fe-fraction} as function of energy. This fraction is higher
than in \cite{AB1} mostly because we used the HiRes data, instead of
AGASA in \cite{AB1}.
\begin{table}[ht]
\begin{center}
\caption{Fraction  of iron nuclei in the total flux as function of
energy.}
\vspace{3mm}
\begin{tabular}{c|c|c|c|c|c}
\hline $E$ (eV) & $1\times 10^{17}$ & $2\times 10^{17}$ & $5\times
10^{17}$
& $7\times 10^{17}$ & $10^{18}$ \\
\hline
$J_{Fe}/J_{\rm tot}$ & $0.97$ & $0.87$ & $0.49$ & $0.26$ & $0.04$ \\
\hline
\end{tabular}
\label{Fe-fraction}
\end{center}
\end{table}

The detailed study of transition for rectilinear propagation for
different values of $E_c$ has been performed in \cite{BGH}. The
fraction of iron-nuclei varies with $E_c$.
\subsubsection{Signatures of AGN}
\label{AGNsignatures} AGN are one of the best candidates for UHECR
sources. They have high luminosities to provide the observed UHECR flux
and have quite efficient mechanisms of acceleration. The predicted
spectra are in good agreement with observations and transition to
galactic cosmic rays is well described using diffusive or quasi-rectilinear
propagation and broken generation spectrum. The density of sources $n_s
\sim (1 - 3)\times 10^{-5}$~Mpc$^{-3}$, found from small-scale
clustering, corresponds well to space density of powerful AGN. However,
in the diffuse spectrum we cannot predict  any specific signature of
AGN, and probably only observations of the direct flux of primary
(e.g.\ protons) or secondary (e.g.\ neutrinos or gamma-rays) radiation
can give the direct evidence for AGN as sources of UHECR.

In this respect the correlations with BL Lacs \cite{corr,TT} is the
direct indication to AGN as UHECR sources. The protons can be the
signal carriers. The important point is that in structured (perforated)
magnetic field, there are directions (holes) with weak magnetic fields
in which BL Lacs can be seen \cite{BGG2}. In MHD simulation
\cite{Dolag} it was found indeed that in many large scale structures
like voids and filaments the magnetic field is weak and protons can
propagate there with small deflections.

Correlation with BL Lacs imply the jet acceleration, because BL Lacs
are AGN with jets directed to an observer. The pinch acceleration
mechanism fits well the picture above, providing  $\gamma_g=2.7$, high
$E_{\rm max}$ and high $E_{\rm min}$, needed for reasonable luminosity
$L_p$.

\subsection{Gamma Ray Bursts}
\label{sec:GRB} GRBs are another candidates for UHECR sources. Like AGN
they have high energy output and can accelerate particles to UHEs, as
first have been proposed in \cite{Usov,acc-GRB-V,acc-GRB-W}

GRB starts with instant energy release $W_{\rm tot}$, which occurs most
probably due to SN explosion. This energy is injected into  a small
volume with initial radius $R_i \sim 10^7$~cm in a form of a jet with
small opening angle $\theta$. The extremely large energy release within
a small volume results in very high temperature of the  gas, which
consists mostly of photons and electron-positron pairs with a small
admixture of initially injected baryons. The pressure of relativistic
$e^+e^-\gamma$ gas accelerates fireball, its temperature falls down and
the Lorentz factor rises. At later stages a fireball becomes the
baryonic jet moving with Lorentz factor $\Gamma_0 = W_{\rm tot}/M_b c^2
\sim 100$, where $M_b$ is initial baryon loading.

The jet is characterized according to observations \cite{Frail} by
small opening angle $\theta \sim 4^{\circ}$ and by corresponding solid
angle $\Omega=\pi\theta^2$. The external observer sees  only part of
the external fireball surface within the angle $1/\Gamma_0 < \theta$.
Because if this fact, the convenient formalism is to consider the fictitious
spherically-symmetric fireball with equivalent isotropic energy output
$W_{\rm tot}$ and equivalent isotropic GRB rate $\nu_{\rm GRB}$ (in
units of Mpc$^{-3}$ yr$^{-1}$). The real energy output and GRB rate are
$\tilde{W}_{\rm tot}=(\Omega/4\pi)W_{\rm tot}$ and $\tilde{\nu}_{\rm
GRB}=(4\pi/\Omega ) \nu_{\rm GRB}$, with the same emissivity
$\mathcal{L}=W_{\rm tot}\nu_{\rm GRB}$.

According to observations  the equivalent isotropic energy output
$W_{\rm GRB} \sim 1\times 10^{53}$~erg, and spherically-symmetric GRB
rate is $\nu_{\rm GRB} \approx 0.5\times 10^{-9}$~Mpc$^{-3}$yr$^{-1}$
with emissivity $\mathcal{L}_{\rm GRB} \approx 0.5\times 10^{44}$~erg
Mpc$^{-3}$yr$^{-1}$. The more detailed estimates differ considerably
from this value: $\mathcal{L}_{\rm GRB} \approx 0.6\times
10^{43}$~erg Mpc$^{-3}$yr$^{-1}$ (M. Schmidt \cite{Schmidta,Schmidtb})
and $\mathcal{L}_{\rm GRB} \approx 1.3\times 10^{44}$~erg
Mpc$^{-3}$yr$^{-1}$ (Frail \etal\ \cite{Frail}). Note, that difference
in spectral intervals used in \cite{Schmidta,Schmidtb} and
\cite{Frail} cannot explain alone the differences between the
above-cited emissivities. We conclude thus that $\mathcal{L}_{\rm GRB}
\approx (0.6 - 13)\times 10^{43}$~ erg Mpc$^{-3}$yr$^{-1}$, to be
compared with higher estimate by Waxman \cite{GRB-W} $\mathcal{L}_{\rm
GRB} \approx (3 - 30)\times 10^{43}$~ erg Mpc$^{-3}$yr$^{-1}$.

There are two sites of acceleration in a jet: mildly relativistic {\em
internal}  shocks inside a fireball \cite{acc-GRB-W} and
ultra-relativistic {\em external} shock \cite{acc-GRB-V}.

\subsubsection{Internal shock acceleration and UHECR}
\label{sec:int-shock} We shall describe here shortly the acceleration
inside fireball, following \cite{acc-GRB-W,GRB-W,Piran}.

Gamma-radiation is produced as synchrotron emission of the
shock-accelerated electrons at the stage when fireball moves with
constant Lorenz-factor $\Gamma_0=W_{\rm tot}/M_bc^2$. Acceleration
occurs due to internal shocks produced by motion of fireball sub-shells
with different Lorentz factors $\Gamma$. This phenomenon is assumed to
be responsible for the observed GRBs and for their time substructure.
In the observer frame the jet looks like a narrow shell with width
$\Delta \sim R/\Gamma_0^2$, where $R$ is a radius, i.e.\ distance from
outer surface to the source (the GRB engine). It is assumed that this
shell consists of mildly relativistic sub-shells, moving relatively
each other in the rest frame of the fireball. The inner sub-shells move
faster than the outer ones, and their collisions produce mildly
relativistic shocks. The GRB spikes arise due to shell substructure.
The range of distances $R_{\gamma}$ where  GRB is produced can be
estimated from the observed minimum duration of spikes,
$\tau_{\rm{sp}}\gsim 1$~ms, and duration of GRB, $\tau_{\rm GRB} \sim 1
- 10$~s. Time between collisions of two sub-shells can be estimated in
the observer frame \cite{Piran} as $\delta t \sim \delta/v \sim
\Gamma_0^2 \delta/c$, where $\delta$ is the distance between two
colliding subshells. Thus, $R_{\gamma}= c\delta t$ is limited between
$R_{\gamma}^{\rm min} \sim \Gamma_0^2 c\tau_{\rm sp} \sim 3\times
10^{11}$~cm and $R_{\gamma}^{\rm max} \sim \Gamma_0^2 c\tau_{\rm GRB}
\sim 3\times 10^{14} - 3\times 10^{15}$~cm. The energetics is described
by $W_{\rm GRB}= f_{\rm GRB} W_{\rm tot}$, with $f_{\rm GRB} < 0.5$ at
least, because half of the total energy is released in afterglow. More
realistically $f_{\rm GRB} \sim 0.2$.

The exit of all particles from the jet occurs only through the front
spherical surface, provided by condition  $\theta > 1/\Gamma_0$.

Acceleration of electrons by internal shocks implies acceleration of
protons. According to \cite{acc-GRB-W,GRB-W} the generation spectrum
$\propto E^{-2}$, $E_{\rm max} \sim 1\times 10^{21}$~eV, $E_{\rm min}
\sim \Gamma_0 m_p c^2 $ and $W_p \sim W_{\rm GRB}$ (in the observer
frame).

As a matter of fact one must write the expressions above as $E_{\rm
min} \sim f_a \Gamma_0 m_pc^2 $ and $W_p \sim f_a W_{\rm GRB}$, where
$f_a$ is a factor of adiabatic cooling. This phenomenon has been
studied in \cite{Rachen} and adiabatic energy losses were found to be
very severe. As was indicated there the efficient mechanism, which diminishes
the influence of adiabatic cooling, is the neutron mechanism of UHECR exit
from expanding fireball. However, some restrictions obtained for this
mechanism make its application rather limited. Waxman \cite{GRB-W}
fought back this criticism by argument that production of HE protons in
the shell at later periods corresponding to a distance $R_{\gamma}^{\rm
max}$ eliminates the problem of adiabatic cooling. We shall argue below
that this problem exists.

There are three stages of adiabatic energy losses. The first one occurs
during acceleration process and it operates at distances
$R_{\gamma}^{\rm min} \lsim R \lsim R_{\gamma}^{\rm max}$. The
adiabatic energy losses are described in the standard
shock-acceleration equation by the term $(p/3)div\; \vec{u}\;\partial
f(r,t,p)/\partial p$, where $\vec{u}$ is hydrodynamical velocity of the
gas flow and $f(r,t,p)$ is the distribution function. In our case the
adiabatic energy losses are large due to expansion of the shell.
Adiabatic energy losses diminish the fraction of the shock energy
transferred to accelerated particles, which results in relation
$W_p=f_a^{(1)} W_{\rm GRB}$, where $f_a^{(1)} <1$ is adiabatic factor.
Qualitatively it can be explained in the following way.  The first
portion of accelerated particles would suffer severe adiabatic energy
losses $f_a \sim R_{\gamma}^{\rm min}/ R_{\gamma}^{\rm max}$, if
further acceleration were absent. In fact, these particles are
re-accelerating and loosing energy simultaneously, as corresponding
diffusion equation describes it. This process diminishes only the
fraction of energy $f_a$ transferred to accelerated particles. It can
be estimated as $f_a \sim R_{\gamma}^{\rm eff}/
R_{\gamma}^{\rm max}$, but for accurate estimate the solution of
diffusion equation is needed. Unfortunately, this model exists only in
the form of estimates.

The second stage starts after acceleration ceases. The end of GRB
radiation signals this moment. Magnetic field in the expanding
shell slowly diminishes, accelerated electrons radiate away their
energy fast, but protons are still confined in the shell, forming
relativistic gas and loosing energy adiabatically. Duration of this
stage is different for protons of different energies and the spectrum
$\propto E^{-2}$ is distorted. The protons with $10^{19} - 10^{21}$~eV
are the first to accomplish this stage and enter the regime of free
expansion (stage 3). This energy range is most interesting for UHECR .
The adiabatic factor $f_a^{(2)}$ depends on the time of diminishing of
equipartition magnetic field. The detailed calculations are needed to
evaluate $f_a^{(2)}$ for the range $10^{19} - 10^{21}$~eV.

At the third stage when the shell of the protons with energies $10^{19}
- 10^{21}$~eV propagates quasi-rectilinearly in the external magnetic
field, the energy losses are also present. They are caused by
the collective effect of transferring proton momenta to the macroscopic
volume of gas due to scattering in magnetic field.

As a toy model, let us consider the spherically symmetric shell of
protons with energies $10^{19} - 10^{21}$~eV, and make an order of
magnitude estimates. At the distance of order of gyroradius $r_L$, a
proton transfers half of its initial (radial) momentum to the gas.
Since protons propagate in the radial direction almost with speed of light,
it follows from causality that volume of gas, which absorbs the
lost momentum cannot be larger than $\sim r_L^3$. The equation of
radial momentum balance reads
\beq %
0.5 W_p/c \sim \Gamma r_L^3 \rho_gc,
\label{Gamma_gas}
\eeq %
where $\rho_g \sim 10^{-24}$ g/cm$^3$ is the
density of the ambient gas and $\Gamma$ is Lorentz factor, obtained by this
gas volume. For the spectrum $\propto E^{-2}$ the total energy of
protons with $ 10^{19} \leq E \leq 10^{21}$~eV is $W_p \approx 0.2
W_{\rm GRB}$. For magnetic field in the presupernova wind at distance
$r \sim 10^{16}$~cm we accept the value of magnetic field $B_0 \sim
1$~G, following the estimate of \cite{Vie-DeM} $0.1 - 10$~G (see also
\cite{GRB-W}). Then for $W_{\rm GRB} \sim 1\times 10^{53}$~erg we
obtain $\Gamma \sim 10^5$ and for $W_{\rm GRB} \sim 5\times
10^{50}$~erg (jet value) $\Gamma \sim 10^3$. In fact for the jet case
the effective mass of the target is much less and $\Gamma$ is larger.
Such gas shell produces the relativistic shock and considered scenario
radically changes.

If considered UHE protons capture electrons from the media with its
frozen magnetic field, the pressure of the beam to the gas further
increases.

In fact, we considered above only one mechanism for energy transfer to
the gas. It is clear, that nature of this phenomenon is very general:
the energy density of proton beam is so large, that part of its energy
is transferred to magnetized ambient plasma. Only when this density
drops to some critical value and collective effects disappear, the
protons can propagate as individual particles, not exciting the ambient
plasma.

Therefore, the total adiabatic factor $f_a$ cannot be of order of 1.

\begin{figure}[ht]
\centering
\includegraphics[width=160mm,height=75mm,clip]{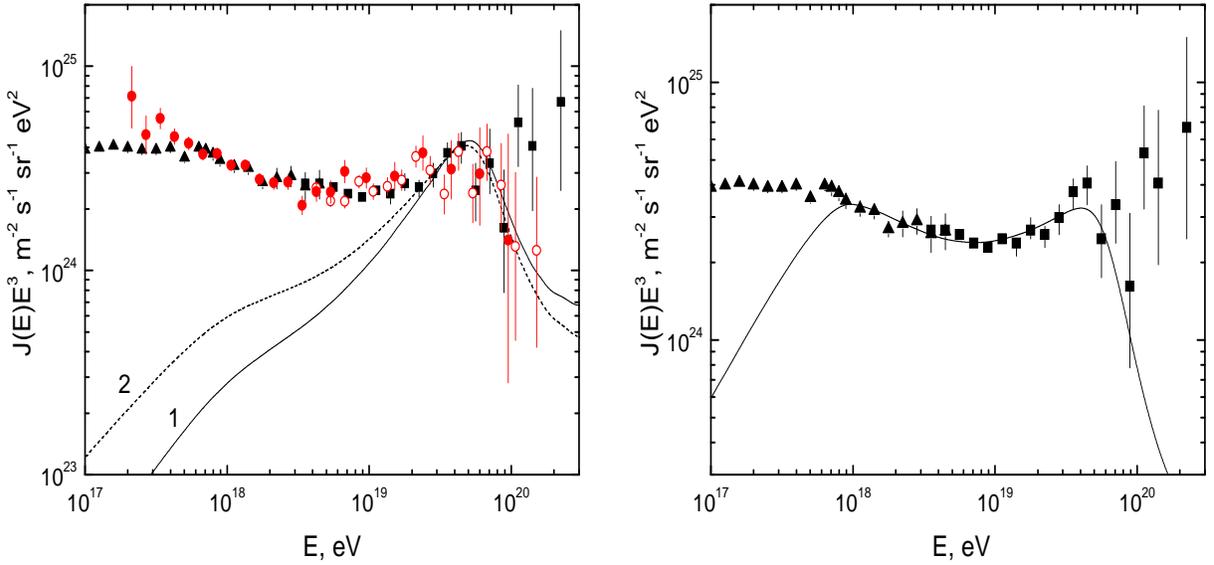}
\caption{\label{GRB} UHECR from GRBs. In the left panel are shown the
fluxes produced by internal shocks (curve 1) with generation spectrum
$\propto E^{-2}$, $E_{\rm max} \sim 1\times 10^{21}$~eV and $E_{\rm
min} \sim 1\times 10^{11}$~eV and by external relativistic shocks
(curve 2) with generation spectrum $\propto E^{-2.2}$, $E_{\rm max}
\sim 1\times 10^{21}$~eV and $E_{\rm min} \sim 1\times 10^{13}$~eV. In
the right panel the spectra are shown with the assumed distribution of
sources over $E_{\rm max}$, like in the case of AGN. The effective generation
spectrum at $E>E_c$ is assumed to be $\propto E^{-2.7}$ with $E_c \sim
1\times 10^{18}$~eV. }
\end{figure}

In the left panel of Fig.~\ref{GRB} the calculated flux of UHE protons
accelerated by internal shocks is presented by curve 1. It corresponds
to generation spectrum $\propto E^{-2}$ with $E_{\rm max}=1\times
10^{21}$~eV and $E_{\rm min}=\Gamma_0 m_pc^2= 1\times 10^{11}$~eV. This
is the case of absence of adiabatic energy losses $f_a=1$. The spectrum
is calculated in the standard way taking into account the energy
losses on CMB radiation. The emissivity needed to fit the observational
data is $\mathcal{L}_{\rm CR}=2.2\times 10^{45}$~erg
Mpc$^{-3}$yr$^{-1}$. It must be equal to $W_p \nu_{\rm GRB}$ and thus
it should not exceed $\mathcal{L}_{\rm GRB}$ given above. As a matter
of fact it is by factor 20 - 300 higher. If adiabatic energy losses for
protons with energies $10^{19} - 10^{21}$~eV is $f_a(>1\times
10^{19})$, the predicted flux in this energy range is additionally
suppressed by this factor and discrepancy increases correspondingly.

If to assume $f_a(E \leq 1\times 10^{19}~{\rm eV}) \ll 1$, while $f_a(E
\geq 1\times 10^{19}~{\rm eV}) \approx 1$, the energy balance is not
changed, because the total energy (including $W_p$ and that
adiabatically lost) remains the same.

In the right panel of Fig.~\ref{GRB} we show the spectrum with the
assumed distribution of sources over $E_{\rm max}$ like in case of AGN.
In this case the required emissivity is higher $\mathcal{L}_{\rm
CR}=3.5\times 10^{46}$~erg Mpc$^{-3}$yr$^{-1}$.

These results have been already presented in \cite{BGG1}. Waxman
\cite{GRB-W}  did not understand that we considered two different cases
(now being put separately in two different panels in Fig.~\ref{GRB})
and argued in fact only against the case shown in the right panel of
Fig.~\ref{GRB}. The left panel shows the same case as considered
by Waxman. Our discrepancy in emissivity is explained as follows. We
normalize our calculations by emissivity $\mathcal{L}_p$ (the energy
released in cosmic ray protons per unit comoving volume and per unit time) and
compare it with $\mathcal{L}_{\rm GRB}=W_{\rm GRB}\nu_{\rm GRB}$.
Waxman normalizes his flux by value $E^2 \dot{n}_{CR}/dE= 0.8\times
10^{44}$~erg/Mpc$^3$yr \cite{GRB-W}. To obtain emissivity, this value
should be multiplied by $\ln(E_{\rm max}/E_{\rm min})$, which is equal
to 23 for $E_{\rm max}=1\times 10^{21}$~eV and $E_{\rm min}=1\times
10^{11}$~eV. The obtained UHECR emissivity roughly agrees with our
value, but it exceeds by an order of magnitude what Waxman cites in
\cite{GRB-W} as the highest value of GRB emissivity ($3\times
10^{44}$~erg Mpc$^{-3}$yr$^{-1}$). Note that the comparison is made for
unrealistic case of absence of adiabatic energy losses.

\subsubsection{External shock acceleration and UHECR}
\label{sec:ext-shock} The flow of ultra-relativistic baryonic jet is
terminated by the external ultra-relativistic shock propagating in
pre-supernova wind. The acceleration by such shocks has been described
(with relevant references) in subsection \ref{sec:acceleration}. As  a
shock propagates in interstellar medium, the protons, experiencing the
different number of $u \to d \to u$ cycles, move together with the
shock ahead of it (upstream) or behind it (downstream). The ``fresh''
particles, which undergo one or a few ($n$) cycles have energy $E \sim
\Gamma_{\rm sh}^2 2^n E_i$, the ``old'' particles approach $E_{\rm max}
\sim e B R_{\rm sh} \Gamma_{\rm sh}$ \cite{GaAch}, where $E_i$ is
initial proton energy upstream, $R_{\rm sh}$ is a radius of the shock
(the distance from GRB origin), and $\Gamma_{\rm sh}\sim 100$ is the
Lorentz-factor of the shock. The acceleration time $t_a \sim E/(e c B
\Gamma_{\rm sh})$ \cite{GaAch} is smaller than the age of the shock $t
\sim R_{\rm sh}/c$ for all particles except those with $E \sim E_{\rm
max}$.

At this stage of our study we are interested only in the total energy
transferred to accelerated protons, disregarding what happens  to these
particles later. In other words we are interested only in energy
balance, using the common assumption $W_p \sim W_{\rm GRB}$ for the
total energy $W_p$ transferred to accelerated protons.

For the typical afterglow distance $R_{\rm sh} \sim 10^{16}$~cm, when
the shock moves with constant Lorentz factor $\Gamma_{\rm sh} \sim
100$, we shall use the following parameters for the upstream HE
protons:  the power-law spectrum $\propto E^{-\gamma_g}$ with the
``standard'' (see subsection \ref{sec:acceleration}) $\gamma_g=2.2$,
with conservative $E_{\rm min} \sim \Gamma_{\rm sh}^2 E_i \sim
10^{13}$~ eV and with $E_{\rm max}= e B R_{\rm sh} \Gamma_{\rm
sh}/\sqrt{3} \approx 1\times 10^{21}$~eV, valid for very large $B
\approx 5$~G. The fraction of the total energy $W_p$ in the form  of
particles with $E_p \geq 1\times 10^{19}$~eV is $f_{\rm UHECR}=0.039$.
In exotic case of existence of pre-accelerated particles in
pre-supernova wind with $E_i \sim 10^{13}$~eV, the minimal energy is 
$E_{\rm min} \approx 10^{17}$~eV and $f_{\rm UHECR}=0.29$.

As a matter of fact, part of accelerated particles are located
downstream. Because they are isotropized in the rest frame, the minimum
energy in the observer frame is shifted to
\beq %
E_{\rm min}=\frac{m_p c^2}{2}\frac{\Gamma_{\rm sh}}{\Gamma'_p} (1+
\Gamma_p^{'2}/\Gamma_{\rm sh}^2),
\label{E_min} %
\eeq %
where $\Gamma'_p$ is the Lorentz factor of proton in the rest system.
The energy (\ref{E_min}) corresponds to a proton moving backward
relative to shock. Additionally the number of particles with maximum
energy diminishes due to angular distribution. It results in
diminishing of $f_{\rm UHECR}$;  we shall not take this effect
into account.

There is one more effect which diminishes  $f_{\rm UHECR}$ for
downstream protons.  The highest energy protons can scatter to backward
direction and escape through the rear surface of the shell, while the
low-energy protons are better confined in the downstream region. In the
laboratory frame the backward moving protons have energy about 1~GeV,
as given by Eq.~(\ref{E_min}). The difference of proton energy (in the
laboratory frame) is transferred to the shell in the process of
magnetic scattering. This energy will go back to particle acceleration,
and most of it will be taken by low-energy particles.

Adiabatic energy losses, not taken into account in the relativistic
shock acceleration,  are discussed in previous subsection.

In realistic models the shock propagates in diminishing magnetic field,
e.g.\ $B(R)=B_0 R_0/R$ \cite{Vie-DeM}, valid for the stellar wind. For
such field it is easy to calculate the deflection angle $\theta$ for
proton propagating from $R_{\rm sh}$ to $R$,
\beq %
\theta \approx \frac{e B_{\rm sh} R_{\rm sh}}{E} \ln R/R_{\rm sh}, %
\label{defl-angle} %
\eeq %
and the deflection angle at which the shock overtakes the proton,
$\theta_c=\sqrt{3}/\Gamma_{\rm sh}$, is the same as for constant
magnetic field.  The maximum acceleration energy $E_{\rm max}$
determined from the condition $\theta=\theta_c$ remains also
approximately the same.  The exit of particles in the interstellar
medium occurs when $\Gamma_{\rm sh}$ diminishes, most notably at the
Sedov phase.

In Fig.~\ref{GRB} the calculated spectrum of UHECR from acceleration by
external shock is given by curve 2 for $\gamma_g=2.2$ and $E_{\rm
max}=1\times 10^{21}$~eV. The UHECR emissivity needed to fit the flux
at $E=1\times 10^{19}$~eV to the observational data is
$\mathcal{L}(>1\times 10^{19}{\rm eV})=4.7\times 10^{44}$~erg
Mpc$^{-3}$ yr$^{-1}$. The lower limit to HE proton emissivity (without
adiabatic energy losses and other effects discussed above) is given by
\beq %
\mathcal{L}_p > \mathcal{L}(>1\times 10^{19}{\rm eV})/f_{\rm
UHECR}= 1.2\times 10^{46}~{\rm erg Mpc}^{-3}{\rm yr}^{-1}
\label{emiss-ext} %
\eeq %
for $E_{\rm min}=1\times 10^{13}$~eV, and
$\mathcal{L}_p > 1.6\times 10^{45}$~erg Mpc$^{-3}$ yr$^{-1}$ for exotic
case with $E_{\rm min}=1\times 10^{17}$~eV. The emissivity
(\ref{emiss-ext}) is 2 - 3 orders of magnitudes higher than the
observed GRB emissivity $(0.6 - 13)\times 10^{43}$~erg
Mpc$^{-3}$yr$^{-1}$. The exotic assumption of $E_{\rm min}=1\times
10^{17}$~eV reduces this discrepancy only by one order of magnitude.

\subsubsection{Discussion and conclusion}
\label{sec:disc-concl} 
The calculations above agree with earlier works
\cite{BGG1,Steckera,Steckerb,Steckerc} (note the emissivities given in the
last work of this list).

The main difference between calculations 
\cite{BGG1,Steckera,Steckerb,Steckerc}) on the one hand, and those by
Waxman (e.g.\ \cite{GRB-W}) and Vietri (e.g.\ \cite{Vie-DeM}), on the
other hand, consists in the following.

While Waxman and Vietri compare GRB emissivity with emissivity of the
observed UHECR (Vietri at $E \geq 1\times 10^{17}$~eV), we compare it
with emissivity, which corresponds to energy transferred to all protons
in the process of their acceleration. Apart from spectral fraction of
energy, it includes also the energy lost by protons before they escape
to extragalactic space (adiabatic energy losses, energy transferred to
the magnetized plasma outside the jet and others discussed above).
Waxman for UHECR energy input uses the value $E^2 d\dot{n}_{\rm CR}/dE
= 0.8\times 10^{44}$~erg Mpc$^{-3}$yr$^{-1}$ \cite{GRB-W}, which after
integration from $E_{\rm min}=1\times 10^{11}$~eV to $E_{\rm max}=
1\times 10^{21}$~eV corresponds to emissivity $\mathcal{L}_p = 1.8
\times 10^{45}$~erg Mpc$^{-3}$yr$^{-1}$ to be compared with the
observed GRB emissivity $(0.6 - 13)\times 10^{43}$~erg
Mpc$^{-3}$yr$^{-1}$. Because of adiabatic energy losses, $\mathcal{L}_p$ 
is expected to be in fact much
higher than the above-mentioned value.

Vietri \etal\ \cite{Vie-DeM} indicate UHECR energy release required by
observations as $1\times 10^{45}$~erg Mpc$^{-3}$yr$^{-1}$ for energy
interval $1\times 10^{17} - 1\times 10^{21}$~eV (compared with our
value $1.6\times 10^{45}$~erg Mpc$^{-3}$yr$^{-1}$ for the same energy
interval). If one includes the energy input to all accelerated
protons, which accompany ultra-relativistic shock, from $E_{min} \sim
\Gamma_{\rm sh}^2 E_i$ to $E_{\rm max}$, this emissivity  increases at
least by factor 7.3, in conflict with observations by factor 60.
There is some difference with our work in numerical calculations:
Vietri \etal\ uses the HiRes data, we -- energy-shifted data by AGASA
and HiRes, which are in good intrinsic agreement (see Fig.~\ref{GRB});
Vietri \etal\ introduce corrections due interaction fluctuations, we do
not need to do it, because we normalize our calculations by
observational data at $E =2\times 10^{19}$~eV, where fluctuations are
absent. The maximum acceleration energy $E_{\rm max}=1\times
10^{21}$~eV at $R_{\rm sh}=1\times 10^{16}$~eV requires according to
our calculations (it is by factor $\sqrt{3}$ less than in \cite{GaAch})
magnetic field in the presupernova wind at this distance $\sim 6$~G,
which is very large and marginally possible only for the SupraNova
model \cite{supranovaa,supranovab}.

Our calculations require the UHECR emissivity $\mathcal{L}(>1\times
10^{19}{\rm eV})= 4.4\times 10^{44}$~erg Mpc$^{-3}$yr$^{-1}$ for
internal shock acceleration ($\gamma_g=2.0$ and $E_{\rm max}=1\times
10^{21}$~eV) and $4.7\times 10^{44}$~erg Mpc$^{-3}$yr$^{-1}$ for
external ultra-relativistic shock acceleration ($\gamma_g=2.2$ and
$E_{\rm max}=1\times 10^{21}$~eV). Including the energy injected in low
energy protons (not necessarily coming outside), this emissivity should
be a minimum of 10 - 30 times higher, and thus contradiction with GRB
emissivity reaches conservatively the factor 30 - 100. A natural
solution to this problem is given by an assumption that accelerated
protons carry $\sim 100$ times more energy than electrons, whose
radiation is responsible for GRBs. However, for $4^{\circ}$ beaming it
increases the required energy output of SN as GRB source from $W_{\rm
SN}\approx 5\times 10^{50}$~erg to $10^{52} - 10^{53}$~erg, which
creates the serious problem for SN models of GRBs. This problem exists
even for the case when one takes into account only ``observed'' UHECR
emissivity $\mathcal{L}(>1\times 10^{19}{\rm eV})$. Then the minimal
energy balance for internal shock model includes $\mathcal{L}_{\rm
GRB}= 1.3\times 10^{44}$~erg Mpc$^{-3}$yr$^{-1}$, $\mathcal{L}_{\rm
afterglow} \approx \mathcal{L}_{\rm GRB}$, $\mathcal{L}(>1\times
10^{19}{\rm eV})$, and  $\mathcal{L}_{\nu} \approx \mathcal{L}(>1\times
10^{19}{\rm eV})$, which results in SN energy output $4\times
10^{51}$~erg. For external shock model the energy balance does not
include neutrinos and it results in $W_{\rm SN} \approx 3\times
10^{51}$~erg.

\section{\label{sec:conclusions} Conclusions}
This work is naturally subdivided into two major parts: {\em (i)} the
model-independent analysis of spectral signatures of UHE proton
interaction with CMB, and {\em (ii)} the model-dependent analysis 
of transition from
extragalactic to galactic cosmic rays and model-dependent study of most
probable UHECR sources: AGN and GRBs.

In the first part the number of assumptions is minimal: we assume the
power-law generation spectrum of extragalactic protons $\propto
E^{-\gamma_g}$ and analyze uniform distribution of the sources with a
possible distortion in the form of large-scale source inhomogeneities
and local overdensity or deficit of the sources. The reference spectrum
is given by the {\em universal spectrum}, calculated for the {\em
homogeneous distribution} of the sources, when separation of sources $d
\to 0$. According to the propagation theorem \cite{AB}, in this case
the spectrum does not depend on propagation mode, e.g.\ it is the same
for rectilinear and diffusive propagation. We analyze also the spectra
with different source separations, giving emphasis to separations $ d
\sim 30 - 50$~Mpc, favorable by clustering of UHECR
\cite{BlMa,KaSe,YNS2}.

Calculation of the universal spectrum is based on equation of
conservation of particles with continuous energy losses of protons
interacting with CMB. We performed the new calculations of energy
losses (first presented in \cite{BGG1}), which agree well with
\cite{BeGrbump,St}, see Fig.~\ref{loss}. The technical element needed
for calculations, the ratio of energy intervals at generation and
observation $dE_g/dE$, is calculated by the new methods (see
Appendix~\ref{app-dEg/dE}). The analysis of spectral features is
convenient to perform with help of {\em modification factor},
$\eta(E)=J_p(E)/J_p^{\rm unm}(E)$, where the spectrum $J_p(E)$ is
calculated with all energy losses included, and unmodified spectrum
$J_p^{\rm unm}(E)$ -  only with account of adiabatic energy losses (red
shift).

There are four major signatures of UHE protons interacting with CMB:
GZK cutoff, bump, dip and the second dip.

{\em GZK cutoff} \cite{GZKG,GZKZK} is the most prominent signature,
which consists in the steepening of the proton diffuse spectrum due to
photopion production. The beginning of the GZK cutoff at $E_{\rm GZK}
\approx 4\times 10^{19}$~eV is difficult to observe. The quantitative
characteristic of the GZK cutoff is given by energy $E_{1/2}$, where
the flux with the cutoff becomes lower by factor 2 in comparison with
power-law extrapolation from lower energies. This quantity is possible
to observe only in integral spectrum. Practically independent of
$\gamma_g$~ $E_{1/2}= 5.3\times 10^{19}$~eV. Data of Yakutsk and HiRes
are compatible with this value (see Fig.~\ref{Ehalf-comp}), while data
of AGASA contradict it. As to spectral shape of the GZK cutoff, we
found it very model-dependent: it depends on separation of sources $d$,
on local source overdensity or deficit, on maximum acceleration energy
$E_{\rm max}$, and being so uncertain, the GZK steepening can be
imitated by $E_{\rm max}$ acceleration steepening.

{\em Bump} is produced by pile-up protons, experiencing the GZK
process. These protons do not disappear, they only shift in energy
towards that where the GZK cutoff begins. The bump is very clearly seen
in the spectra of individual sources (see left panel of
Fig.~\ref{bump}), but they disappear in the diffuse spectrum (right
panel of Fig.~\ref{bump}), because individual bumps are located at
different energies.

{\em Dip} is the most robust signature of UHE protons interacting with
CMB. It has very specific shape (see Fig.~\ref{mfactor}), which is
difficult to imitate by other processes, unless one has a model with
many free parameters. Dip is produced by pair production on the CMB
photons $p+\gamma \to p+e^++e^-$ and is located in energy interval
$1\times 10^{18} - 4\times 10^{19}$~eV. Dip has two flattenings (see
Fig.~\ref{mfactor}), one at energy $E \sim 1\times 10^{19}$~eV, which
automatically reproduces the ankle, well seen in the observational data
(see Fig.~\ref{dips}), and the other at $E \sim 1\times 10^{18}$~eV,
which provides transition from extragalactic to galactic cosmic rays.
The prediction of the dip shape is robust. It is not noticeably
modified by many phenomena included in calculations: by different
distances between sources (Fig.~\ref{discretness}), by changing the
rectilinear propagation of protons to the diffusive
(Fig.~\ref{dip-diff}), by different $E_{\rm max}$
(Fig.~\ref{diff-Emax}), by local overdensity and deficit of the sources
(Fig.~\ref{overdensity}), by large-scale source inhomogeneities
and fluctuations in $p\gamma$-interactions (Fig.~\ref{fluct}).

The dip is very well confirmed by data of Akeno-AGASA, HiRes, Yakutsk
and Fly's Eye (see Fig.~\ref{spectra} for the latter), while data of
Auger at this stage do not contradict the dip (Fig.~\ref{dips}). The
agreement with the data of each experiment Akeno-AGASA, HiRes and
Yakutsk is characterized by $\chi^2 \approx 19 - 20$ for about 20
energy bins, and with only two free parameters, $\gamma_g$ and the flux
normalization constant. The best fit is reached at $\gamma_g=2.7$, with
the allowed range 2.55 - 2.75.

The dip is used for energy calibration of the detectors, whose
systematic energy errors are up to $20\%$. For each detector
independently the energy is shifted by factor $\lambda$ to reach the
minimum $\chi^2$. We found $\lambda_{\rm Ag}=0.9$,~ $\lambda_{\rm
Hi}=1.2$ and $\lambda_{\rm Ya}=0.75$ for AGASA, HiRes and Yakutsk
detectors, respectively. Remarkably, that after this energy shift the
fluxes and spectra of all three detectors agree perfectly, with
discrepancy between AGASA and HiRes at $E> 1\times 10^{20}$~eV being
not statistically significant (see Fig.~\ref{Ag-Hi-Ya}).
The AGASA excess over predicted GZK cutoff might have the statistical 
origin combined with systematic errors in energy determination, and
the GZK cutoff may exist. 

The difference in energy shifts between fluorescent energy measurement
(HiRes) and on-ground measurements (AGASA, Yakutsk) with $\lambda > 1$
for HiRes and $\lambda < 1$ for AGASA and Yakutsk may signal a
systematic difference in energy determination by these two methods.

The excellent agreement of the dip with observations should be
considered as  confirmation of UHE proton interaction with CMB.

For astrophysical sources of UHECR the presence of nuclei in primary
radiation is unavoidable. Whatever a source is, the gas in it must
contain at least helium of cosmological origin with ratio of densities
$n_{\rm He}/n_{\rm H}=0.079$. Any acceleration mechanism operating in
this gas must accelerate helium nuclei, if they are ionized. The
fraction of helium in the primary radiation depends critically on the
mechanism of injection.  Fig.~\ref{dip-mix} shows that while 10\%
mixing of helium with hydrogen in the primary radiation leaves good
agreement of dip with observations, 20\% mixing upsets this agreement.
One may conclude that good agreement of dip with observations indicates
a particular acceleration mechanism, more specifically -- an
injection mechanism. One may also note that the shape of the dip
contains the information about chemical composition of primary
radiation (see Fig.~\ref{dip-nucl}).

The low-energy flattening of the dip is seen for both rectilinear and
diffusive propagation (see Fig.~\ref{trans}). It is explained by
transition from adiabatic energy losses to that due to $e^+e^-$-pair
production. Low-energy flattening of the extragalactic spectrum
provides transition to more steep galactic component, as one can see it
in the left panel of Fig.~\ref{transition}. Note, that when spectrum is
multiplied to $E^{2.5}$ the flat spectrum looks like one raising with
energy. In Fig.~\ref{transition} the transition is shown for the
diffusive spectrum. This model of transition implies that at energy
below $E_{\rm tr}\approx 5\times 10^{17}$~eV the galactic iron is the
dominant component, while at the higher energy extragalactic protons
with some admixture of extragalactic helium dominate. The observational
spectrum feature, which corresponds to this transition, is the second
knee. The prediction of this model, the dominance of proton component
at $E > 1\times 10^{18}$~eV, is confirmed by data of HiRes
\cite{Sokola,Sokolb}, HiRes-MIA \cite{HiRes-MIA}, by Yakutsk
\cite{Glushkov2000} and does not contradict to the Haverah Park data at
$E >(1 - 2)\times 10^{18}$~eV \cite{HP}. However, the
Akeno\cite{Akeno-mass} and Fly's Eye \cite{FE} data favor the mixed
composition dominated by heavy nuclei.

{\em The second dip} is produced by interplay between pair production
and photopion production. It is a narrow feature at energy
$E_{\rm 2dip}=6.3\times 10^{19}$~eV, which can be
observed by detectors with good energy resolution. If detected, it
gives the precise mark for energy calibration of a detector.

The UHECR sources have to satisfy two conditions: they must be very
powerful and must accelerate particles to $E_{\rm max}\geq 1\times
10^{21}$~eV. There is one more restriction \cite{DTT,FK,YNS1,Sigl1,
YNS2,BlMa,KaSe}, coming from observation of small-scale clustering: the
space density of the sources should be $(1 - 3)\times
10^{-5}$~Mpc$^{-3}$, probably with noticeable uncertainty in this
value. Thus, these sources are more rare, than typical representatives
of AGN, the Seyfert galaxies, whose space density is $\sim 3\times
10^{-4}$~Mpc$^{-3}$. The sources could be the rare types of AGN, and
indeed the analysis of \cite{corr} show statistically significant
correlation between directions of particles with energies $(4 -
8)\times 10^{19}$ eV and directions to AGN of the special type -- BL
Lacs. The acceleration in AGN can provide the maximum energy of
acceleration to $E_{\rm max}\sim 10^{21}$~eV. The appropriate
mechanisms can be acceleration by relativistic and ultra-relativistic
shocks (see section \ref{sec:acceleration}). These mechanisms typically
provide spectra flatter than one with $\gamma_g \approx 2.6 - 2.8$,
needed for agreement of the dip with data. The distribution of the
sources over $E_{\rm max}$ suggested in \cite{KaSe} solve this problem:
starting from some energy $E_c$ the spectrum becomes steeper. In this
case the diffuse spectrum below $E_c$ has canonical generation index
$\gamma_g=2.0$ or $\gamma_g=2.2$, and above this energy $\gamma_g
\approx 2.6 - 2.8$ to fit the observations. The energy $E_c$ is a free
parameter of the models, and we keep it typically as $10^{17} -
10^{18}$~eV. Such broken generation spectrum relaxes the requirements
for source luminosities. For example, for $\gamma_g=2.7$ and
$E_c=1\times 10^{18}$~eV, the typical emissivity needed to fit the
observed flux is $\mathcal{L}_0=n_s L_p \sim 3\times 10^{46}$~erg
Mpc$^{-3}$yr$^{-1}$, which is held for the source density $n_s=2\times
10^{-5}$~Mpc$^{-3}$, estimated from the small-angle clustering
\cite{BlMa,KaSe}, and for the source luminosity $L_p \sim 6\times
10^{43}$~erg/s, very reasonable for AGN. The calculated spectra are in
a good agreement with observations (see Fig.~\ref{spectra}). The
transition from galactic to extragalactic cosmic rays in this model
occur at the second knee (see left panel of Fig.~\ref{transition}).

Another potential UHECR sources are GRBs. They have large energy output
and can accelerate particles to UHEs. Acceleration occurs at the
multiple inner shocks \cite{acc-GRB-W,UHECR-W} and at
ultra-relativistic external shock \cite{acc-GRB-V,Vie-DeM}. The maximum
acceleration energy can be as high as $ \sim 1\times 10^{21}$~eV. In
case of ultra-relativistic shock with Lorentz factor of shock
$\Gamma_{\rm sh}\sim 100$ and at distance $R_{\rm sh} \sim 1\times
10^{16}$~cm, typical for afterglow region, the maximum energy  $E_{\rm
max}=(1/\sqrt{3})e B R_{\rm sh} \Gamma_{\rm sh}$ reaches $1\times
10^{21}$~eV in case of very strong external magnetic field $B \sim
6$~G. Such strong magnetic field can be marginally provided by two
combined effects: strong field in presupernova wind (possible only in
the model of SupraNova \cite{supranovaa,supranovab}) combined with
preshock amplification of the field.

The energy output of GRBs necessary to provide the observed UHECR flux
imposes the serious problem \cite{Steckera,Steckerb,Steckerc,BGG1}. It
is usually assumed that energy output in the form of accelerated
protons is of order of magnitude of that observed in GRB photons. The
observed emissivity of GRBs is
\beq %
\mathcal{L} \sim (0.6 - 13)\times
10^{43}~{\rm erg} {\rm Mpc}^{-3}{\rm yr}^{-1}, %
\label{GRBemissivity} %
\eeq %
where the lower value is from works by Schmidt
\cite{Schmidta,Schmidtb} and the large value - from work by Frail
\etal\ \cite{Frail}. Both authors used in their estimates the similar
energy interval of photons, (10 - 1000~keV) and (0.2 - 2000~keV).

Our calculations of UHECR spectra for internal shocks (see
Fig~\ref{GRB} curve 1) with $\gamma_g=2.0$ and $E_{\rm max}=1\times
10^{21}$~eV give UHECR emissivity $\mathcal{L}(\geq 1\times
10^{19}~{\rm eV})= 4.4\times 10^{44}~{\rm erg}{\rm Mpc}^{-3}{\rm
yr}^{-1}$, by factor 3.4 - 73 higher than (\ref{GRBemissivity}). If to
include $\mathcal{L}_{\nu} \sim \mathcal{L}_{\rm UHECR}$ contradiction
with (\ref{GRBemissivity}) increases to 6.8 - 150. If to include in
balance the protons with energies down to $E_{\rm min} \sim 1\times
10^{11}$~eV (not necessarily coming out) the contradiction increases by
another factor 5. Adiabatic energy losses further increase this
discrepancy.

Our calculations for external shock acceleration (see curve 2 
in Fig~\ref{GRB}) with $\gamma_g=2.2$ and $E_{\rm max}=1\times 10^{21}$~eV
results in UHECR emissivity $\mathcal{L}(\geq 1\times 10^{19}~{\rm
eV})= 4.7\times 10^{44}~{\rm erg}{\rm Mpc}^{-3}{\rm yr}^{-1}$, i.e.\ by
factor 3.6 - 78 times higher than (\ref{GRBemissivity}). It increases
by factor 26, if to include protons with energies down to $E_{\rm min}
\sim \Gamma_{\rm sh}^2 E_i\sim 10^{13}$~eV, or by factor 3.5 for
unrealistic case of $E_{\rm min} \sim 1\times 10^{17}$~eV. Adiabatic
energy losses are less than in case of internal shock acceleration.

This energy crisis can be resolved, assuming that accelerated protons
in GRBs take away $\sim 100$ times more energy than electrons
responsible for GRB photons. However, this assumption dramatically
affects the status of SN origin for GRBs , even in the case of narrow
$4^{\circ}$ beaming.

\subsection{Predictions for the Auger measurements}
\label{sec:predictions} 
What do we predict for the Auger observations
on the basis of our analysis?
\begin{figure}[ht]
\begin{minipage}[h]{8cm}
\centering
\includegraphics[width=76mm,height=76mm,clip]{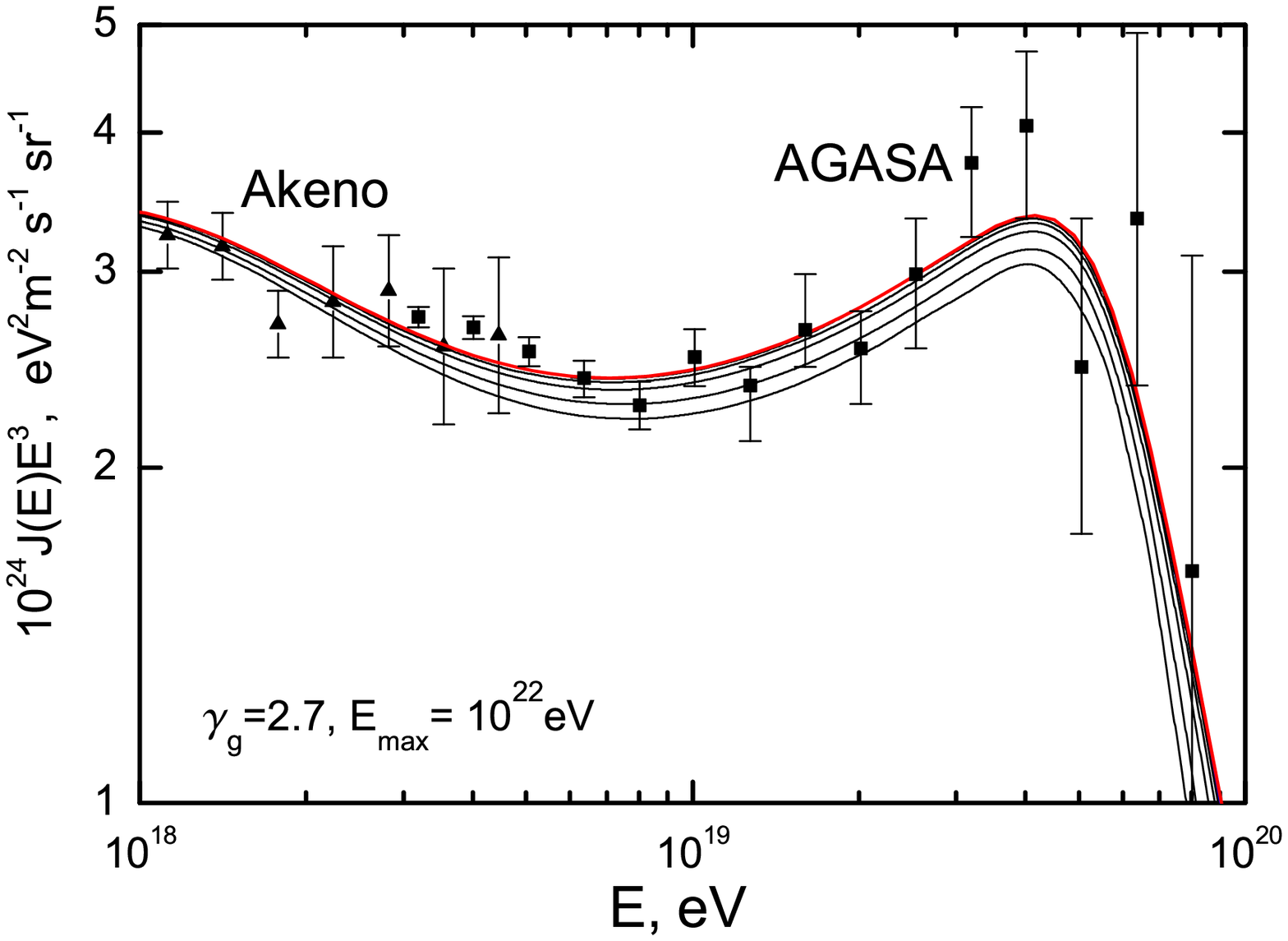}
\end{minipage}
\hspace{5mm}
\begin{minipage}[h]{8cm}
\centering
\includegraphics[width=76mm,height=76mm,clip]{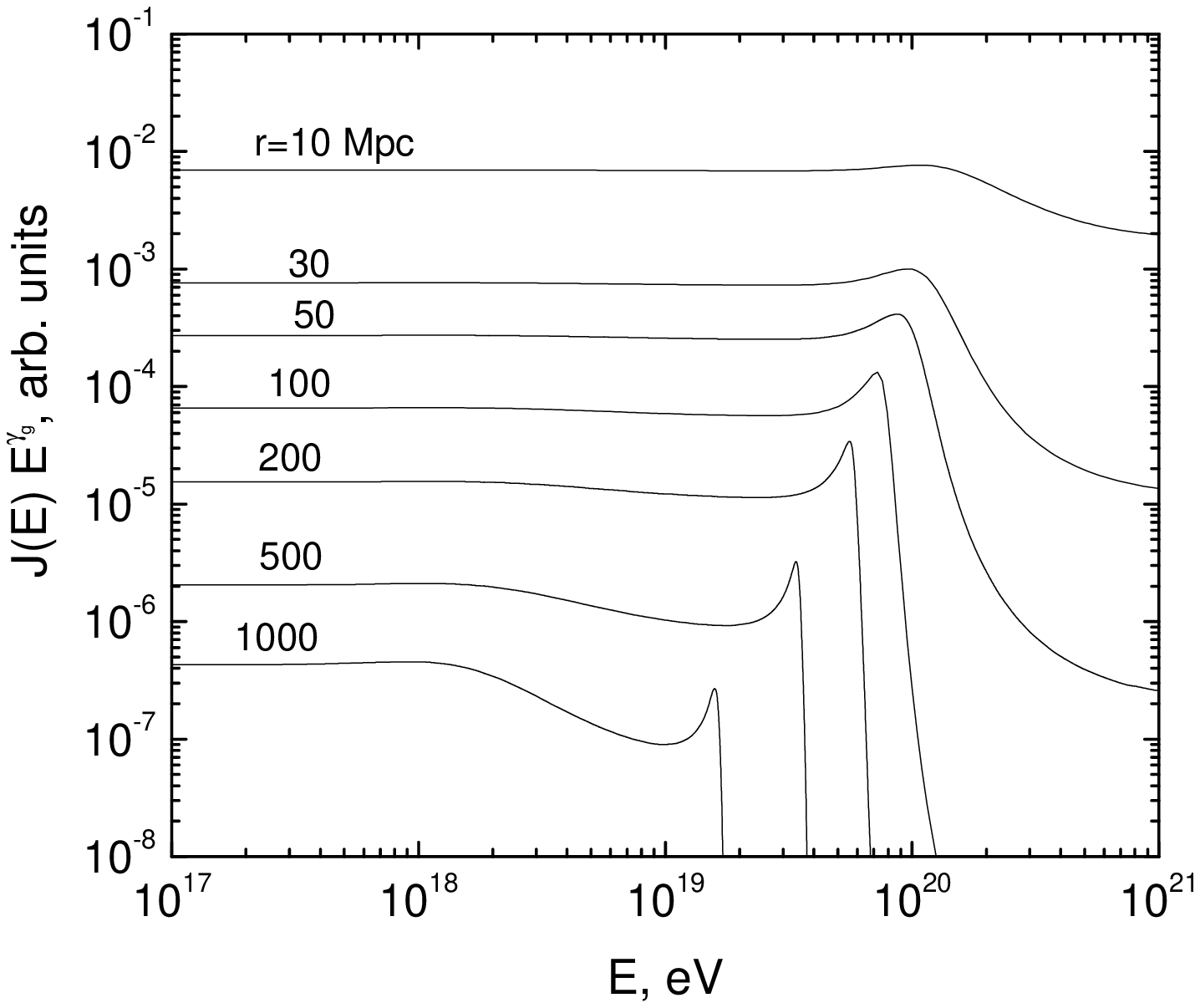}
\end{minipage}
\caption{ Predictions for the Auger detector. In the left panel the 
diffuse spectrum
prediction is displayed for different distances $d$ between sources,
which provide the largest uncertainties. The spectrum is normalized by
the energy-shifted AGASA data which coincides well with HiRes and
Yakutsk data (see Fig.~\ref{Ag-Hi-Ya}). In energy range $1\times
10^{18} - 7\times 10^{19}$~eV the uncertainties in the predictions are
small and Auger should observe the beginning of the GZK cutoff at $E
\geq 5\times 10^{19}$~eV. In the right panel the spectra with bumps
from the nearby sources are displayed ($\gamma_g$ is the generation
index). } \label{prediction}
\end{figure}
\begin{itemize}
\item
We predict the dip at energy $1\times 10^{18} - 4\times 10^{19}$~eV
not only as the spectrum shape, but with the absolute values of flux,
based on the agreement of AGASA-HiRes-Yakutsk fluxes after energy
calibration by the dip (see Fig.~\ref{Ag-Hi-Ya}). This spectrum is
displayed in the left panel of Fig.~\ref{prediction}. The distortion of
the shape of the predicted spectrum can occur only due to presence of
extragalactic nuclei.
\item
The above-mentioned distortion is important as a tool for measuring the
chemical composition of UHECR. The chemical composition is the most
difficult problem for UHECR experiments, in which little progress has
been reached during last 30 years. The measurement of the dip shape
gives the precise method of determination of 10 - 15 \% of admixture of
helium in primary UHECR (see Fig.~\ref{dip-nucl}). The distortion of
the proton dip most noticeably occurs at energy $E \gsim 3\times
10^{19}$~eV, (see Fig.~\ref{dip-mix}) where precise measurements of
Auger are expected.
\item
Normalization of energies measured by  AGASA, Yakutsk and HiRes
detectors with the help of the dip (see section
\ref{sec:Ag-Hi-discrep}) implies that energy measured by fluorescent
method is higher than that by on-ground method (the energy shift for
AGASA and Yakutsk is $\lambda < 1$, while for HiRes  $\lambda > 1$).
With help of the hybrid events Auger can resolve this problem.
\item
The second dip (see Fig.~\ref{fluct-ratio}) may be marginally observed
and used for energy calibration of the detector.
\item
Yakutsk and HiRes data give evidence for the numerical characteristic
of the GZK cutoff $E_{1/2}=5.3 \times 10^{19}$~eV in the integral
spectrum (see Fig.~\ref{Ehalf-comp}). In the Auger data this value can be
measured more reliably.
\item
The predicted shape of the GZK steepening has many uncertainties (see
section \ref{sec:GZKcutoff} for the discussion). However, beginning of
the GZK cutoff in the narrow energy interval $(4 - 7)\times 10^{19}$~eV
is predicted with the high accuracy (see left panel in
Fig.~\ref{prediction}). The precise spectrum measurement in this energy
range can give the decisive proof of the GZK cutoff.
\item
The mass composition at 
$1\times 10^{18} \lsim E \lsim 1\times 10^{19}$~eV measured by the 
fluorescent technique can distinguish between models of the
second knee transition from galactic to extragalactic cosmic rays 
(protons with 10 - 20\% of helium) and the ankle  transition (iron dominance)
\item
The bumps in the spectrum of single sources, if detected, can be be
observed (see Fig.~\ref{prediction}).
\end{itemize}

\begin{acknowledgments}
We acknowledge the participation of R. Aloisio and P. Blasi in some
parts of this work. B. Hnatyk, D. De Marco and M. Kachelriess are thanked for
useful discussions. Participation of M. I. Levchuk at the initial stage
of pair-production energy loss recalculation is gratefully acknowledged. We
thank ILIAS-TARI for access to the LNGS research infrastructure and for
the financial support through EU contract RII3-CT-2004-506222. The
work of S.G. is partly supported by grant LSS-5573.2006.2 . 
\end{acknowledgments}

\appendix
\section{\label{app-enloss} Calculations of energy losses}
We consider here some details of calculations of energy-losses due to 
production of $e^+e^-$-pairs and pions. 

Pair production energy loss of ultrahigh-energy protons in low-temperature
photon gas, e.g. CMB,
\beq %
p + \gamma_{CMB} \rightarrow p + e^+ + e^-, %
\label{eep} %
\eeq %
has been previously discussed in many papers. The differential
cross-section for this process in the first Born approximation was
originally calculated in 1934 by Bethe and Heitler \cite{BetHeit} and
Racah \cite{Racah}. In 1948 Feenberg and Primakoff \cite{FeenPrim}
obtained the pair production energy loss rate using the extreme
relativistic approximation for the differential cross-section. In
1970 the accurate calculation was performed by Blumenthal
\cite{Blumenthal}. Later some analytical approximations to differential
cross-sections were obtained in Ref.~\cite{Chodor}.

All authors neglected the recoil energy of proton, putting
$m_p\rightarrow\infty$, the effect being suppressed by a factor of
$m_e/m_p \approx 5 \times 10^{-4}$.

In spite of the fact that all calculations actually used the same
Blumenthal approach, there are noticeable discrepancies in the results
of different authors; they are displayed in Fig.~1b of the
Ref.~\cite{stanev}.

To clarify the situation we recalculate the pair production energy
loss of high-energy proton in the low-temperature photon gas. In
contrast to Ref.~\cite{Blumenthal} we use the first Born
approximation approach of Ref.~\cite{BergLinder} taking into
account the finite proton mass. The exact non-relativistic
threshold formula with corrections due to different Coulomb
interactions of electron and positron with the proton (see e.g.\
Ref.~\cite{LL}) is used. No series expansions of
$\sigma(\varepsilon_r)$, where $\varepsilon_r$ is the photon
energy in the proton rest system, are involved in our
calculations.
\begin{figure}[ht]
\begin{minipage}[h]{8cm}
\centering
\includegraphics[width=77mm,height=77mm,clip]{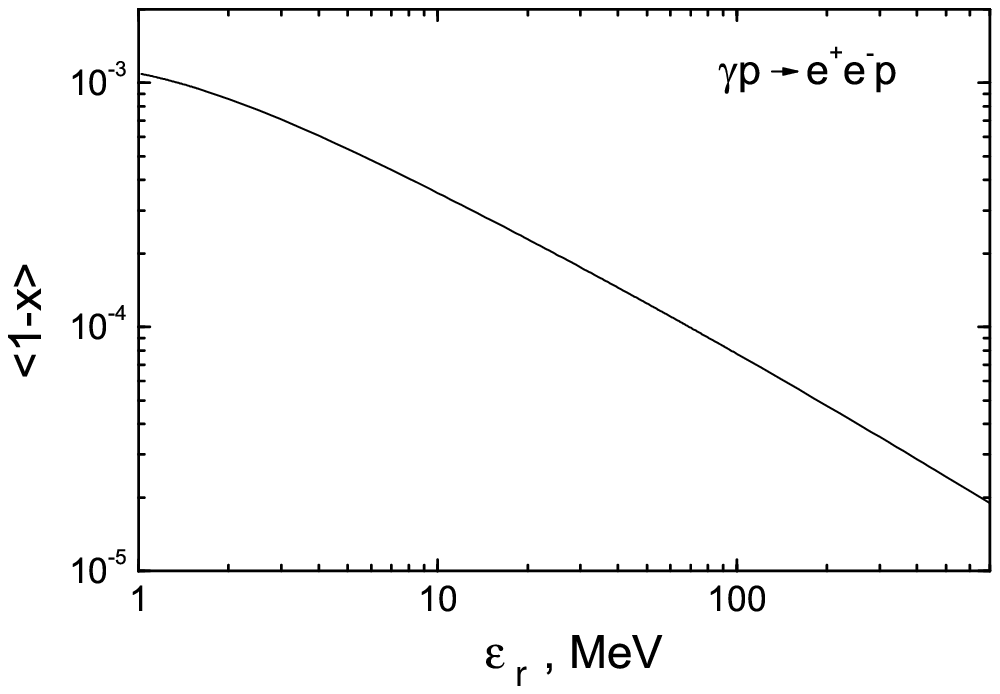}
\end{minipage}
\hspace{5mm}
\begin{minipage}[h]{8cm}
\vspace{-1mm} \centering
\includegraphics[width=75mm,height=75mm,clip]{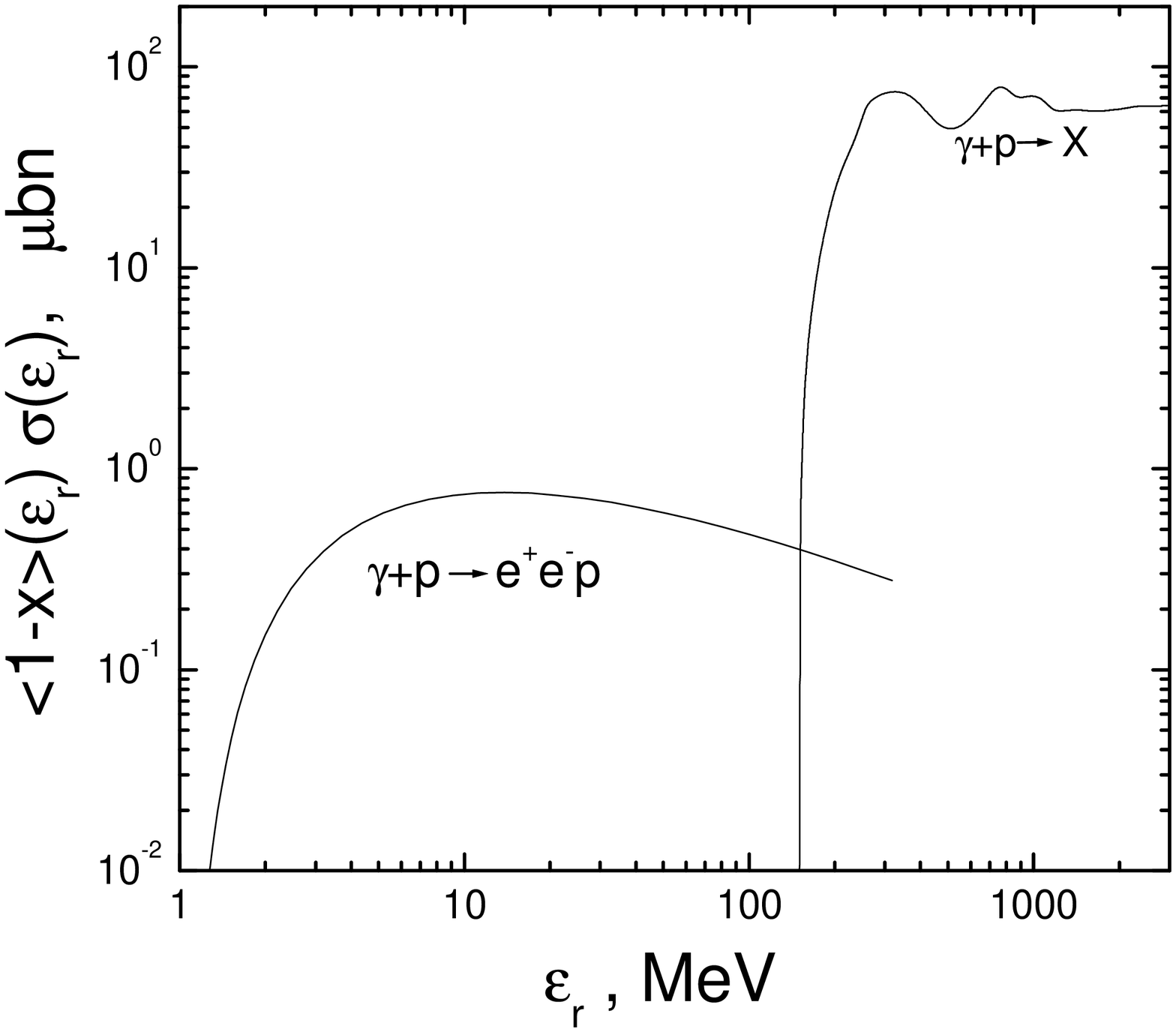}
\end{minipage}
\caption{ In the left panel is shown the average fraction of the
incident proton energy $E_p$ carried away by $e^+e^-$ pair as a
function of the photon energy $\varepsilon_r$ in the proton rest
frame. The right panel displays the product of fraction of energy
lost, $\langle 1-x \rangle$, and the cross-section for pair
production, $\gamma p \rightarrow e^+e^-p$, or photopion
production $p+\gamma \rightarrow X$ as function of $\varepsilon_r$.} %
\label{comp} %
\end{figure}
The average fraction of proton energy loss in one collision with a
photon is defined as
\beq %
\label{averfraction} %
\langle 1-x \rangle(\varepsilon_r) = \frac 1
{\sigma(\varepsilon_r)} \int \limits
_{x_{min}(\varepsilon_r)}^{x_{max}(\varepsilon_r)}  dx \; (1-x)
\frac{d\sigma(\varepsilon_r,x)}{dx}, %
\eeq %
where $x=E^\prime_p/E_p$, with $E_p$ and $E^\prime_p$ being the
incident and final proton energies, respectively, in the
laboratory system. The upper and lower limits on fractions $x$ are
given by
\beq %
x_{_{\min}^{\max}}(\varepsilon_r)=x_c(\varepsilon_r) \pm
y_c(\varepsilon_r),
\eeq %
\beq %
x_c(\varepsilon_r)= \frac 12 \left( 1+
\frac{m_p^2-m_\pi^2}{m_p^2+2m_p \varepsilon_r} \right); \;\;\;
y_c(E_c)= \sqrt{x_c^2(\varepsilon_r) - \frac{m_p^2}{m_p^2+2m_p
\varepsilon_r}}.
\eeq %

Our strategy was to calculate $\langle 1-x \rangle(\varepsilon_r)$
by performing the direct fourfold integration of the exact matrix
element over the phase space. It should be noted, that direct
numerical integration, especially at high energies, is difficult
in this case because of forward-backward peaks in the
electron-positron angular distributions. To overcome this problem,
we performed two integrations over polar and azimuth angles in the
$e^+e^-$ subsystem analytically. This was facilitated by using of
the \emph{MATHEMATICA 4} code. The residual two integrations over
energy and scattering angle in the initial $p\gamma$ subsystem
were carried out numerically. We calculate simultaneously the
total cross-section for pair production. The accuracy of our
calculations was thus controlled by comparison of the calculated
total cross-section with the well-known Bethe-Heitler
cross-section.

The average fraction of proton energy lost in one collision with a
photon is plotted in Fig.~\ref{comp} as a function of the photon energy
in the proton rest system.

The product of this fraction and the total cross-section for pair
production is shown in Fig.~\ref{comp}. This function should be
integrated over the photon spectrum  to obtain the average energy loss
due to pair production in the photon gas with this spectrum.

The comparison of our calculations with  Ref.~\cite{stanev} shows
the negligible difference (see Fig.~\ref{loss}).

Calculations of photopion energy losses are described in 
section \ref{sec:en-loss}. In the right panel of Fig.~\ref{comp} we show 
$\langle 1-x \rangle \sigma$ for these energy losses as function of 
$\varepsilon_r$.
\section{\label{app-dEg/dE} Connection between energy intervals at epochs of
production and observation }
If we consider the protons with energy $E$
in the interval $dE$ at the epoch with red-shift $z=0$, what will be
the corresponding interval of generation $dE_g$ at epoch $z$, when
energy of a proton was $E_g(z)$? The connection between these two
intervals is given by Eq.~(36) of Ref.~\cite{BeGrbump}. Here we shall
confirm this formula using a different, more simple derivation. Note,
that intermediate formulae (40) and (41) used for derivation of final
Eq.~(36) in Ref.~\cite{BeGrbump} have a misprint: the correct power of
$(1+z)$ term there is 3, not 2.

Regarding the proton  energy losses of a proton on CMB at
arbitrary epoch with red-shift $z$, we shall use, as in Section
\ref{sec:en-loss}, the notation $b(E,z)=-dE/dt$ and $\beta(E,z)=-
(1/E)dE/dt$ with $ b(E,z)=(1+z)^2 b_0[(1+z)E]$. Here and
henceforth $b_0(E)$ and $\beta_0(E)$  are used for energy losses
at $z=0$. The energy loss due to red-shift at the epoch $z$ and
the Hubble parameter $H(z)$ are given by
\beq %
\left (dE/dt\right )_{\rm r-sh}=- E H(z),~ ~ ~
H(z)=H_0 \sqrt{\Omega_m(1+z)^3+\Omega_{\Lambda}}. %
\label{r-sh} %
\eeq %
The energy of a proton at epoch $z$, 
\beq E_g(z)= E+
\int_t^{t_0}dt\left [ \left (dE/dt\right )_{{\rm r-sh}} +\left
(dE/dt\right )_{{\rm CMB}}\right], \label{E-g,t} 
\eeq 
can be easily
rearranged as \beq E_g(z)=E+\int_0^z\frac{dz'}{1+z'}E_g(z')+\int_0^z
dz' \frac{1+z'}{H(z')}b_0\left [(1+z')E_g(z')\right ], \label{E-g,z}
\eeq using $dt/dz=-H^{-1}(z)/(1+z)$ and Eq.~(\ref{b_z}).

Differentiating Eq.~(\ref{E-g,z}) with respect to $E$,  one finds for
energy interval dilation $y(z) \equiv dE_g(z)/dE$: \beq
y(z)=1+\int_0^z\frac{dz'}{1+z'}y(z')+\int_0^z dz'
\frac{(1+z')^2}{H(z')}y(z')\left (\frac{db_0(E')}{dE'}\right
)_{E'=(1+z')E_g(z')}. \label{eq-z} \eeq Corresponding differential
equation is
\beq %
\frac{1}{y(z)}\frac{dy(z)}{dz}=\frac{1}{1+z}+\frac{(1+z)^2}{H(z)} \left
(\frac{db_0(E')}{dE'}\right )_{E'=(1+z)E_g(z)}. %
\label{eq-z,diff} %
\eeq %
The solution of Eq.~(\ref{eq-z,diff}) is
\beq %
y(z)\equiv \frac{dE_g(z)}{dE}=(1+z)\exp\left[\frac{1}{H_0}\int_0^z dz'
\frac{(1+z')^2}{\sqrt{\Omega_m(1+z')^3+\Omega_{\Lambda}}}
\left(\frac{db_0(E')}{dE'}\right )_{E'=(1+z')E_g(z')}\right], %
\label{dE_g/dE} %
\eeq %
where $E_g(z)$ is an energy at epoch z. In case of the highest energies
the intergrand in Eq.~(\ref{dE_g/dE}) can be expanded in powers of $z'$,
and after simple transformations one obtains 
$$
dE_g/dE \approx b(E_g)/b(E),
$$
valid with 10\% accuracy at $E\gsim 4\times 10^{19}$~eV.

Eq.~(\ref{dE_g/dE}) coincides with Eq.~(36) from \cite{BeGrbump}.

We shall give now another derivation of Eq.~(\ref{dE_g/dE}) based on
the exact solution to the kinetic equation for propagation of UHE
protons with continuous energy losses. This solution automatically
includes $dE_g/dE$ term which coincides with that given by
Eq.~(\ref{dE_g/dE}). This equation for density of UHE protons,
$n_p(E,t)$, reads
\beq %
\label{eq:kineq} %
\frac{\partial}{\partial t}n_p(E,t) - \frac{\partial}{\partial E} \left
[ \tilde b(E,t) n_p(E,t) \right ] - Q_g(E,t) = 0, %
\eeq %
with $\tilde b(E,t) = E H(t) + b(E,t)$, where $E H(t)$ describes
adiabatic energy loss and $b(E,t)$ -- those due interaction with CMB.

Written in the equivalent form
\beq %
\label{eq:kineq1} %
\frac{\partial}{\partial t}n_p(E,t) - \tilde
b(E,t)\frac{\partial}{\partial E}n_p(E,t) - n_p(E,t)\frac{\partial
\tilde b(E,t)}{\partial E} - Q_g(E,t) = 0, %
\eeq %
this equation can be solved with help of auxiliary characteristic
equation $dE/dt = - \tilde b(E,t)$, which solution is $\mathcal{E}
(t)=E_g(E_i,t_i,t)$, where like in Section \ref{sec:universal}
$E_g$ is the generation energy at age $t$, if $E=E_i$ at $t_i$.
When energy of a proton is taken on the characteristic $E =
\mathcal{E}(t)$, the second term in the l.h.s.\ of
Eq.~(\ref{eq:kineq1}) disappears and the solution found with help
of integration factor is
\beq %
n_p(\mathcal{E},t) = \int_{t_{\min}}^t dt' Q_g \left [
\mathcal{E}(t'),t' \right ]
 \exp \left [ \int_{t'}^t dt''
\frac{\partial}{\partial \mathcal{E}} b \left(\mathcal{E}(t''),t''
\right ) \right ]. %
\label{eq:n_p(t)} %
\eeq %
Introducing the red-shift $z$ as a variable, $dt/dz = -
H^{-1}(z)/(1+z)$, we obtain for present epoch ($z=0$):
\beq %
\label{eq:n_p(z)} %
n_p(\mathcal{E}) = \int_0^{z_{\max}} \frac{dz'}{(1+z')H(z')} Q_g \left
[ \mathcal{E}(z'),z' \right ] \exp \left [ \int_0^{z'}
\frac{dz''}{(1+z'')H(z'')} \frac {\partial b(
\mathcal{E},z'')} {\partial \mathcal{E}} \right ] %
\eeq %
Using
\[
\frac{\partial b(\mathcal{E},z)}{\partial \mathcal{E}} = H(z) + (1+z)^3
\left [ \frac{\partial b_0(E')}{\partial E'} \right ]
_{E'=(1+z)\mathcal{E}(z)}
\]
one finally gets
\beq %
n_p(\mathcal{E}) = \int \limits_0^{z_{\max}}
\frac{dz'}{(1+z')H(z')} Q_g \left [ \mathcal{E}(z'),z' \right ] (1+z')
\exp  \left [\frac 1 {H_0} \int \limits_0^{z'} dz''
\frac{(1+z'')^2}{\sqrt{\Omega_m (1+z'')^3 + \Omega_\Lambda}} \left(
\frac {\partial b_0(E')} {\partial E' }\right
)_{E'=(1+z'')\mathcal{E}(z'')} \right ] %
\label{eq:n_p(z)_final} %
\eeq %

Since $\mathcal{E}(z') = E_g(z')$, one easily recognizes inside the
integral the term $dt' Q_g(E_g,t')$ and $dE_g/dE$ as given by
Eq.~(\ref{dE_g/dE}).

\section{\label{app-fluct} Kinetic equation solution vs continuous
loss approximation: physical interpretation}
We shall compare here
the kinetic equation solution for proton spectra,
$n_{kin}(E)$, with that obtained in continuous-energy-loss (CEL)
approximation, $n_{cont}(E)$, and study the physical meaning of
their difference.

Additionally to the full kinetic equation (\ref{flucteqn}) we shall
consider here also the stationary kinetic equation with  the photopion
production only. Such equation describes realistically the spectrum at
$E > 1\times 10^{20}$~eV, where pion energy losses dominate, and
effects of CMB evolution are negligible. However, we formally consider
this equation at the lower energies, too, to study the difference
between kinetic equation  solution and CEL approximation.

The general kinetic equation is reduced to the stationary one with
photopion interaction only, putting  in Eq.~(\ref{flucteqn}) 
$\partial n/\partial t=0$,  $H(t)=0$, $T=T_0$ and $b_{pair}=0$.

Then the stationary kinetic equation reads
\begin{equation}
-P(E)n_p(E)+\int \limits_E^{E_{max}} dE' P(E',E) n_p(E') + Q_{gen}(E)
=0 , 
\label{app-kin}
\end{equation}
with $P(E)$ and $P(E',E)$ given by Eqs.~(\ref{P(E,t)}) and
(\ref{P(E',E,t)}). Introducing variable $x=E/E'$, we rearrange the
regeneration term into
\begin{equation}
\frac{1}{E}\int \limits_0^1 dx \left ( \frac{E}{x} \right )
P(E/x,x)n_p(E/x) . 
\label{reg-term}
\end{equation}
Expanding the integrand into the Taylor series in powers of $(1-x)$ one
observes the  exact compensation of the zero power term with
$-P(E)n(E)$. The first power term results in the stationary continuous
energy loss equation
\beq %
\label{app-CEL} %
\frac{\partial}{\partial E}\left[ b(E) n_p(E) \right] + Q_{gen}(E)=0,
\eeq and account of $(1-x)^2$ term gives the stationary Fokker-Planck
equation \beq \frac{\partial}{\partial E}\left[ b(E) n_p(E) +
\frac{\partial}{\partial E}\left(E^2D(E)n_p(E) \right) \right] +
Q_{gen}(E)=0  ,
\label{app-FP} %
\eeq %
where the term with diffusion
coefficient $D(E)$ in momentum space describes fluctuations.

This result reproduces the general proof (see e.g.\ \cite{LL10}, p.~89)
for the case when all processes go with a small energy transfer.

Let us now consider the analytic solution to Eq.~(\ref{app-kin}) with
$E_{\rm max} \rightarrow \infty$ in the asymptotic limit $E \gg
E_{GZK}$. We shall use here ({\em and only here}) a toy model with
simplified assumptions about interactions, namely with one-pion
production $p+\gamma \rightarrow \pi + N$, with isotropic pion
distribution in c.m.\ system and with asymptotic cross-section
$\sigma_{p\gamma} \rightarrow \mbox{const}$, when $E_c \rightarrow
\infty$. With these assumptions we readily obtain
\beq %
P(E) \rightarrow c \sigma_{p\gamma}n_\gamma, \;\;\;\; P(E',E)
\rightarrow c \sigma_{p\gamma}n_\gamma/E' ,
\eeq %
and using $Q_{gen}(E) = Q_0 E^{-\gamma_g},$ we arrive at
\beq %
n_{kin}(E) = \frac{\gamma_g}{\gamma_g -1}~
\frac1{\sigma_{p\gamma}n_\gamma c}~ Q_0 E^{-\gamma_g} .
\eeq %

For continuous energy loss equation (\ref{app-CEL}), using 
$b(E') = \langle E'- E \rangle c \sigma_{p\gamma}n_\gamma $ 
and $\langle E' - E \rangle = \frac 12 E'$, valid
for isotropic pion distribution in c.m.\ system, we obtain
\beq %
n_{cont}(E) = \frac 2{c\sigma_{p\gamma}n_\gamma}Q_0 E^{-\gamma_g} .
\eeq %

Therefore, the asymptotic ratio is
\beq %
\frac{n_{kin}(E)}{n_{cont}(E)} = \frac{\gamma_g}2  , 
\label{asymp-ratio} %
\eeq %
valid also for $\gamma_g =2$. This result has been obtained earlier in
\cite{BGZ}. Of course, at the considered energies nothing but
fluctuations are responsible for ratio (\ref{asymp-ratio}). This ratio,
being asymptotic, can be used, however, as an estimate of role of
fluctuations at  $E >> E_{GZK}$.

Let us now come over to the case of finite $E_{max}$ and consider $E$
very close to $E_{max}$.

The regeneration term in Eq.~(\ref{app-kin}) tends to zero and
\beq %
n_{kin}(E) \rightarrow \frac{Q_0 E_{max}^{-\gamma_g}}{P(E_{max})} ,
\eeq %
having non-zero value at $E=E_{max}-\epsilon$ and being zero at
$E=E_{max}+\epsilon$, where $\epsilon$ is infinitesimally small energy.

For particle density at $E=E_{max}-\epsilon$ in continuous loss
approximation one obtains from Eq.~(\ref{app-CEL})
\beq %
n_{cont}(E_{max}) = \frac 1{b(E_{max})} \int_{E_{max}-E}^{E_{max}} Q_0
E^{-\gamma_g}dE \sim \frac{\epsilon}{E_{max}} \rightarrow 0 .
\eeq %

Therefore, $n_{kin}(E)/n_{cont}(E) \rightarrow \infty$ as $E
\rightarrow E_{max}$.

This result has a clear physical meaning. In our model sources are
homogeneously distributed and from nearby sources the particles with
energy $E_{max}$ can arrive at the observational point with the same
energy due to fluctuations. But in CEL approximation they cannot do it,
since they loose energy at any distance traversed, whatever small it
is. The described effect influences the ratio $r(E) =
n_{kin}(E)/n_{cont}(E)$ at all energies close enough to $E_{max}$.\\
\begin{figure}[ht]
\includegraphics[width=85mm,height=75mm]{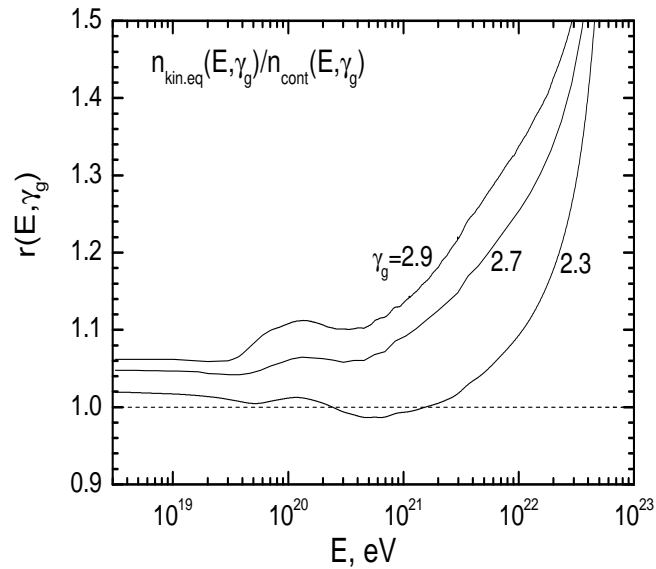}
\caption{Ratio of solutions of kinetic equation (\ref{app-kin}) and
CEL equation (\ref{app-CEL}) for $E_{\rm max}=1\times 10^{23}$~eV as 
a function of
energy in case of realistic $p\gamma$-interaction  and for different
values of generation indices $\gamma_g$, shown by numbers on the
curves. } \label{rat_fl_cont}
\end{figure}
Let us now come over to the solution of Eq.~(\ref{app-kin}) with
interaction given by the detailed calculations as described in Section
\ref{sec:fluctuations}. The calculated
ratio $r(E)=n_{kin}(E)/n_{cont}(E)$ is plotted in
Fig.~\ref{rat_fl_cont} for $E_{\rm max}=1\times 10^{23}$~eV and for
three values of $\gamma_g$ equal to 2.3,~ 2.7 and 2.9. One may observe
in Fig.~\ref{rat_fl_cont} some features described by our toy model: the
ratio $r(E)$ increases with $\gamma_g$ as it should according to
Eq.~(\ref{asymp-ratio}) and it tends to $\infty$ at $E_{max}$. For
spectra of interest for the present work with $2.3 \lsim \gamma_g \lsim
2.8$ the ratio $r(E)$ at  $1\times 10^{19} \leq E \leq 1\times 10^{21}$
eV does not exceed  $10$ \%. Note that in Fig.~\ref{fluct} it is
smaller at low energies because the non-fluctuating pair-production
process is included there.

The ratio of the solutions for kinetic equation and for CEL equation
includes fluctuations, but it includes also some other effects. The
kinetic equation tends to CEL equation (\ref{app-CEL}) in the limit
of small $1-x$. But in fact for all reasonable energies $1-x$ is
limited by  minimum value $m_\pi/(m_\pi+m_p)\approx 0.13$, which cannot
be considered as small value for the discussed accuracy. The same
argument is valid for comparison of Monte-Carlo simulation with CEL
approximation: why they should coincide if the energy transfer process
is discrete?  However, we do see from Fig.~\ref{rat_fl_cont} that the
ratio $r(E)$ differs from 1 just by a few percent, when energy becomes
less than $1\times 10^{19}$~eV.

Appendix~\ref{app-2dip} gives an example when kin/CEL ratio at energy
close to $1\times 10^{20}$~eV is explained not by fluctuations. At
energy $E \approx 6\times 10^{19}$~eV
the  kin/CEL ratio is affected by an interplay between pair- and
photopion production.

We shall now discuss the comparison of our calculations with
Monte-Carlo (MC) simulations.

As was emphasized in Section \ref{sec:fluctuations} the kinetic
equation solution must coincide exactly with MC simulation when
the number of the MC tests tends to infinity, the distribution of
sources is homogeneous, $\gamma_g$ and $E_{\rm max}$ are the same
and interactions are included in the identical way. D. De Marco
and M. Kachelriess run their MC programs specially for comparison
with our calculations using $\gamma_g=2.7$, $E_{\rm max}=1\times
10^{23}$~eV and homogeneous distribution of sources. They used
SOPHIA interaction model \cite{stanev}, which is not identical to
our interaction model especially at the highest energies $E \geq
1\times 10^{21}$~eV, but the average energy losses coincide well
in both models (see Fig.~\ref{loss}). The comparison of MC/CEL and
kin/CEL ratios are given in the Table below. The data of MC are
kindly provided by D. De Marco (D) and M. Kachelriess (K).
\begin{table}[ht]
\begin{center}
\caption{Comparison of kin/CEL and MC/CEL ratios} \vspace{3mm}
\begin{tabular}{c|c|c|c}
\hline
$E$ (eV) & $1\times 10^{20}$ & $1\times 10^{21}$ & $1\times 10^{22}$\\
\hline
$kin/CEL$ & $1.05$ & $1.1$ & $1.3$\\
\hline
$MC/CEL~ (D) $& $1.16$ & $1.16$ & $1.25$\\
\hline
$MC/CEL~ (K) $& $1.1$ & $1.3$ & $ - $ \\
\end{tabular}
\label{r(E)}
\end{center}
\end{table}

From Table~\ref{r(E)} one can see the tendency of smaller kin/CEL
ratios in comparison with MC/CEL, which can be due to different
interaction models used in these calculations. We are planning the
joint work with D. De Marco and M. Kachelriess on the detailed study of
the discussed effects by methods of kinetic equation and Monte-Carlo.

The main conclusion of our study is that fluctuations in photopion
production modify only weakly the  spectra with different $\gamma_g$
at energies up to $1\times 10^{21}$~eV (see Fig.~\ref{rat_fl_cont}).

\section{\label{app-2dip} The second dip}
As explained qualitatively in Section~\ref{sec:2dip} the second dip 
appears  due to breaking of compensation between particle exit and
regeneration given by the first and second terms in
Eq.~(\ref{app-kin}). This is a narrow feature near the energy 
$E_{\rm eq2}$ which is produced because of the sharp (exponential) 
increase of photopion energy losses with energy, seen in Fig.~\ref{loss}.  
At energy $E$ just slightly below $E_{\rm eq2}$ the continuous (adiabatic and 
pair-production) energy losses dominate, and spectrum is universal. At 
energy slightly above $E_{\rm eq2}$ the continuous 
energy losses are small, and spectrum is determined by  
Eq.~(\ref{app-kin}) with the high accuracy compensation between the absorption,
$-P(E)n_p(E)$, and regeneration terms. At $E \sim E_{\rm eq2}$ the continuous 
energy losses break this compensation, increasing the absorption term,
and triggering thus the appearance of the dip. 
We shall study here this {\em triggering mechanism} quantitatively.

Coming back to the basic kinetic equation (\ref{flucteqn}), we
rearrange the terms connected with expansion of the universe
(proportional to $H(t)$) and with pair-production energy losses,
including them in $P_{eff}(E,t)$. We obtain
\begin{equation}
\frac{\partial n_p(E,t)}{\partial t}= -P_{eff}(E,t)n_p(E,t)+\int
\limits_E^{E_{max}} dE' P(E',E,t) n_p(E',t)+ Q_{gen}(E),
\label{app-flucteqn}
\end{equation}
where $P_{eff}(E,t)=P(E,t)+P_{cont}(E,t)$, with $P(E,t)$ given as before
by Eq.~(\ref{P(E,t)}) and
\beq %
P_{cont}(E,t)=2H(t)- \left [\beta_{pair}(E,t)+H(t) \right ] \frac{\partial \ln
n(E,t)}{\partial \ln E}- \frac{\partial b(E,t)}{\partial E}
\label{P-cont} %
\eeq %
As was described above, $P_{cont}(E)$ breaks the compensation between
$-P(E)n_p(E)$ and the regeneration term in Eq.~(\ref{app-flucteqn}), triggering
thus the modification of $n_{kin}(E)$. It is convenient to
introduce the auxiliary {\em trigger function} defined at $t=t_0$ as
\begin{equation}
T(E)=\left\{ \begin{array}{ll}
P_{eff}(E)/P_{cont}(E) ~ & {\rm at}~~ E \leq E_c \\
P_{eff}(E)/P(E)        ~ & {\rm at}~~ E \geq E_c ,
\end{array}
\right.
\end{equation}
with $E_c=6.1\times 10^{19}$~eV, determined from equation $P(E)=P_{cont}(E)$
and being equal to $E_{\rm eq2}$.
\begin{figure}[ht]
\vspace{-1mm}
\begin{minipage}[h]{8cm}
\vspace{-2mm} \centering
\includegraphics[width=74mm,height=71mm,clip]{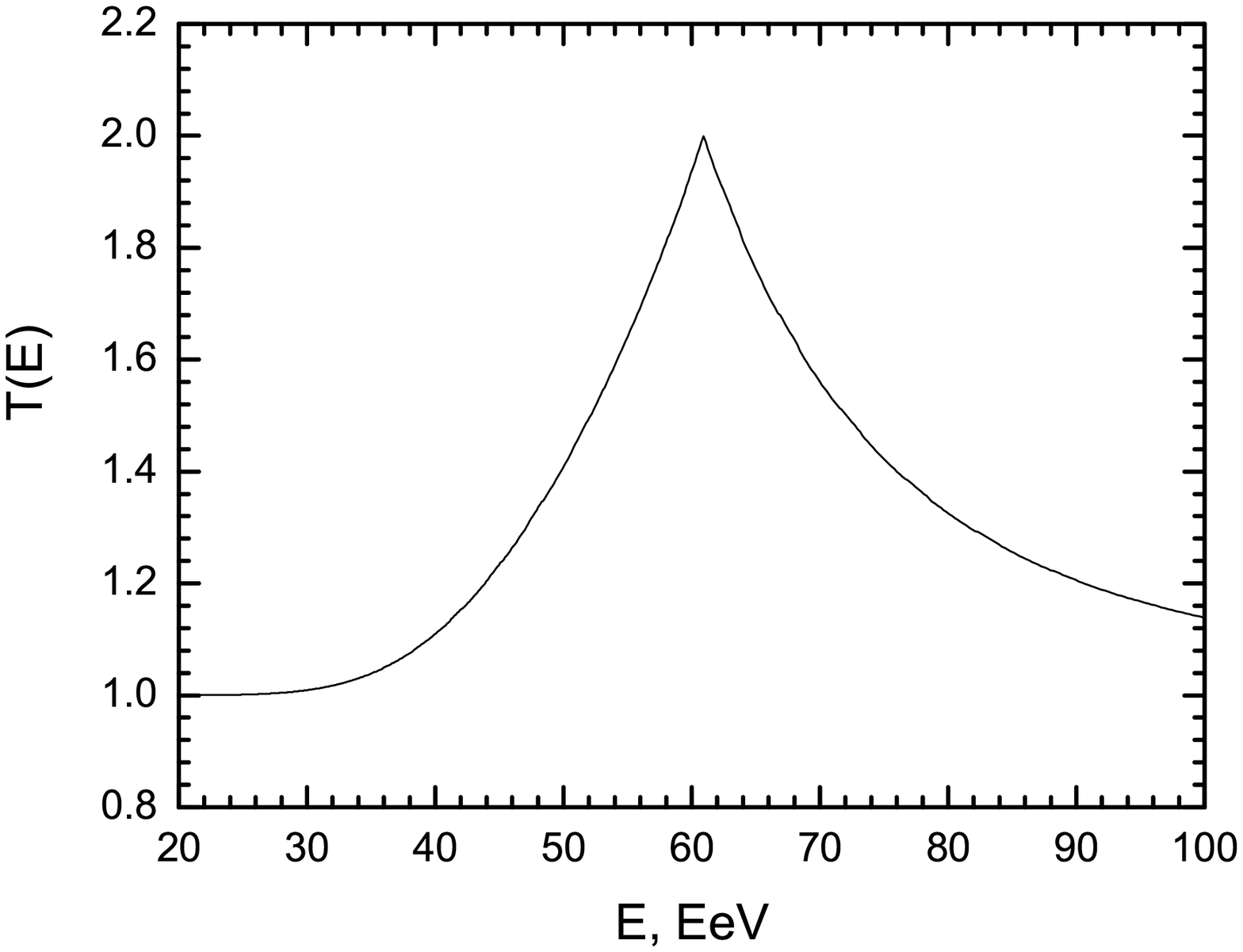}
\end{minipage}
\hspace{5mm}
\begin{minipage}[h]{8cm}
\centering
\includegraphics[width=74mm,height=72mm,clip]{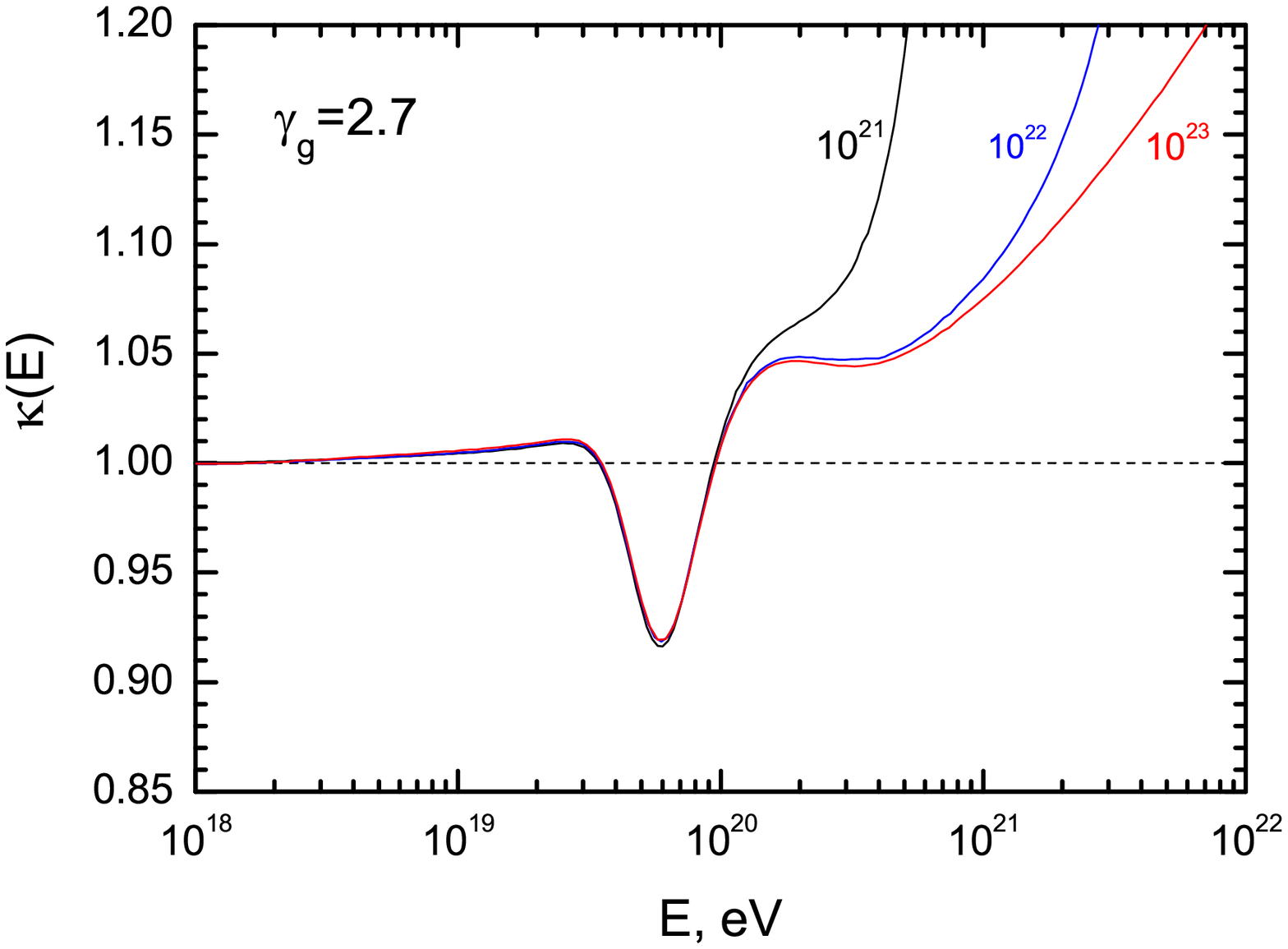}
\end{minipage}
\caption{In the left panel the trigger function $T(E)$ is
displayed as function of energy. In the right panel the ratio 
$\kappa(E)=n_{kin}(E)/n_{cont}(E)$ is plotted for 
three values of $E_{\rm max}$. Proton density $n_{kin}(E)$ is
found as a solution of the total kinetic equation (\ref{flucteqn}),
and $n_{cont}(E)$ is the universal spectrum calculated in the 
continuous energy loss approximation for homogeneous distribution of
the sources. The position of the second
dip, $E=6.3\times 10^{19}$~eV, coincides well with the maximum of
trigger function $E_c=6.1\times 10^{19}$~eV, and widths of both
features are similar.}
\label{fluct-ratio} %
\end{figure}
The trigger function describes how $P_{eff}(E)$ in
Eq.~(\ref{app-flucteqn}) is changing from $P_{cont}(E)$ at $E \ll
E_c$, where $T=1$, to $P(E)$ at $E \gg E_c$, where
$T=1$ as well. The calculated trigger function is plotted
in the left panel of Fig.~\ref{fluct-ratio}. In the calculations
we used for $n_p(E)$ the universal spectrum, because as
Fig.~\ref{fluct} shows, its distortion is small. The trigger
function $T$ reaches 1 at $E \lsim 3\times 10^{19}$~eV, and
therefore at these energies $P_{eff}(E)\approx P_{cont}(E)$;
the regeneration term is small at these energies, too, and thus 
$n_{kin}(E)\approx n_{cont}(E)$ holds. At $E \gsim 1\times 10^{20}$~eV~~ 
$T \approx 1$,~~$P_{eff}(E) \approx P(E)$ and $n_{kin}(E)$ becomes
the solution of Eq.~(\ref{app-kin}) with photopion energy losses
only and with $n_{kin}(E)/n_{cont}(E)$ as shown in
Fig.~\ref{rat_fl_cont}. The maximum deflection of $n(E)$ from
these two regimes reached at $E_c=6.1\times 10^{19}$~eV.  For the
steep spectrum $n(E)$, which is the realistic case, the second
term in Eq.~(\ref{P-cont}) is positive and  large in comparison with 
other terms, $P_{cont}(E)>0$, the
effective exit probability $P(E)+P_{cont}(E)$ increases, and thus
$n_{kin}(E)$ decreases. The triggering mechanism predicts that
$E_c$ does not depend on $\gamma_g$ and $E_{\rm max}$.

The exact calculations for $n_{kin}(E)/n_{cont}(E)$ are shown in the
right panel of Fig.~\ref{fluct-ratio}. In fact, this is just the ratio
of the modification factors from Fig.~\ref{fluct} ('fluct.' and
'cont.loss.' curves), shown with  a strong multiplication in order to
demonstrate the fine structure of this function. One can observe the
narrow dip at $E_{\rm 2dip} = 6.3\times 10^{19}$~eV, which is only
marginally seen in  Fig.~\ref{fluct}. The second dip appears in the
solution of the Fokker-Planck equation, too.

The minimum of the dip coincides well with the  maximum of the
trigger function $E_c=6.1\times 10^{19}$~eV, and the small
difference between them is caused by the fact that $E_c$ is
calculated for $t=t_0$, while for $E_{\rm 2dip}$ the contribution
from epochs with $t<t_0$ is present. The widths of both features
are numerically similar. As the exact calculations show, the
position of the dip does not depend on $\gamma_g$ and $E_{\rm max}$, 
in accordance with triggering mechanism.

We would like to emphasize the similarity of the first and the second dip.
The first dip starts at energy $E_{\rm eq1}$, where pair-production
energy losses become equal to the adiabatic energy losses. The second
dip occurs near energy $E_{\rm eq2}$, where pair-production and 
pion-production energy losses are equal. Both dips are the faint
features, not seen well when measured spectrum is exposed in a natural 
way $J_{\rm obs}(E)$ vs $E$. The first dip is clearly seen in ratio 
of measured spectrum to unmodified spectrum, 
$J_{\rm obs}(E)/J_{\rm unm}(E)$,  the second dip is
predicted to be visible in the ratio of the measured spectrum to the 
smooth universal spectrum,  $J_{\rm obs}(E)/J_{\rm univ}(E)$. The
latter is predicted as ratio $n_{kin}(E)/n_{cont}(E)$ shown in 
Fig.~\ref{fluct-ratio}.

\section{
\label{app-Emax} %
The generation function for case of distribution of sources
over maximal energies} %
We calculate here analytically the generation function
$Q_{gen}(E)$, which gives the number of produced particles per
unit comoving volume of universe and per unit time at $z=0$, in
case the sources are distributed over maximum of acceleration
$\varepsilon \equiv E_{max}$.

We consider first the acceleration by non-relativistic shock,
assuming the power-law spectrum of accelerated particles $\sim E^{-2}$
with exponential suppression $\exp(-E/\varepsilon)$ at $E \geq
\varepsilon $. Namely, we take the generation function for a single
source, $q_{gen}(E)$, which gives the number of particles with energy
$E$ produced per unit time, in the form
\begin{equation}
\label{D1} q_{gen}(E,\varepsilon)=\frac{L_p(\varepsilon)}
{2+\ln(\varepsilon/E_{min})} \varphi_{gen}(E,\varepsilon),
\end{equation}
where $E_{min} \sim 1$ GeV is the minimum energy in acceleration
spectrum, $L_p(\varepsilon)$ is the luminosity of a source in the form
of accelerated particles with energy $E$, and
\begin{equation}
\label{D2}
 \varphi_{gen}(E,\varepsilon)=\left \{
\begin{array}{ll}
E^{-2} & \mbox{~~~~~~~at~~~} E \leq \varepsilon,    \\
\varepsilon^{-2} \exp(1 - E/\varepsilon) & \mbox{~~~~~~~at~~~} E \geq
\varepsilon .
\end{array}
\right.
\end{equation}
Then the generation function per unit volume $Q_g(E)$ is
\begin{equation}
\label{D3} Q_{gen}(E)=\int_{\varepsilon_{min}}^{\varepsilon_{max}}
d\varepsilon
\frac{n_s(\varepsilon)L_p(\varepsilon)}{2+\ln(\varepsilon/E_{min})}
\varphi_{gen}(E,\varepsilon),
\end{equation}
where $n_s(\varepsilon) \equiv n_s[\varepsilon,L_p(\varepsilon)]$ is
the space density of the sources.

We  introduce also the spectral emissivity,
$\mathcal{L}(\varepsilon)= n_s(\varepsilon) L_p(\varepsilon)$ and the
total emissivity $\mathcal{L}_0 = \int \mathcal{L} (\varepsilon) d
\varepsilon$.

$Q_{gen}(E)$ can be evaluated from Eqs.~(\ref{D3}) and (\ref{D2})
as follows: \\
for $E < \varepsilon_{min}$,
\begin{equation}
\label{D4} Q_{gen}(E) \sim E^{-2} ;
\end{equation}
for $\varepsilon_{min} \leq E \leq \varepsilon_{max}$ the main
contribution to $Q_{gen}(E)$ is given by
\begin{equation}
\label{D5}
 Q^{(1)}_{gen}(E) \sim E^{-2} \int_E^{\varepsilon_{max}}
d\varepsilon \mathcal{L}(\varepsilon),
\end{equation}
while the integral from $\varepsilon_{min}$ to $E$ results in
\begin{equation}
\label{D6}
 Q^{(2)}_{gen}(E) \sim E^{-1} \mathcal{L}(E) ;
\end{equation}
and for $E>\varepsilon_{max}$
\begin{equation}
\label{D7} Q_{gen}(E) \sim \exp(-E/\varepsilon_{max}).
\end{equation}

The physically interesting regime is given by $\varepsilon_{min} < E <
\varepsilon_{max}$. To reproduce $Q_{gen}(E) \sim E^{-2.7}$ needed for
the best fit of the dip, $\mathcal{L}(\varepsilon)$ from (\ref{D5}) and
(\ref{D6}) must be power-law function $\sim E^{-\beta}$ with
$\beta=1.7$.

In this case we obtain
\begin{equation}
\label{D8} Q_{gen}(E) \sim \left \{
\begin{array}{ll}
E^{-2} & \mbox{~~~~~~~at~~~}  E < \varepsilon_{min}    \\
E^{-2.7} & \mbox{~~~~~~~at~~~}  \varepsilon_{min} \leq E \leq \varepsilon_{max}    \\
\exp(-E/\varepsilon_{max})  & \mbox{~~~~~~~at~~~}  E >
\varepsilon_{max},
\end{array}
\right.
\end{equation}
which coincides with the phenomenological broken generation spectrum 
(\ref{broken}).

The function $Q_{gen}(E)$ from (\ref{D8}) can be easily normalized
using the condition $\int E Q_{gen}(E) dE = \mathcal{L}_0$.

One may generalize this derivation for the case
\begin{equation}
\label{D9} \varphi_{gen}(E,\varepsilon) = \left \{
\begin{array}{ll}
E^{-(2+\alpha)} & \mbox{~~~~~~~at~~~}  E < \varepsilon   \\
E^{-(2+\alpha)} f(\frac E{\varepsilon}) & \mbox{~~~~~~~at~~~} E \geq
\varepsilon,
\end{array}
\right.
\end{equation}
where $f(x)$ is a function which diminishes with $x$ like $e^{-x}$ or
faster and $f(1)=1$.

Similar to (\ref{D1}) and (\ref{D3}) we have
\begin{eqnarray}
\label{D10} q_{gen}(E,\varepsilon) & = & \alpha E_{min}^\alpha
L_p(\varepsilon) \varphi_{gen}(E,\varepsilon), \\
\label{D11} Q_{gen}(E,\varepsilon)& = & \alpha E_{min}^\alpha \int_{
\varepsilon_{min}}^{ \varepsilon_{max}} d \varepsilon
\mathcal{L}(\varepsilon) \varphi_{gen}(E,\varepsilon).
\end{eqnarray}

Using the properties of $f(x)$ we can prove again that the power-law
function $Q_{gen}(E) \sim E^{-\gamma_g}$ in the interval
$\varepsilon_{min} < E < \varepsilon_{max}$ needs the power-law
function $\mathcal{L}(\varepsilon) \sim \varepsilon^{-\beta}$.

Then using again the properties of $f(x)$ we obtain
\begin{equation} %
\label{D12} %
Q_{gen}(E) \sim \left \{
\begin{array}{ll} %
E^{-(2+\alpha)} & \mbox{~~~~~~~at~~~}  E \leq \varepsilon_{min}    \\
E^{-(1+\alpha+\beta)} & \mbox{~~~~~~~at~~~}  \varepsilon_{min} \leq E \leq \varepsilon_{max} \\
f(E/\varepsilon_{max})  & \mbox{~~~~~~~at~~~}  E > \varepsilon_{max},
\end{array} %
\right.
\end{equation}

We are interested in particular in case of acceleration at relativistic
shock, when $\alpha \approx 0.2 - 0.3$. The case of $\gamma_g =
1+\alpha+\beta = 2.7$ results then in $\beta = 1.7 - \alpha \approx 1.4
- 1.5$.

\bibliography{abbrev,uhecr}
\bibliographystyle{apsrev}

\end{document}